\documentclass[reprint,10pt, twocolumn, superscriptaddress, notitlepage,nofootinbib, floatfix, longbibliography]{revtex4-1}
\usepackage{amsmath,amsfonts, amssymb, amsthm, dsfont}
\usepackage[T1]{fontenc}
\usepackage{yfonts}
\usepackage{bm}
\usepackage{mathrsfs}
\usepackage{graphicx}
\usepackage{verbatim}
\usepackage[bookmarks=true,colorlinks=true,linkcolor=blue, urlcolor=blue, citecolor=blue]{hyperref}
\hypersetup{linktocpage}
\usepackage{wasysym}
\usepackage{xcolor}
\usepackage{bbold}
\usepackage{braket}
\usepackage{diagbox} %diagonal slash through cells in a table
\usepackage{environ} %define eqs
\usepackage{relsize} % increase the size of math symbols with \mathlarger
\usepackage{quiver} %Commutative diagram maker
\usepackage[caption=false]{subfig} %subfigs

\NewEnviron{eqs}{%
\begin{equation}\begin{split}
    \BODY
\end{split}\end{equation}
}

\newcommand{\comments}[1]{}
\newcommand{\mb}[1]{\mathbf{#1}}
\renewcommand{\cal}[1]{\mathcal{#1}}

\newcommand{\op}[1]{\ket{#1}\!\bra{#1}}

\renewcommand{\set}[1]{\{#1\}}
\renewcommand{\op}[1]{\ket{#1}\!\bra{#1}}

\def\U{\mathrm{U}(1)}
\def\H{\mathcal{H}}
\def\Z{\mathbb{Z}}

\def\C{\mathcal{C}}

\def\cZ{\mathcal{Z}}

\def\id{\mathds{1}}

\DeclareMathOperator{\Tr}{Tr}
\DeclareMathOperator{\Rep}{Rep}

\theoremstyle{plain}
\newtheorem*{theorem*}{Theorem}
\newtheorem{theorem}{Theorem}

\newtheorem{definition}{Definition}%[section]

\definecolor{tyler}{rgb}{1,.3,0}

\newcommand{\add}[1]{{{#1}}}

\makeatletter
\def\l@subsubsection#1#2{}
\makeatother

\newcommand{\AddFig}[1]{\vcenter{\hbox{\includegraphics[scale=1]{#1}}}}

\begin{document}

\title{Towards a classification of mixed-state topological orders in two dimensions}
\author{Tyler D. Ellison}
\author{Meng Cheng}
\affiliation{Department of Physics, Yale University, New Haven, Connecticut  06511-8499, USA}

\date{\today}

\begin{abstract}
The classification and characterization of topological phases of matter is well understood for ground states of gapped Hamiltonians that are well isolated from the environment. However, decoherence due to interactions with the environment is inevitable – thus motivating the investigation of topological orders in the context of mixed states. Here, we take a step toward classifying mixed-state topological orders in two spatial dimensions by considering their (emergent) generalized symmetries. We argue that their 1-form symmetries and the associated anyon theories lead to a partial classification under two-way connectivity by quasi-local quantum channels. This allows us to establish mixed-state topological orders that are intrinsically mixed, i.e., that have no ground state counterpart. We provide a wide range of examples based on topological subsystem codes, decohering $G$-graded string-net models, and ``classically gauging’’ symmetry-enriched topological orders. One of our main examples is an Ising string-net model under the influence of dephasing noise. We study the resulting space of locally-indistinguishable states and compute the modular transformations within a particular coherent space. Based on our examples, we identify two possible effects of quasi-local quantum channels on anyon theories: (1) anyons can be incoherently proliferated – thus reducing to a commutant of the proliferated anyons, or (2) the system can be ``classically gauged'', resulting in the symmetrization of anyons and an extension by transparent bosons. Given these two mechanisms, we conjecture that mixed-state topological orders are classified by premodular anyon theories, i.e., those for which the braiding relations may be degenerate. 
\end{abstract}

\maketitle

\tableofcontents

\section{Introduction}
Quantum many-body systems showcase a remarkably diverse range of quantum phases of matter. The ground states of gapped Hamiltonians, in particular, can exhibit topological order (TO), where the wave function is long-range entangled and cannot be smoothly deformed into a product state without encountering a phase transition. TO leads to a variety of intriguing phenomena, including localized excitations with unusual braiding statistics, and topologically-protected ground state degeneracies on a torus. These features endow TO with intrinsic robustness against local perturbations and make them of great promise for applications in fault-tolerant quantum information processing. 

In the past two decades, significant progress has been made in classifying and characterizing TOs in gapped ground states~\cite{Kitaev:2005hzj, Wen_2015, Johnson_Freyd_2022, WenRMP}. 
\add{By now, mathematical frameworks have been established to classify TOs in (2+1)$d$: it is widely accepted that TOs are completely determined (up to invertible phases of matter) by their associated anyon theories, which capture the universal properties of the localized quasiparticle excitations. }

However, the majority of the existing studies~\cite{wen2004quantum, simon2023topological, zeng2019quantum} assume the TO is in a well-isolated system, and thus is described by pure states. 
In reality, a physical system is influenced by its environment and is best captured by a mixed state. This is particularly relevant for applications in quantum information, as the resilience to environmental noise is a key requirement for a quantum memory. Therefore, understanding TOs in mixed states is of both fundamental importance and a timely issue. 

It is well-understood that TOs in (2+1)$d$ are not stable against coupling to thermal baths~\cite{HastingsFiniteT, 
CastelnovoPRB2007, Nussinov2008thermalselfcorrecting, Lu:2019owx, Sang:2023rsp}: they can be smoothly connected to infinite-temperature Gibbs states without undergoing any thermal phase transition. This agrees with the strong belief that there is no self-correcting quantum memory at finite temperature in (2+1)$d$~\cite{Poulin2013thermal, Brown:2014idi}. On the other hand, the very fact that a topological quantum memory can exist suggests~\cite{Dennis:2001nw} that TOs are robust against local noise, and thus, should be well-defined for mixed states. 

A useful theoretical setup to investigate these problems is a many-body ground state subject to quasi-local quantum channels (QLCs) and measurements. Examples recently studied in this kind of setup include quantum critical states~\cite{Garratt:2022ycp,Yang:2023dol,Sun:2023alk, Lee:2023fsk,Zou:2023rmw, Ma:2023tmy}, symmetry-protected topological (SPT) phases~\cite{deGroot2022, Lee:2022hog, Ma:2022pvq, ZhangQiBi2022, Moligini2023topological,
 Ma:2023rji, Ma:2024kma, guo2024locally, Xue:2024bkt}, and topological states~\cite{Fan:2023rvp,Bao:2023zry,  Wang:2023uoj, Chen:2023vxo, Li:2024rgz}. In general, it has been found that decoherence can lead to distinct mixed-state phases of matter. For example, it was shown in Refs.~\cite{Bao:2023zry, Li:2024rgz} that there are distinct error-induced phases that emerge from noisy TOs, which can be characterized by different topological boundary conditions in the replicated Hilbert space representation.  

\subsection{Summary of main results}

In this work we systematically study mixed-state TOs in two dimensions arising from decohering ground states of gapped Hamiltonians and develop a general framework with the goal of classifying TOs in mixed states.

We begin by giving a working definition of mixed-state TO in Section~\ref{sec: generalities}, i.e., we specify the class of mixed states considered in this work and define an equivalence relation on them such that the equivalence classes correspond to distinct mixed-state phases. We comment on the fact that, similar to ground state TOs, mixed-state TOs exhibit locally indistinguishable states on manifolds of nontrivial topology. We review the toric code (TC) under bit-flip noise, as a first example.  

We then consider general topological Pauli stabilizer states subject to Pauli noise in Section~\ref{Pauli}. We show that the theory of subsystem codes provides a natural framework for studying such systems. We define the associated subsystem code by a ``gauge group'', which is generated by the original stabilizer group and the noise operators. In the limit of maximal decoherence, we show that the effect of noise is to completely decohere the gauge subsystem of the subsystem code, leaving the logical subsystem intact. 

We focus on a special class of Pauli noise, for which the associated subsystem code is topological (in the sense of Ref.~\cite{Ellison:2022web}). Such mixed states are associated with an Abelian anyon theory, which intuitively speaking, describes the ``strong'' 1-form symmetries of the mixed state. Unlike the ground state case, the Abelian anyon theory is not required to be modular, i.e., it may possess nontrivial anyons that braid trivially with all other anyons. Such anyon theories are said to be ``premodular''. This suggests a classification of mixed-state TOs that is more diverse than the pure state classification for gapped ground states. We study how the anyon theory is affected by a QLC, which leads to an algebraic equivalence relation between premodular Abelian anyon theories. We further define a topological invariant, which gives a partial classification of mixed-state TOs. 

In Section~\ref{sec: stringnet}, we move beyond the Pauli stabilizer formalism and discuss mixed-state TOs characterized by non-Abelian anyon theories. We start by considering mixed states constructed from non-Abelian string-net models by adding local noise. Our primary example is a mixed state constructed from the Ising string-net model by incoherently proliferating bosons. The state is characterized by a strong 1-form symmetry associated to an anyon theory that is both non-Abelian and non-modular.  

We generalize the construction to string-net models with a $G$-graded fusion category as an input. We then further extend the result in Section~\ref{sec: classical gauging} to symmetry-enriched TOs (which may or may not admit a string-net model), and build a mixed state by ``classically'' gauging the symmetry. 
In Sections~\ref{sec: walkerwang}, we give the most general construction of mixed states based on a premodular anyon theory, using the Walker-Wang model. Finally, in Section~\ref{sec: general intrinsic} we discuss algebraic equivalence relations among premodular anyon theories induced by QLCs, and comment on the resulting mixed-state TOs that have no pure state counterpart, i.e., that are intrinsically mixed-state TOs. 

\section{Generalities} \label{sec: generalities}

\subsection{Locally-correlated mixed states}

To define TO in the context of pure states, we restrict ourselves to the ground states of gapped local Hamiltonians, which we refer to as gapped ground states (GGSs). It is believed that a GGS encodes all of the characteristic data of the TO, including the universal behavior of the localized excitations of a parent Hamiltonian (e.g. the fusion and braiding of the excitations).\footnote{Excited states with zero energy density can be described in terms of these localized excitations, which are only weakly coupled, and can be thought of as ground states of the same Hamiltonian but with ``pinning potentials''. We consider these states as GGS as well.} 
For the purpose of defining TO, we also restrict to GGS that are short-range correlated, i.e., the connected correlator of any pair of local operators decays rapidly with their separation. In particular, this rules out long-range correlated states associated with spontaneous symmetry breaking (e.g. the GHZ states).

\begin{figure}[t] 
\centering
\includegraphics[width=.45\textwidth]{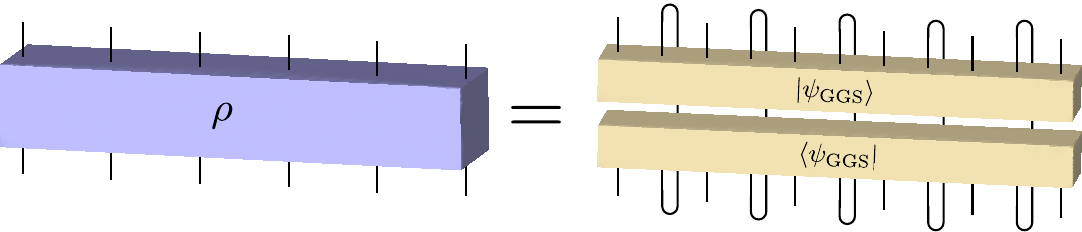}
\caption{Graphical representation of a purification into a~GGS. We restrict the discussion to mixed states that can be purified into GGSs, i.e., $\rho = \Tr_A[\op{\psi_{\rm GGS}}]$, for some GGS $\ket{\psi_{\rm GGS}}$ and subsystem $A$. We also require that $\rho$ has short-ranged R\'enyi-1 and -2 correlations, according to Definition~\ref{def: short range renyi correlations}. We refer to states that satisfy these properties as locally correlated mixed states.}
\label{fig: GGSpurification}
\end{figure}

For a generic mixed state, there is no clear notion of a parent Hamiltonian. Therefore, the class of mixed states that should be considered in defining TOs is more subtle. Inspired by short-range correlated GGSs, in this work, we consider mixed states with the following two properties: (1) they can be purified into a GGS, as shown in Fig.~\ref{fig: GGSpurification}, and (2) they have local correlations. 

Below, we clarify the sense in which the mixed states are required to have local correlations. 
We begin by defining the R\'enyi-1 and R\'enyi-2 expectation values.
\begin{definition}
    For a mixed state $\rho$ and an operator $M$, the R\'enyi-1 and R\'enyi-2 expectation values are
\begin{align}
    E^{(1)}_\rho(M) &= \Tr[M\rho]\\
    E^{(2)}_\rho(M) &= \frac{\Tr[M\rho M^\dagger \rho]}{\Tr[\rho^2]}.
\end{align}   
\end{definition}

\noindent Here, the R\'enyi-1 expectation value is the usual expectation value. The R\'enyi-2 expectation value, on the other hand, can be understood using the Choi-Jamiołkowski representation of $\rho$, defined in the doubled Hilbert space. From this perspective, the expectation value $E_\rho^{(2)}(M)$ is the ordinary expectation value of $M\otimes M^\dag$ for the doubled state. We also point out that, if $\rho=\op{\psi}$ is a pure state, then the R\'enyi-2 expectation value reduces to $|\braket{\psi|M|\psi}|^2$. In a similar fashion, one can define a R\'enyi-$n$ expectation value for $n$ replicas of the Hilbert space.

We can now define the connected correlators that correspond to the R\'enyi-$n$ expectation values, for $n=1,2$.
\begin{definition}
    For a mixed state $\rho$ and operators $M_i$ and $M_j$, the R\'enyi-$n$ connected correlator ($n =  1,2$) is
    \begin{align}
    C^{(n)}_\rho(M_i,M_j) &= E^{(n)}_\rho(M_iM_j) - E^{(n)}_\rho(M_i)E^{(n)}_\rho(M_j).
\end{align}
Similarly one can define fidelity connected correlator using Eq.~\eqref{eq: fidelity}.
\end{definition}

\noindent The R\'enyi-2 correlator can again be interpreted as an ordinary connected correlator within the doubled Hilbert space. 

Finally, we can define R\'enyi-$n$ locally-correlated mixed states, for $n=1,2$. 
\begin{definition} \label{def: short range renyi correlations}
    A mixed state $\rho$ is R\'enyi-$n$ locally correlated ($n=1,2$), if for any operators $M_i$ and $M_j$ localized near the sites $i$ and $j$, we have
    \begin{align}
        C^{(n)}_\rho(M_i,M_j) &= O(|i-j|^{-\infty}),
    \end{align}
    where $O(|i-j|^{-\infty})$ is a function that decays faster than any power law in $|i-j|$. 
\end{definition}

\noindent Note that R\'enyi-1 locally correlated states are short-range correlated states in the usual sense. 

More formally, we define mixed-state TOs in this work in terms of mixed states $\rho$ with the following three properties:
\begin{enumerate}
    \item $\rho$ can be purified into a GGS.
    \item $\rho$ is R\'enyi-1 locally correlated.
    \item $\rho$ is R\'enyi-2 locally correlated.
\end{enumerate}
In a slight abuse of nomenclature, we refer to these mixed states simply as ``locally-correlated mixed states''.

The first condition generalizes the notion of short-range entangled (SRE) mixed states proposed in Ref.~\cite{Ma:2022pvq} (see also Ref.~\cite{Ma:2023rji}). Namely, a mixed state is SRE, if there is a purification into a SRE GGS. Here, we require that topologically ordered mixed states can be purified into GGSs more generally. The second condition rules out spontaneous symmetry breaking and long-range correlations, similar to the case for pure-state TOs. 

The third condition is motivated by recent progress in understanding spontaneous symmetry breaking order in mixed states. In particular, it was found that observables that are nonlinear in the density matrix are necessary to characterize certain phases and phase transitions in mixed states~\cite{Lee:2022hog, Bao:2023zry, Ma:2023rji}. For example, the phenomenon known as strong-to-weak symmetry breaking can be detected by long-range order in the R\'enyi-2 correlations of local order parameters~\cite{Lee:2022hog, Ma:2023rji, Lessa:2024gcw}.

We emphasize that this notion of locally-correlated mixed states allow us to give a working definition of TO. We do not claim that the conditions on mixed states above are the most exhaustive or the most general. Ultimately, one may want to consider a class of mixed states with no reference to Hamiltonians or replica Hilbert spaces. We comment further on this point in Section~\ref{sec: discussion}. \add{It also may be advantageous to replace the R\'enyi-2 expectation values with those based on the fidelity:}
\begin{align} \label{eq: fidelity}
    \add{E^{F}_\rho(M)=F(\rho, M\rho M^\dagger) = \Tr\Big[\sqrt{\sqrt{\rho} M\rho M^\dagger \sqrt{\rho}}\Big]}.
\end{align}
\add{The fidelity expectation value was proposed in the context of strong-to-weak spontaneous symmetry breaking in Ref.~\cite{Lessa:2024gcw} and was shown to correctly capture the decoding transition in the bit-flip decohered toric code state. However, in general the fidelity correlations are much more difficult to compute, so we will use R\'enyi-2 correlators mostly in this work.}

We note that a broader class of mixed states are those that can be decomposed into a convex sum of pure GGSs. All the examples of mixed states considered in this work can be represented as such a convex sum, but the converse is not true.  The simplest counterexample is the thermal state of a classical Hamiltonian tuned to a finite-temperature critical point. 
Such a state contains purely classical long-range correlations and thus can not be purified into a GGS. Even assuming that correlation functions of local operators are all short-range, we can still find fully separable mixed states which do not admit a purification into a GGS~\cite{Ma:2023rji}.

An interesting question is whether a thermal state is locally correlated. Since a thermal state can always be purified into a thermofield double state, the question becomes whether the thermofield double state is the ground state of a gapped local Hamiltonian. To the best of our knowledge, the general case remains open, although Ref.~\cite{Cottrell:2018ash} proposed parent Hamiltonians for thermofield double states and presented evidence that the Hamiltonians are (quasi-)local.    
It was shown in Ref.~\cite{Lucia:2021orn} that thermofield double states for 2D Kitaev's quantum double models are SRE. The same is true for thermal states of 1D local Hamiltonians. 

\subsection{Equivalence relation on mixed states}
\label{sec:equiv relation}

For ground states, a gapped phase is defined as an equivalence class of short-range correlated GGS, where the equivalence relation is given in terms of quasi-local unitary circuits (QLUCs), with at most $\rm polylog$ depth in the system size. Namely, two GGSs belong to the same phase if and only if they can be mapped to each other by a QLUC. Here, QLUC serves as a model for quasi-adiabatic evolution generated by a gapped local Hamiltonian.

\begin{figure}[t] 
\centering
\includegraphics[width=.45\textwidth]{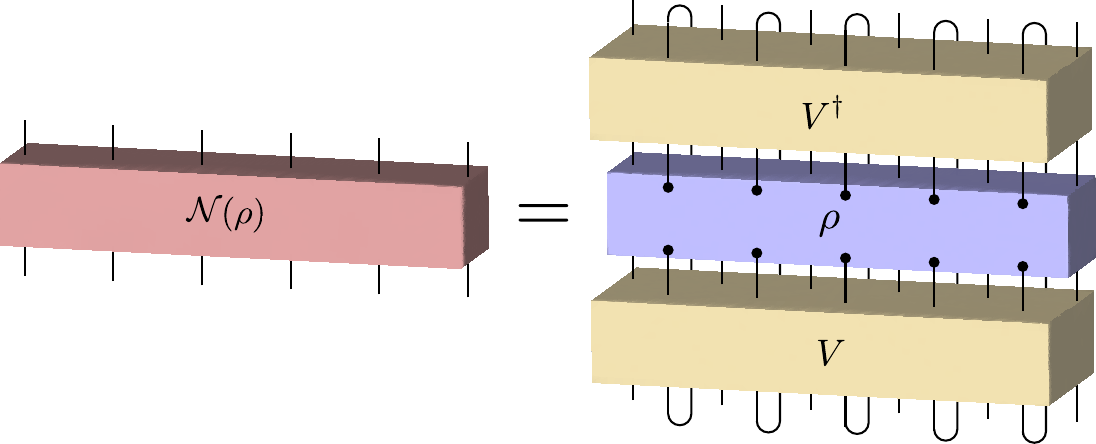}
\caption{Graphical representation of a QLC $\cal{N}$. The action of the QLC $\cal{N}$ on a density matrix $\rho$ is equivalent to, in sequence, adding ancilla to $\rho$, conjugating by a QLUC $V$, and tracing out the ancilla (see Definition~\ref{def: QLC}).}
\label{fig: QLC}
\end{figure}

A natural generalization of QLUCs to mixed states is a quasi-local quantum channel (QLC), e.g., a finite time evolution generated by a local Lindbladian. However, because quantum channels are in general non-invertible, in order to define an equivalence relation, it becomes necessary to consider two-way connectedness by QLCs, which we take as the definition of mixed state phase~\cite{Ma:2022pvq, Sang:2023rsp}. Below, we first formalize the definition of a QLC (depicted in Fig.~\ref{fig: QLC}).
\begin{definition} \label{def: QLC}
A quantum channel $\mathcal{N}$ is a QLC if it can be purified into a circuit $V$, whose depth scales at most as $\mathrm{polylog}(L)$ with the linear system size $L$, acting on $\mathcal{H} \otimes \mathcal{H}_A$. Here, $\mathcal{H}$ and $\mathcal{H}_A$ are the physical and ancillary Hilbert spaces, respectively. 
The action of $\cal{N}$ on a mixed state $\rho$ is thus given by $\Tr_\add{B}[V^\dagger (\rho \otimes | 0 \rangle \langle 0 | )V]$, where $| 0 \rangle$ represents a many-body product state in $\mathcal{H}_A$, \add{and $B$ is some subsystem, which may include $A$}. 
\end{definition}

Following Refs.~\cite{Ma:2022pvq, Ma:2023rji, Sang:2023rsp}, we define mixed-state TOs using the following equivalence relation:  
\begin{definition}
    Two locally-correlated mixed states $\rho_1$ and $\rho_2$ are equivalent, or belong to the same mixed-state TO, if and only if they are two-way connected by QLCs. Namely, there exists two QLCs $\mathcal{N}_{12}$ and $\mathcal{N}_{21}$ such that $\rho_1=\mathcal{N}_{12}(\rho_2)$ and $\rho_2=\mathcal{N}_{21}(\rho_1)$.
\end{definition}
\noindent We point out that this definition is closely related to the one in Ref.~\cite{Coser2019classificationof}, which essentially amounts to replacing QLCs with fast evolution by local Lindbladians.  That is, evolution for time that grows sub-linearly (e.g. polylog) with the system size. \add{We also note that, due to the definition of a QLC, one is free to add or remove ancilla without changing the mixed-state TO.}

% \add{Adding ancillas}

It is natural to define the trivial phase as the unique equivalence class containing the product states. 
In all known examples, a trivial mixed state can be written as a convex sum of SRE states, however the converse is not necessarily true. According to this definition, the maximally mixed state also belongs to the trivial phase. It can be constructed from a product state by applying depolarizing noise and the product state can be constructed from it by tracing it out and tensoring with the product state. Similar to the ground state case, we are allowed to freely stack trivial states, as adding unentangled ancilla is part of the definition of QLCs.
We also note that all bosonic invertible GGSs, e.g., the $E_8$ state in (2+1)$d$, belong to the trivial mixed-state phase~\cite{Ma:2022pvq}. This in particular means that the chiral central charge is no longer a well-defined invariant for mixed state TOs.

Let us conclude the section with some physical examples of QLCs. 
\add{An important class of examples is the time evolution of open quantum systems governed by Lindbladian equations. If the Lindbladian is local, then finite-time evolution under the Lindbladian corresponds to a QLC on the initial state. }

\add{We also point out that, intuitively, one expects that thermal states of a local Hamiltonian at different positive temperature belong to the same mixed-state phase if there is no thermal phase transition in between. This has been proven in 1D. Namely, all thermal states of local Hamiltonians in 1D with positive temperature can be two-way connected by QLCs~\cite{Kato:2016pgk}.}

\add{It is also natural to expect that a topological code with a small amount of noise, which can be modeled by applying a finite-depth quantum channel close to the identity to the pure state, belongs to the same phase as the pure state. Thus there must be a ``recovery channel" to map the decohered state back to the pure state. The recovery channel should be a QLC in order for the code to be error-correctable. This has been demonstrated explicitly in the example of a $\Z_2$ TC with bit-flip noise in \cite{Sang:2023rsp} (see also \cite{Sang:2024vkl} and \cite{Bauer:2024qpc}). }

\subsection{Locally indistinguishable states}

A hallmark of TO for GGSs is that there is a topological degeneracy when the system is put on a torus -- with the dimension of the ground state subspace being equal to the number of anyon types. The degeneracy for a higher-genus surface can also be determined from the anyon theory. Moreover, the ground states are locally indistinguishable, meaning that any two ground states $\ket{\psi_1}$ and $\ket{\psi_2}$ have the property 
\begin{equation}
 \braket{\psi_1|M|\psi_1}-\braket{\psi_2|M|\psi_2}=O(L^{-\infty}),  
\end{equation}
for any quasi-local operator $M$. Here, $L$ is the system size, and $O(L^{-\infty})$ is a function that decays faster than any power law of $L$, e.g. $e^{-(L/\xi)^\alpha}$ for any $\alpha>0$.

The notion of local indistinguishability naturally generalizes to mixed states~\cite{Nussinov2009thermalTQO}. We say that two mixed states $\rho_1$ and $\rho_2$ are locally indistinguishable, if they satisfy 
\begin{eqs} \label{eq: mixed indistinguishability}
    \Tr[M \rho_1] - \Tr[M\rho_2 ] = O(L^{-\infty}),
\end{eqs}
for any quasi-local operator $M$.  

In contrast to the pure-state case, the collection of locally indistinguishable mixed states do not form a vector space. Rather 
they form a convex manifold. As pointed out in Ref.~\cite{Li:2024rgz}, it is insightful to consider the extremal submanifold, i.e., the submanifold of extremal points. In general, this submanifold contains several connected components. Each connected component can have one of the following two possibilities:
\begin{enumerate}
    \item It is a single point. In this case, the state is completely ``classical''.
    \item There is a continuum of extremal points forming a connected manifold of dimension $d$. Physically, this manifold should be isomorphic to the manifold of pure states in a $d$-dimensional Hilbert space. We refer to this space as a ``coherent space'' of dimension $d$. Note that an isolated extremal point can be thought of as a 0-dimensional coherent space.
\end{enumerate}

We also note that two locally indistinguishable states $\rho_1$ and $\rho_2$ remain so under an arbitrary QLC. Explicitly, for an arbitrary QLC $\cal{N}$ and a quasi-local operator ${M}$, we can compute 
\begin{eqs} \label{eq: SDLQC indistinguishability}
    \Tr[M \cal{N}(\rho_1)] &= \Tr[\cal{N}^*(M) \rho_1 ],\\
                           &= \Tr[\cal{N}^*(M) \rho_2 ], \\
                           &= \Tr[M \cal{N}(\rho_2)],
\end{eqs}
where $\cal{N}^*$ is the dual channel,\footnote{If the representation of $\cal{N}$ in terms of Kraus operators is $\cal{N}(\rho) = \sum_i K_i \rho K_i^\dagger$, then the action of the dual channel on an operator $M$ is $\cal{N}^*(M) = \sum_i K_i^\dagger M K_i$.} 
which preserves quasi-locality.
Therefore, the states $\cal{N}(\rho_1)$ and $\cal{N}(\rho_2)$ are also locally indistinguishable for any QLC $\cal{N}$. 

However, the discussion above does not mean that the convex manifold of locally indistinguishable states must be invariant under QLCs, for two reasons. First, two locally distinguishable states may become indistinguishable under the QLC. This can happen, for example, if $\cal{N}^*(M)=0$ for all quasi-local operators $M$ that distinguished the two states. An example that illustrates this point is discussed in Section~\ref{sec: decohered TC}. On the other hand, it can also happen that two different states that are locally indistinguishable become identical under a QLC. A simple example is a swap channel that takes any (pure state) TO to a trivial product state, under which the space of locally indistinguishable states is completely erased. In either case, the QLC is degenerate, i.e., it has a nontrivial kernel when viewed as a linear map on the space of operators.

In fact, the convex manifold of locally indistinguishable states is not an invariant for a mixed-state phase, as illustrated by the example in Section~\ref{sec: decohered TC}. However, mixed-state TOs still give rise to coherent spaces of locally indistinguishable states on closed oriented manifolds, which depend on the topology of the manifold -- similar to ground state TOs. We conjecture that there is a subspace within the coherent space that is robust to perturbations of the mixed state, and that this subspace provides an invariant for the mixed-state phase.

Lastly, one could consider the R\'enyi-2 (R\'enyi-$n$) expectation values to distinguish between states. It is possible that these higher-R\'enyi expectation values are able to distinguish between states that are otherwise locally indistinguishable according to the R\'enyi-1 expectation values. One can thus formulate different notions of local indistinguishability based on observables that are nonlinear in $\rho$, as recently proposed in Ref.~\cite{Li:2024rgz}. Although, we do not consider such notions of local indistinguishability any further in this work.

\subsection{Example: decohered 2D toric code} \label{sec: decohered TC}

To exemplify the general discussion above, we consider decohering a (2+1)$d$ TC state with bit-flip noise~\cite{Dennis:2001nw, Fan:2023rvp}. We recall that the TC state is defined on a square lattice with a qubit on each link, and with stabilizers given by 
\begin{equation}
    A_v=\prod_{v\in e}X_e, \quad B_p = \prod_{e\in p}Z_e.
\end{equation}
A ground state $\ket{\psi}$ is defined by $A_v\ket{\psi}=\ket{\psi}, B_p\ket{\psi}=\ket{\psi}$ for all $v$ and $p$. The corresponding density matrix is denoted by $\rho_0=\op{\psi}$. Following the common convention, we call a vertex violation $A_v=-1$ an $e$ particle at the vertex $v$, and a plaquette violation $B_p=-1$ an $m$ particle in the plaquette $p$.

Now, suppose that the TC is subjected to noise described by the bit-flip channel $\rho_X=\cal{N}_X(\rho_0)$, defined as 
\begin{equation} \label{eq: NX}
    {\cal N}_X = \bigotimes_e {\cal N}_{X,e}, \quad {\cal N}_{X,e}(\rho)=(1-p)\rho + pX_e \rho X_e.
\end{equation}
Here, $p$ parameterizes the strength of the decoherence and satisfies $0\leq p\leq 1/2$. For small $p$, the TO still persists up to some critical value $p=p_c\approx 0.109$. For $p>p_c$, the system enters a new error-induced phase where the TO is lost. This transition can be characterized using quantum information-theoretic measures~\cite{Fan:2023rvp}. For example, for small $p$ we can view the TC as a quantum memory that encodes two logical qubits on a torus. For $p>p_c$, no coherent quantum information can be stored on a torus anymore. The same transition can be detected by topological entanglement negativity, which is $\ln 2$ before the transition and $0$ after~\cite{Fan:2023rvp}.  

To understand the error-induced phase, it is particularly instructive to consider the strong-decoherence limit, i.e., $p=1/2$. There are a number of equivalent ways to represent the density matrix in this limit. If we work in the eigenbasis of the $X_e$ operators, and use the loop picture, where $X_e=-1$ means the link $e$ is occupied by a $\Z_2$ string, then the density matrix becomes a classical ensemble of $X$ loops:
\begin{equation}
    \rho_X\propto \sum_C \op{C}.
\end{equation}
Here, $C$ denotes closed loops on the lattice. 

One can think of this mixed state as a classical $\Z_2$ gauge theory and the $C$ as the electric field lines, defined by the Gauss law, or the closed loop condition. To some extent, the ensemble represents a classical TO~\cite{Castelnovo2007}: when put on a torus, there are four different ensembles distinguished by the winding number mod 2 of loops around the two non-contractible cycles. These four ensembles can not be distinguished by local observables, but they cannot form coherent superposition. Instead, one can form a classical mixture (i.e., a convex sum) of the classical states. In other words, the space of locally indistinguishable states consists of four isolated extremal points (see below for a more detailed discussion) \cite{Li:2024rgz}.

Another way to write the state is to directly expand the bit-flip channel:
\begin{equation}
    \rho_X = \frac{1}{2^{N_e}} \sum_{n_e=0,1}\Big(\prod_{e}X_{e}^{n_e}\Big)\op{\psi}\Big(\prod_e X_e^{n_e}\Big).
\end{equation}
Note that the state $\prod_{e}X_{e}^{n_e}\ket{\psi}$, in general, contains a number of $m$ particles. Fixing a particular configuration of $m$ particles, there are $2^{N_v-1}$ many ways to create it (the $-1$ is because $\prod_v A_v=1$ on a torus), so the state can also be written as 
\begin{equation}
    \rho_X = \frac{1}{2^{N_e-N_v+1}} \sum_{\mb{m}}\op{\mb{m}}.
\end{equation}
Here, $\mb{m}$ denotes a configuration of $m$ anyons, with the constraint that, in total, there should be even number of them, and $\ket{\mb{m}}$ is the state with the corresponding configuration of $m$ anyons. Therefore, the density matrix describes an ``incoherent'' proliferation of $m$ particles. This should be contrasted with a ``coherent'' proliferation:
\begin{equation}
    \ket{\psi_m}\propto \sum_{\mb{m}}\ket{\mb{m}}=\prod_e \frac{1+X_e}{2}\ket{\psi},
\end{equation} 
which is a product state with $X_e=1$ everywhere.

A useful fact is that for $p>p_c$, the decohered TC state can be purified into a SRE state. This is most easily seen at $p=1/2$, where one can start from the product state $\ket{X_e=1}$, and apply the following quantum channel:
\begin{equation}
    \mathcal{E}=\prod_p \mathcal{E}_p, \quad \mathcal{E}_p = \frac12(\rho + B_p\rho B_p).
\end{equation}
Ref. \cite{Chen:2023vxo} showed that the same is true for $p_c<p<1/2$. This allows one to show that the entire $p>p_c$ phase is trivial, in the sense defined in Section~\ref{sec:equiv relation}. For $p=1/2$, we have already found a channel that maps a product state to a the decohered TC state. To show two-way connectedness, we need to find a channel to map the decohered TC state to a product state. This can be done by tracing out the decohered state and appending a product state. 

Finally, let us discuss some subtleties related to the space of locally indistinguishable states for the system defined on a sphere. For simplicity, let us focus on those states which are locally identical to the decohered TC ground state. We first notice that the full Hilbert space of the system on a sphere can be labeled by the eigenvalues of $A_v$ and $B_p$.\footnote{Note that the TC can be defined on an arbitrary triangulation. We continue to refer to the vertex and plaquette terms as $A_v$ and $B_p$.} 

Let us now consider the subspace $\H_m$ of states that satisfy $A_v=1$, for every vertex $v$. In other words, $\H_m$ is the space of states with only $m$ anyon excitations. The most relevant local operator is $B_p$ here, which detects whether there is an $m$ anyon. The plaquette operators and the Pauli $X$ operators, in fact, generate all operators that keep the subspace $\H_m$ invariant.
For a state $\ket{\psi}\in \H_m$, we have
\begin{align}
    \Tr \big[\cal{N}_X(\op{\psi})B_p \big]&= \Tr \big[\op{\psi}\cal{N}_X(B_p)\big] \\
    &= (1-2p)^4 \braket{\psi|B_p|\psi}.
\end{align}
Curiously, if $p=1/2$, then the expectation value of $B_p$ is always 0 for any $\ket{\psi}\in \H_m$. Thus the dimension of locally indistinguishable states is $2^{N_p-1}$, where $N_p$ is the number of plaquettes.  This is an example of the phenomenon mentioned earlier, i.e., that locally distinguishable states can become indistinguishable under a QLC.\footnote{We remark that the states of the form $\cal{N}_X(\op{\psi})$ can often be distinguished by their R\'enyi-2 expectation values instead. This is because $B_p \cal{N}_X(\op{\psi}) B_p$, which appears in the R\'enyi-2 expectation value is not equal to $\cal{N}_X(\op{\psi})$, if $\ket{\psi}$ has a superposition of $\mb{m}$ configurations. }

However, if $p \neq 1/2$, clearly the result is very different. For example, if $\ket{\psi}$ contains $m$ anyons at fixed locations, $\cal{N}_X(\op{\psi})$ is still locally distinguishable from the decohered ground state. Thus, identifying locally indistinguishable states amounts to finding states in the subspace $\H_m$ such that the expectation value of $B_p$ is sufficiently close to $1$. One example is to consider a state $\ket{p_1,p_2}$ with a pair of ${m}$ anyons, say at plaquettes $p_1,p_2$, and then superpose all such states with two ${m}$ anyons:
\begin{equation}
    \ket{\Psi} = \sqrt{\frac{2}{N_p(N_p-1)}}\sum_{p_1\neq p_2}\ket{p_1,p_2}.
\end{equation}
We find that $\braket{\Psi |B_p|\Psi}=1-O(L^{-2})$.
However, this does not meet the criterion for local indistinguishability in Eq.~\eqref{eq: mixed indistinguishability}, as we require that the correction should decay super-polynomially in $L$. In fact, this is the generic situation: in order for $\braket{B_p}$ to be close to 1, the density of $m$ anyons must be 0. In other words, there are $O(1)$ such anyons. Suppose the typical number of $m$ anyons is $n$. To get a uniform state one has to superpose states with different configurations, so the weight of a state with a definition configuration is on average $O(L^{-n})$. Then the expectation value $\braket{B_p}$ is $1-O(L^{-2})$, which does not satisfy the criterion for local indistinguishability. 
We thus find that the space of locally indistinguishable states on a sphere consists of just a single classical state. 

\add{Now, we consider locally indistinguishable states on a torus. To avoid the subtleties of $p=1/2$, we take $p$ to be slightly below $1/2$.
% the kind of exponentially large 
% number of indistinguishable states created by populating $m$ anyon excitations.
% This is strong evidence that the mixed state belongs to the trivial phase. On the other hand, we also
We know that for $p_c<p<1/2$ the manifold of locally indistinguishable states has four extremal points, labeled by the eigenvalues of the non-contractible Wilson loops of $m$ anyons. This is different from a product state, but does not preclude the decohered state from having trivial mixed-state TO.}

\add{In summary, below $p_c$, there is a continuum of locally indistinguishable states on a torus, signifying that the noise can be decoded and the system can be used as a quantum memory. On the other hand, above $p_c$ and with $p\neq 1/2$, the space of locally indistinguishable states on a torus has four extremal points, representing the fact that the system serves as a classical memory that stores two bits.}

\section{Decohered Pauli stabilizer states }
\label{Pauli}
In this section we study general topological stabilizer states, under Pauli decoherence channels. More explicitly, we consider 
a Pauli stabilizer state $\ket{\psi}$ defined by the Pauli stabilizer group ${\cal S}_0$. Namely, $\ket{\psi}$ satisfies
\begin{equation}
    S\ket{\psi}=\ket{\psi}, \,\, \forall S\in {\cal S}_0.
\end{equation}
We assume that ${\cal S}_0$ admits a set of local generators and that, on an infinite plane, the only Pauli operators that commute with every element of $\cal{S}_0$ are the elements of $\cal{S}_0$. These are the conditions for a stabilizer code to be topological~\cite{Bravyi:2010ida, Haah2021classification}.

Then, suppose that we apply a QLC ${\cal N}$ of the form:
\begin{equation}
    {\cal N}=\prod_i {\cal N}_{i}, \quad \: {\cal N}_{i}(\rho) = \sum_{r=1}^n p_rP_{i,r} \rho P^\dag_{i,r}.
    \label{defN}
\end{equation}
where $i$ labels lattice sites, $P_{i,r}$ is a local Pauli operator, and $p_1,p_2,\dots, p_{n}$ satisfy $p_r\geq 0$ and $\sum_{r=1}^{n} p_r=1$. Here, we assume the channel is translation-invariant.
\add{Note that because the $P_{i,r}$ are Pauli operators,  they always commute up to a phase factor.}  Therefore, the channels actually commute:
 \begin{equation}
    {\cal N}_{i}\circ {\cal N}_{j}={\cal N}_{j}\circ {\cal N}_{i}.
 \end{equation}

Denote by ${\cal G}$ the algebra of local operators generated by the $P_{i,r}$'s and ${\cal S}_0$.
In order to analyze the effect of the quantum channel, it turns out to be convenient to think of the problem as a subsystem code, which we now briefly review.

\subsection{Topological subsystem codes}

\begin{figure}[t] 
\centering
\includegraphics[width=.26\textwidth]{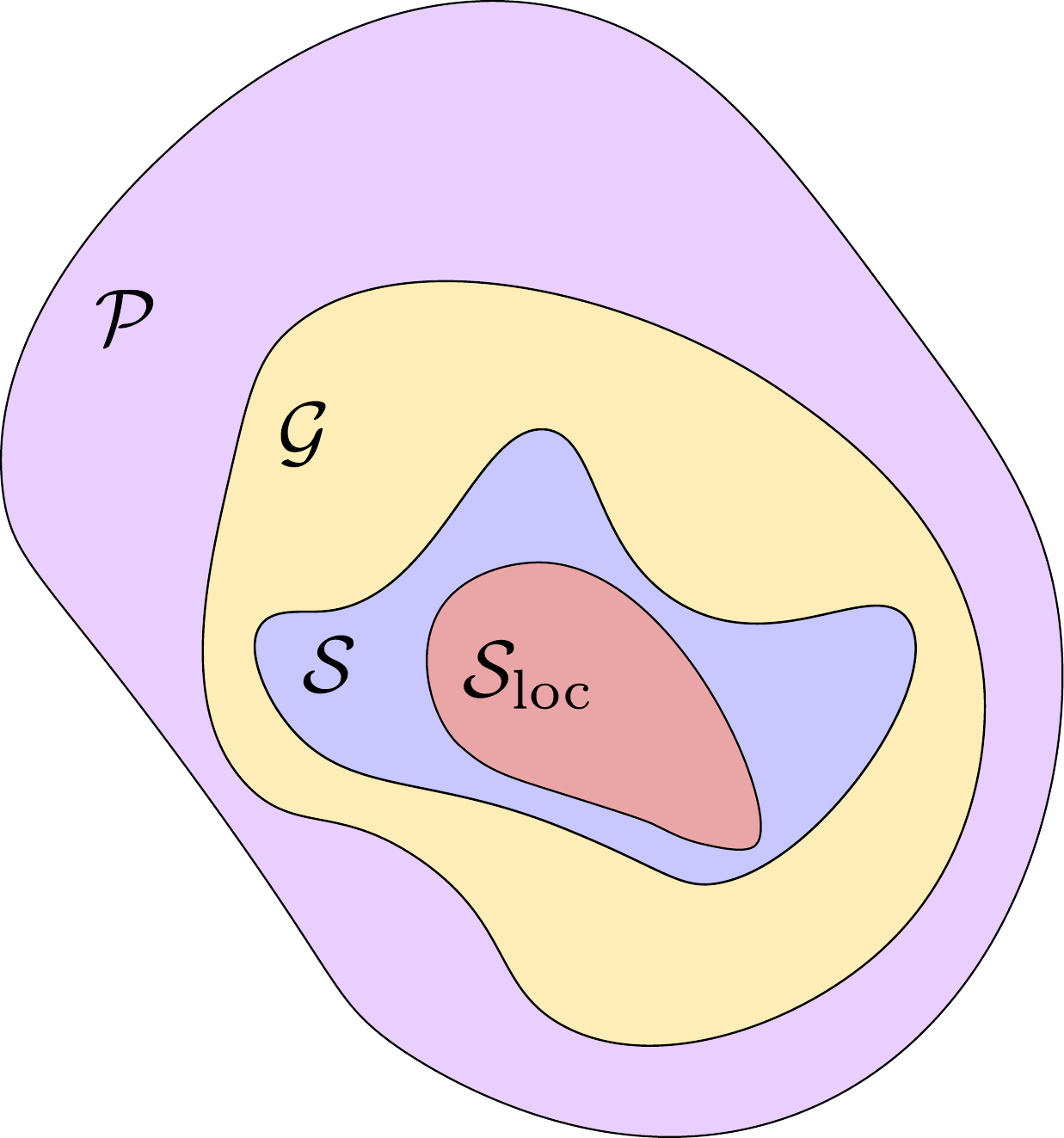}
\caption{Containment of the groups for defining a topological subsystem code. The gauge group $\cal{G}$ (yellow) is an arbitrary subgroup of the Pauli group $\cal{P}$ (purple). The stabilizer group $\cal{S}$ (blue) is the center of $\cal{G}$ up to roots of unity. Finally, the subgroup of locally generated stabilizers $\cal{S}_{\rm loc}$ (red) is contained in $\cal{S}$. If the subsystem code is topological, then $\cal{G}$ admits a set of local generators, and on an infinite plane, there are no logical operators and $\cal{S}_{\rm loc} = \cal{S}$.}
\label{fig: subsystemcontainment}
\end{figure}
 
Let us briefly review the theory of Pauli subsystem codes~\cite{PoulinPRL2005, BombinPRB2009, BombinTSC, Ellison:2022web}. The starting point is the ``gauge group'' ${\cal G}$, which is a group of Pauli operators. The elements of ${\cal G}$ are referred to as gauge operators.
We then define the stabilizer group ${\cal S}$ of the subsystem code as the center of ${\cal G}$, denoted as ${\cal Z}(G)$:
 \begin{equation}
    {\cal S}\propto {\cal Z}({\cal G}).
    \label{defS}
 \end{equation}
Here, the proportionality symbol means that ${\cal S}$ is defined up to roots of unity. We return to this issue later. The stabilizer group defines the code space ${\cal H}_C$:
\begin{equation}
    {\cal H}_C = \{\ket{\psi}: S\ket{\psi}=\ket{\psi}, \forall S\in {\cal S}\}.
\end{equation}
By Eq.~\eqref{defS}, the gauge operators preserve the code space. We show the containment of the various groups of Pauli operators in Fig.~\ref{fig: subsystemcontainment}.

Unlike ordinary stabilizer codes, in a subsystem code, quantum information is only stored in a subsystem of ${\cal H}_C$, known as the logical subsystem. More precisely, with ${\cal G}$ and ${\cal S}$, we have the following decomposition of the Hilbert space:
 \begin{equation}
    {\cal H}={\cal H}_C\oplus {\cal H}_C^\perp, \quad \: {\cal H}_C={\cal H}_G\otimes {\cal H}_L.
 \end{equation}
Here, 
${\cal H}_G$ is the gauge system, such that ${\cal G}/{\cal S}$ acts on ${\cal H}_G$ faithfully and irreducibly as the Pauli algebra. ${\cal H}_L$ is the logical subspace, on which the gauge operators act as the identity. When ${\cal G}$ is proportional to ${\cal S}$, the gauge subsystem ${\cal H}_G$ is trivial and the subsystem code is equivalent to a stabilizer code defined by ${\cal S}$.  

Similar to stabilizer codes, a logical operator is a Pauli operator that preserves the code space. All logical operators form the group ${\cal Z}_{\cal P}({\cal S})$. Here, ${\cal Z}_{\cal P}({\cal S})$ denotes the centralizer of ${\cal S}$ in the Pauli group ${\cal P}$, i.e., the subgroup of Pauli operators that commute with every element of $\cal{S}$. \add{Note that the centralizer can contain elements outside of the gauge group, so ${\cal Z}({\cal G})\subset{\cal Z}_{\cal P}({\cal S})$.} A nontrivial logical operator, in particular, should act nontrivially on ${\cal H}_L$; this is given by the group ${\cal Z}_{\cal P}({\cal S})/{\cal G}$

\add{So far, our definition of subsystem codes is for a generic quantum system without any notion of locality. We further say a subsystem code is \emph{topological} if (1) $\cal{G}$ admits a set of local generators, and (2) on an infinite plane, there are no logical operators and ${\cal S}$ can be generated by local operators. Note that the condition on the stabilizer group in (2) does not have to hold for arbitrary topologies -- i.e., on a torus, there may be stabilizers that cannot be generated by local stabilizers~\cite{Ellison:2022web}.}

% So far, our definition of subsystem codes is for a generic quantum system without any notion of locality. For a local quantum system, we further require that ${\cal G}$ is generated by {local} Pauli operators. Notice that this does not have to be true for ${\cal S}$: it may contain nonlocal generators. For simplicity, we also assume that ${\cal G}$ is translation invariant.

% A subsystem code is \emph{topological} if (1) $\cal{G}$ admits a set of local generators, and (2) on an infinite plane, there are no logical operators and ${\cal S}$ can be generated by local operators. We primarily consider topological subsystem codes in the discussion below.

In order to study the decohered stabilizer code on torus (or higher-genus surfaces), it is useful to introduce further structure, since the stabilizer group ${\cal S}$ may contain nonlocal generators. We therefore split ${\cal S}$ into a locally-generated subgroup ${\cal S}_{\rm loc}$ and ${\cal S}_T$, such that ${\cal S}/{\cal S}_{\rm loc}={\cal S}_T$. Here, the subgroup ${\cal S}_{\rm loc}$ is generated by geometrically local stabilizers. We define a ``local'' code space according to ${\cal S}_{\rm loc}$:
\begin{equation}
    {\cal H}={\cal H}_{lC}\oplus {\cal H}_{lC}^\perp.
\end{equation} 
Namely, ${\cal H}_{lC}$ is the subspace with $S=1$ for every $S\in {\cal S}_{\rm loc}$.

Then, we further split ${\cal H}_{lC}$ according to eigenvalues of ${\cal S}_T$. Denote by ${t}$ a group homomorphism from ${\cal S}_T$ to $\U$, such that $t(T)$ is an eigenvalue of $T$. Indexing the generators of $\cal{S}_T$ by $i$, we can define a vector $\mb{t}$ whose $i$th entry is $\mb{t}_i = t(T_i)$. With this, $\cal{H}_{lC}$ decomposes as 
\begin{equation}
    {\cal H}_{lC}=\bigoplus_{\mb{t}} {\cal H}_{lC}^{\mb{t}},
    \label{splittingHlC}
\end{equation}
where the subspace ${\cal H}_{lC}^{\mb{t}}$ is labeled by a vector $\mb{t}$. Every state in ${\cal H}_{lC}^{\mb{t}}$ is an eigenstate of $T\in {\cal S}_T$ with eigenvalue $\mb{t}_i = t(T)$.
We take ${\cal H}_{lC}^{\mb{t}=\mb{1}}$ to be the code space ${\cal H}_C$. Each ${\cal H}_{lC}^{\mb{t}}$ can be further factorized as
\begin{equation}
    {\cal H}_{lC}^{\mb{t}}={\cal H}_G^{\mb{t}}\otimes {\cal H}_L^{\mb{t}}.
\end{equation}
For each $\mb{t}$ subspace, we can define a stabilizer group ${\cal S}^{\mb{t}}$, which is generated by ${\cal S}_{\rm loc}$ and $t(T_i)^{-1}T_i$. 

\subsection{Decohered topological stabilizer states}
\label{sec:decoheredSS}

Now, we return to the decohered stabilizer code. The initial (local) stabilizer group of the code is ${\cal S}_0$. The Pauli errors $P_{i,r}$ generate a group $E$.  It is natural to assume that any element $e$ of $E$ at least fails to commute with some stabilizer in ${\cal S}_0$, otherwise, by the assumption of ${\cal S}_0$ being topological, $e$ must belong to ${\cal S}_0$ as well. 

We consider the subsystem code defined by the gauge group ${\cal G}=\braket{{\cal S}_0, E}$.  
We assume that the subsystem code defined by this gauge group is {topological}. We emphasize that this is a nontrivial condition. Indeed in Section~\ref{sec: Ydecohered TC}, we discuss an example where this condition is not met. 
We note that, in particular, the topological condition guarantees that the decohered stabilizer state is locally correlated, \add{in the sense of Section~\ref{sec: generalities}}. First of all, the decohered state admits a purification into a GGS, since it is constructed by adding Pauli noise to a topological stabilizer state. Furthermore, we show in Appendix~\ref{app: Renyi2 topological codes} that it is R\'enyi-1 and -2 locally correlated. 

We now consider the maximum decoherence limit and argue that the topological stabilizer state becomes maximally mixed on the subsystem ${\cal H}_G^{\mb{t}}$. In agreement with the the subsystem code literature, the subsystem ${\cal H}_L^{\mb{t}}$ defines a noiseless subsystem~\cite{Knill2000noiseless,Kribs2006operator}. 

We first observe that, for a state $\rho_0$ in the code space of ${\cal S}_0$, we can write 
\begin{equation} \label{eq: gauge group noise}
    {\cal N}(\rho_0)=\sum_{g\in {\cal G}}p_g g\rho g^\dag,
\end{equation}
where the sum is over all elements of the gauge group ${\cal G}$. Unlike Eq.~\eqref{defN}, the expression for the action of $\cal{N}$ on $\rho_0$ includes the stabilizers of $\cal{S}_0$. Since $\rho_0$ is invariant under the elements of $\cal{S}_0$, this just adds an overall constant factor, which is absorbed into the normalization. The maximal decoherence limit then corresponds to taking $p_g=\frac{1}{|{\cal G}|}$. We denote the channel in this limit by ${\cal N}_m$. As we saw in the $\Z_2$ TC example, the physics of the error-induced phase is well captured by this limit.

We would now like to argue that $\cal{N}_m$ acting on an arbitrary pure state $\rho_0 = \op{\psi}$ in the code space of $\cal{S}_0$ creates a maximally-mixed state in the gauge subsystem. That is, we would like to show that
\begin{eqs} \label{eq: maximally mixed on gauge}
    {\cal N}_m(\rho_0) = \sum_{\mb{t}}\frac{1}{\dim {\cal H}_G^{\mb{t}}}\id_{{\cal H}_G^{\mb{t}}}\otimes \rho_{L}^{\mb{t}},
\end{eqs}
for some state $\rho_{L}^{\mb{t}}$ on the logical subsystem. This follows immediately from the fact that the gauge group generates the full matrix algebra on the gauge subsystem~\cite{Knill2000noiseless,Zanardi2004Tensor, Kribs2006operator}. Therefore, the channel in Eq.~\eqref{eq: gauge group noise} behaves like the depolarizing noise channel within the gauge subsystem. Nonetheless, we find it instructive to demonstrate Eq.~\eqref{eq: maximally mixed on gauge} explicitly. 

We begin by noticing that the state $\rho_0 = \op{\psi}$ belongs to the code space $\H_{lC}$, since ${\cal S}_{\rm loc}\subset {\cal S}_0$. 
According to Eq.~\eqref{splittingHlC}, this means that we can decompose $\ket{\psi}$ as
\begin{equation}
    \ket{\psi}=\sum_{\mb{t}}\ket{\psi^{\mb{t}}}, \quad \ket{\psi^{\mb{t}}}\in \H_{lC}^{\mb{t}},
\end{equation}
where $\ket{\psi^{\mb{t}}}$ is
\begin{equation}
    \ket{\psi^{\mb{t}}}=\frac{1}{|\mathcal{S}_T|}\sum_{T\in {\cal S}_T} t(T)^{-1} T \ket{\psi}.
\end{equation}
Given the factorization of the code space, we can further Schmidt decompose $\ket{\psi_\mb{t}}$ to obtain
\begin{equation}
    \ket{\psi^\mb{t}}=\sum_\alpha \lambda_{\mb{t},\alpha}\ket{\psi^{\mb{t}}_{G,\alpha}}\otimes \ket{\psi^{\mb{t}}_{L,\alpha}}.
\end{equation}
Here, the states $\ket{\psi^{\mb{t}}_{G/L,\alpha}}$ are orthonormal for different indices $\alpha$, and ${\sum_\alpha\lambda_{\mb{t},\alpha}^2=1}$.

With this notation, the action of $\cal{N}_m$ on $\rho_0$ can be written as
\begin{multline}
     {\cal N}_m(\rho_0) = \\
     \sum_{\mb{t},\mb{t}'}\sum_{\alpha,\beta}\lambda_{\mb{t}\alpha}^*\lambda_{\mb{t}'\beta}{\cal N}_m(\ket{\psi_{G,\alpha}^{\mb{t}}}\!\bra{\psi_{G,\beta}^{\mb{t}'}})\otimes\ket{\psi_{L,\alpha}^{\mb{t}}}\!\bra{\psi_{L,\beta}^{\mb{t}'}}.
\end{multline}
We have used here that, by definition, the gauge operators act as the identity on the logical subsystem. We see that, to understand the effects of the channel $\cal{N}_m$, we only need to consider its effects on $\ket{\psi_{G,\alpha}^{\mb{t}}}\!\bra{\psi_{G,\beta}^{\mb{t}'}}$.

By explicit calculation, we find that $\cal{N}_m(\ket{\psi_{G,\alpha}^{\mb{t}}}\!\bra{\psi_{G,\beta}^{\mb{t}'}})$ is
\begin{multline}
    {\cal N}_m(\ket{\psi^\mb{t}_{G,\alpha}}\!\bra{\psi^{\mb{t}'}_{G,\beta}})
    = \frac{1}{|\mathcal{G}|}\sum_{g\in {\cal G}} g\ket{\psi^\mb{t}_{G,\alpha}}\!\bra{\psi^{\mb{t}'}_{G,\beta}}g^\dag \\
    = \sum_{\tilde{g}\in {\cal G}/{\cal S}_T} \tilde{g}\ket{\psi^\mb{t}_{G,\alpha}}\!\bra{\psi^{\mb{t}'}_{G, \beta}}\tilde{g}^\dag\sum_{T\in {\cal S}_T} t(T){t'}(T)^{-1}.
\end{multline}
To obtain the second line, we split the gauge operator $g$ into $g = \tilde{g}T$, and used $T\ket{\psi_G^{\mb{t}}}=t(T)\ket{\psi_G^{\mb{t}}}$. Next, we use the following identity:
\begin{equation}
    \add{\sum_{T_i\in {\cal S}_T} \mb{t}_i(\mb{t}'_i)^{-1} = |{\cal S}_T|\delta_{\mb{t},\mb{t}'}.}
\end{equation}
% \begin{equation}
%     \sum_{T\in {\cal S}_T} t(T){t'}(T)^{-1} = |{\cal S}_T|\delta_{\mb{t},\mb{t}'}.
% \end{equation}
This allows us to write
\begin{eqs}
    {\cal N}_m(\ket{\psi^\mb{t}_{G, \alpha}}\!\bra{\psi^{\mb{t}'}_{G,\beta}}) = \frac{|{\cal S}_T|}{|{\cal G}|}\delta_{\mb{t},\mb{t}'}\sum_{\tilde{g}\in {\cal G}/{\cal S}_T}\tilde{g}\ket{\psi^\mb{t}_{G,\alpha}}\!\bra{\psi^{\mb{t}'}_{G,\beta}}\tilde{g}^\dag.
\end{eqs}

Now, we only need to prove that
\begin{equation}
    {\cal N}_m(\ket{\psi_{G,\alpha}^{\mb{t}}}\!\bra{\psi_{G,\beta}^{\mb{t}}})=\delta_{\alpha\beta} \frac{1}{\dim {\cal H}_G^{\mb{t}}}\id_{{\cal H}_G^{\mb{t}}}.
    \label{Npsiapsib}
\end{equation}
We prove this for $\mb{t}=\mb{1}$. For other $\mb{t}$, we can just replace ${\cal S}$ with ${\cal S}^{\mb{t}}$. For brevity, we drop the $\mb{t}$ superscript. 
For any Pauli operator $O$ in the gauge subsystem, we have 
\begin{equation}
\begin{split}
     {\cal N}_m(O)=\frac{1}{|{\cal G}/{\cal S}|}\sum_{\tilde{g}\in {\cal G}/{\cal S}} \tilde{g}O\tilde{g}^\dag
     =\begin{cases}
        O, & \text{if } O\propto\id\\
        0, & {\rm otherwise}.
    \end{cases}
\end{split}
\end{equation}
This follows from the fact that ${\cal G}/{\cal S}$ is isomorphic to the Pauli algebra on the gauge subsystem. Then, for a general operator $O$ in the gauge subsystem, it follows that 
\begin{equation}
    {\cal N}_m(O)=\Tr O \cdot \frac{1}{\dim {\cal H}_G}\id.
\end{equation}
Given that $\Tr(\ket{\psi_{G,\alpha}^{\mb{t}}}\!\bra{\psi_{G,\beta}^{\mb{t}}})=\delta_{\alpha\beta}$, we arrive at Eq.~\eqref{Npsiapsib}.

Putting this all together, we finally find that the maximally decohered state $\rho_m$ is
\begin{align*}
    {\cal N}_m(\rho_0)&=\sum_{\mb{t}}\frac{1}{\dim {\cal H}_G^{\mb{t}}}\id_{{\cal H}_G^{\mb{t}}}\otimes\sum_{\alpha}\lambda_{\mb{t},\alpha}^2\ket{\psi_{L,\alpha}^{\mb{t}}}\!\bra{\psi_{L,\alpha}^{\mb{t}}} \\
    &= \sum_{\mb{t}}\frac{1}{\dim {\cal H}_G^{\mb{t}}}\id_{{\cal H}_G^{\mb{t}}}\otimes \rho_{L}^{\mb{t}}.
\end{align*}

Physically, we have shown that after the maximal noise channel, the gauge subsystems are left in the maximally mixed state, but the state in the logical space ${\cal H}_L^{\mb{t}}$ is unaffected. Furthermore, there is no quantum coherence between logical subspaces with different $\mb{t}$: one can only form convex sum of $\rho_L^{\mb{t}}$'s. In a sense, ${\cal S}_T$ encodes classical information. Therefore, coherent spaces are labeled by the eigenvalues $\mb{t}$.

We remark that when the logical subsystem is trivial (e.g., when the model is placed on a sphere), the gauge subsystem is the code space, and we simply find a maximally mixed state in the code space. The density matrix can then be written as a stabilizer state~\cite{fattal2004entanglement}: 
\begin{equation}
    \rho_m = \frac{1}{|{\cal S}|}\sum_{S\in {\cal S}}S.
\end{equation}

\subsection{Abelian anyon theories} \label{sec: anyon theories of subsystem codes}

To identify interesting examples of decohered topological stabilizer states, beyond the decohered TC state in Section~\ref{sec: decohered TC}, we find it valuable to introduce the language of anyon theories. In general, topologically ordered states in (2+1)$d$ are characterized by anyon theories, which are defined by abstract mathematical data consisting of a set of anyon types, their fusion rules, $F$-symbols, and $R$-symbols.\footnote{For a thorough exposition of this data, we refer to Appendix~E of Ref.~\cite{Kitaev:2005hzj}.} For topological Pauli stabilizer states, however, the anyon theories are Abelian, meaning that the data simplifies greatly. To specify an Abelian anyon theory we need (1) an Abelian group $\cal{A}$ of anyons whose product represents the fusion of anyons, and (2) a function $\theta: \cal{A} \to \U$ that determines the exchange statistics of the anyons. We note that a formal definition of anyon types for topological Pauli subsystem codes is given in Ref. \cite{Ellison:2022web}.

The exchange statistics of the anyons can be determined by first identifying the string operators that move the anyons around the system, using, for example, the construction in Refs.~\cite{Ellison:2022web, BombinPoulin2012Universal}. The string operators can then be used to compute the exchange statistics following Refs.~\cite{Levin2003fermion, Kawagoe2020, Ellison:2022web}. The identity element $1 \in \cal{A}$, is a boson, so it satisfies $\theta(1) = 1$. Furthermore, 
$\theta(a)$ gives a quadratic form over the anyon group. This is to say that $\theta(na)=\theta(a)^{n^2}$, for any anyon type $a$. The $T$ matrix of the anyon theory is defined as
\begin{eqs}
    T_{ab} = \theta(a) \delta_{ab},
\end{eqs}
for $a,b \in \cal{A}$.

Using $\theta$, we can also define a bilinear form $B(a,b)$ over ${\cal A}$, which captures the braiding relations between anyon types $a$ and $b$:
\begin{equation}
    B(a,b)=\frac{\theta(ab)}{\theta(a)\theta(b)}.
\end{equation}
An anyon theory is called ``modular'' if, for every anyon type $a$, there exists an anyon type $b$ such that $B(a,b)\neq 1$. Otherwise, the anyon theory is premodular (or non-modular).
The $S$ matrix of the anyon theory is defined as 
\begin{equation}
    S_{ab}=\frac{1}{\sqrt{|\cal A|}}B(a,b).
\end{equation}
If the anyon theory is modular, then the $S$ matrix is unitary. 
It is proven in Ref.~\cite{Ellison:2022web} that the anyon theory of a Pauli topological state is always modular. For Pauli topological subsystem codes, on the other hand, the anyon theory may be premodular.

If an anyon theory is premodular, then there exists at least one ``transparent'' anyon type $a$, such that $B(a,b)=1$, for every anyon type $b \in \cal{A}$. The subgroup of transparent anyons ${\cal T}$ is defined as
\begin{equation}
    {\cal T}=\{a\in {\cal A}\,|\, B(a,b)=1, \, \forall b\in {\cal A}\}.
\end{equation}
For a modular theory, ${\cal T}$ consists of only the trivial anyon. In general, transparent anyons can be either bosons or fermions. We define $\cal{T}_b$ as the subgroup of the transparent anyons that have bosonic statistics. 

One can form the quotient group $\cal{A}^{\rm{min}}={\cal A}/{\cal T}_b$. This defines another anyon theory, which can be interpreted as the anyon theory obtained from condensing the bosons in $\cal{T}$. If $\cal{T}=\cal{T}_b$, then, $\cal{A}^{\rm{min}}$ is modular. Otherwise,  $\cal{A}^{\rm min}$ takes the form $\cal{A}^{\rm min} = \C \boxtimes \Z_2^{(1)}$, where $\C$ is modular, $\boxtimes$ denotes the operation of stacking two independent anyon theories, and $\Z_2^{(1)}$ is the anyon theory generated by an order 2 transparent fermion (defined below). 

To make the discussion more explicit, let us introduce a family of Abelian premodular anyon theories with a single generator. These anyon theories appear in the examples in the next section. Following \cite{Bonderson2012}, the anyon theories are denoted by $\Z_N^{(p)}$, where $\Z_N$ indicates the fusion group, and $p$ is an integer for odd $N$ and a half-integer when $N$ is even. The group elements are denoted by $[a]$ with $a=0,1,\cdots, N-1$ defined mod $N$. The basic data is given by 
\begin{align}
    [a]\times [b]=[a+b]\\ \label{eq: ZN exchange}
    \theta([a])=e^{\frac{2\pi i p}{N}a^2},\\ \label{eq: ZN braiding}
    B([a],[b])=e^{\frac{4\pi i p}{N}ab},
\end{align}
where addition is taken mod $N$.
Intuitively, when $p$ is an integer, $\Z_N^{(p)}$ is the anyon theory generated by $em^{p}$ in a $\Z_N$ TC.

The transparent subgroup $\cal{T}$ can be determined from the data above. Here, we list a few common cases:
\begin{enumerate}
    \item If $N$ is odd and $p\in \Z$, then the transparent subgroup is $\Z_{(p, N)}^{(0)}$. In particular, if $p$ is coprime with $N$ the theory is modular.
    \item If $N$ is even and $p\in \Z$, the transparent subgroup is $\Z_{2(N/2,p)}$ generated by $[\frac{N/2}{(N/2,p)}]$. These theories are always non-modular.  
    \item If $N$ is even and $p\in\Z+\frac12$, then the transparent subgroup $\Z_{(N,2p)}^{(0)}$ is generated by $[\frac{N}{(N,2p)}]$.
\end{enumerate}

Let us now comment on the connection between anyon theories and the discussion of topological subsystem codes in the previous sections. As shown in Ref.~\cite{Ellison:2022web}, the nonlocal stabilizers ${\cal S}_T$ on a torus (or any higher-genus surface) are generated by the string operators of the transparent anyons ${\cal T}$ along non-contractible cycles. Once the values of ${\cal S}_T$ are fixed, the logical operators for the quantum coherent subspace ${\cal H}_L^{\mb{t}}$ are string operators of the modular part of $\cal{A}^{\rm{min}}$ along non-contractible cycles.

Let us also revisit the locally indistinguishable states of the decohered topological stabilizer codes, using the language of Abelian anyon theories. We consider the case when the system is put on a torus. Before adding noise, a basis for the ground state space can be chosen to be $\ket{a}_x$ where $a$ is the anyon label, and $\ket{a}_x$ is the eigenstate of Wilson loops around the $y$ direction of the torus. That is, for an arbitrary $a$, the state $\ket{a}_x$ satisfies 
\begin{equation}
    W_y(b)\ket{a}_x=\frac{S_{ab}}{S_{Ia}}\ket{a}_x, \quad W_x(b)\ket{a}=\ket{a \times b}_x.
\end{equation}

The coherent subspaces are eigenspaces of $W_x(t)$ and $W_y(t)$ for all $t\in \mathcal{T}$. Let us consider the $W_x(t)=W_y(t)=1$ subspace. This coherent subspace, in particular, can be obtained by decohering the following ground state subspace on a torus:
\begin{equation}
    \text{span}\Big\{ 
    \frac{1}{\sqrt{|\mathcal{T}|}}\sum_{t\in \cal{T}} \ket{a\times t}_x\:\big|\: a\in \cal{A}\Big\}. 
\end{equation}

The basis states of the coherent space are in one-to-one correspondence with elements of $\cal{A}/\cal{T}$.
Thus, the dimension of the coherent space is equal to $|\cal{A}/\cal{T}|$. 
When $\cal{T} = \cal{T}_b$, within the coherent space, the action of string operators for $\tilde{a}\in \cal{A}^{\rm min}$ that wrap around the torus is identical to that of a pure state TO with $\cal{A}^{\rm min}$ as the anyon theory.
   
\subsection{Examples}
\label{sec:decohered ss examples}

\subsubsection{$\Z_2$ toric code with $X$ noise}
The first example is again the $\Z_2$ TC under bit-flip errors, \add{but here we study the model using the language of subsystem codes}. In this case, the stabilizer group ${\cal S}_0$ is generated by the vertex $A_v$ and plaquette $B_p$ stabilizers of the TC. The gauge group defined with bit-flip noise is ${\cal G}=\langle {\cal S}_0, X\rangle$. The group ${\cal S}_{\rm loc}$ is generated by the $A_v$. The group ${\cal S}_T$ is then generated by the string operators of $X$ along non-contractible loops on the dual lattice. Physically, ${\cal S}$ are the closed string operators of the $m$ anyon. Therefore the associated Abelian anyon theory is ${\cal A}=\{1,m\}$, \add{denoted as $\Z_2^{(0)}$ using the notations introduced in Sec. \ref{sec: anyon theories of subsystem codes}.} The anyon theory consists only of transparent anyons (i.e., ${\cal T}={\cal A}$).

We have shown in Section~\ref{sec: decohered TC} that, in the maximal decoherence limit, the density matrix becomes a diagonal classical ensemble of closed loops. On a torus, the mod 2 winding numbers of the loops around two directions are topological invariants. Thus, we have four classical states:
\begin{equation}
    \rho_X^{w_xw_y}\propto\sum_{C\in \mathbf{C}_{w_xw_y}} \op{C}, \quad \:w_{x/y}=0,1.
\end{equation}
Here, $w_{x/y}$ indicates the even/odd parity of the number of loops wrapping around $x$ or $y$ directions, and $\mathbf{C}_{w_xw_y}$ is the corresponding set of loop configurations. These four extremal states are precisely distinguished by the eigenvalues of the nonlocal stabilizers, as expected from the general result. 
%A general density matrix is a convex sum of the four density matrices. In other words, there are four isolated extremal points in the space of locally indistinguishable states on a torus.  
The $e$ string operator of the TC, namely product of $Z$ operators along a path on the lattice, can toggle between the four extremal points. 

\subsubsection{$\Z_2$ toric code with ``fermionic'' noise}
\label{fermion-TC}

As described in Section~\ref{sec: decohered TC}, bit-flip noise applied to the TC state can be interpreted as incoherently proliferating $m$ anyons. Here, we consider instead an incoherent proliferation of $f=e\times m$ particles. Note that since $f$ is a fermion, it can not be condensed coherently. However, an incoherent proliferation is still possible.

Since the fermion is a bound state of $e$ and $m$, one has to specify the relative positions. It is useful to fix an orientation, where the $m$ is always at the plaquette $p$ to the ``northeast'' of $e$ at the vertex $v$. We say that such a bound state is a fermion at the plaquette $p$.

\begin{figure*}[t]
\centering
\subfloat[\label{fig: ystabilizers}]{\includegraphics[width=.25\textwidth]{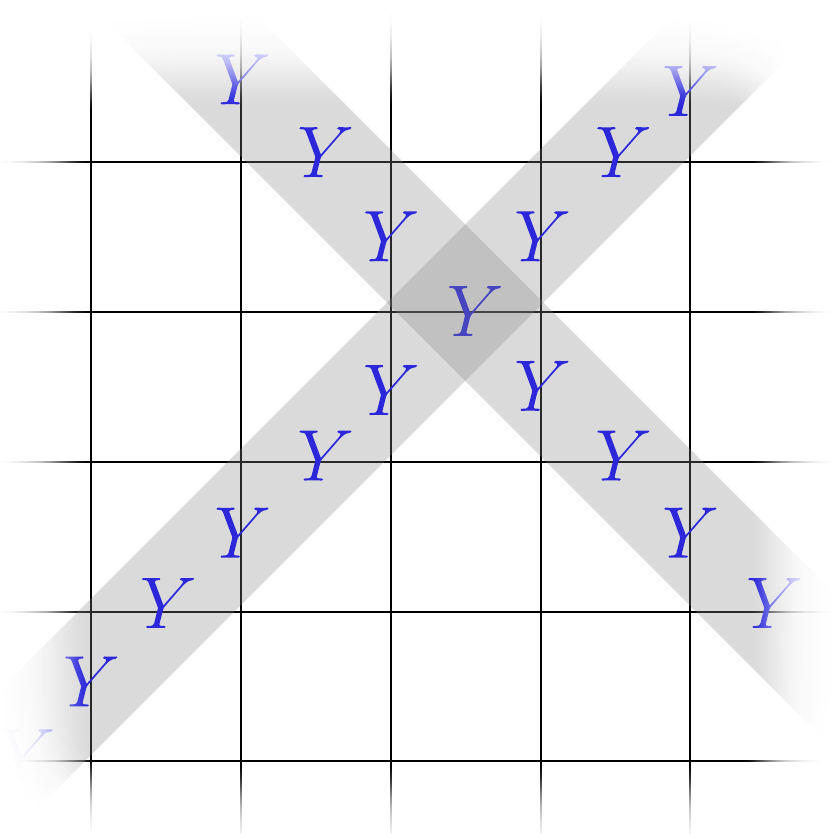}} \quad
\subfloat[\label{fig: vertexproduct}]{\includegraphics[width=.25\textwidth]{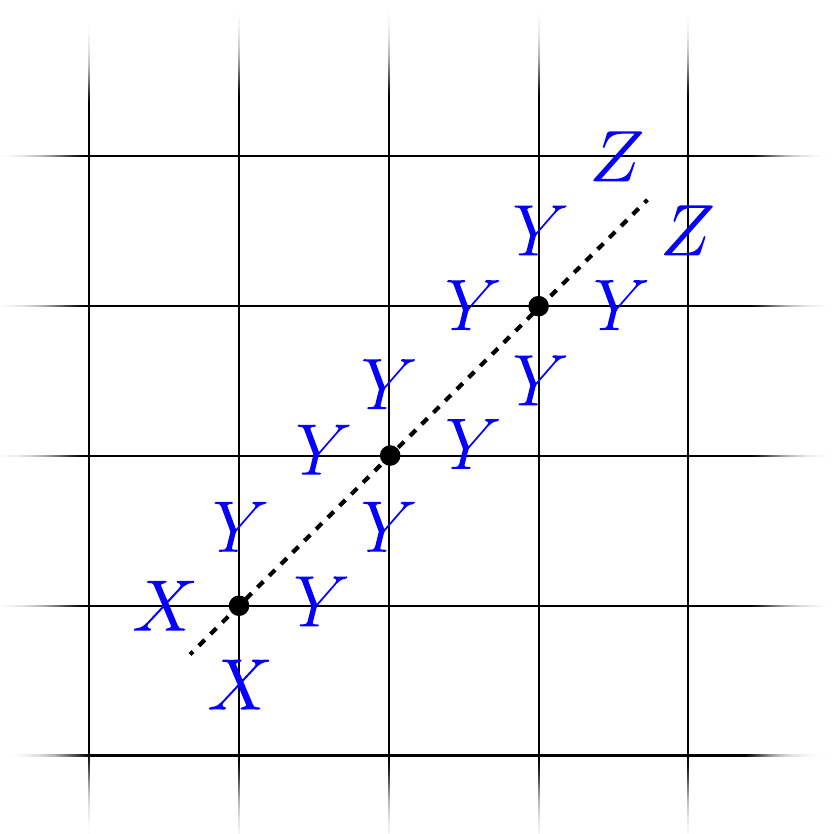}} \quad
\subfloat[\label{fig: renyiorderparams}]{\includegraphics[width=.25\textwidth]{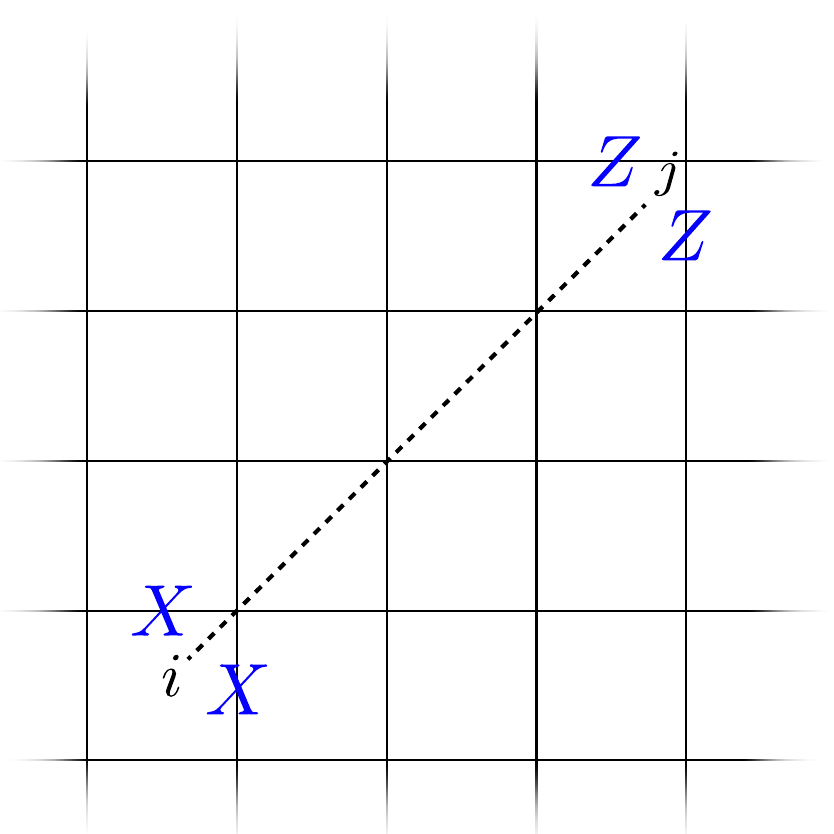}}
\caption{The $Y$-decohered TC state and its R\'enyi-2 order parameters. (a) The stabilizer group of the $Y$-decohered TC subsystem code is generated by products of Pauli $Y$ operators along diagonals. (b) A product of $A_vB_{\rm{NE}(v)}$ operators along a diagonal path (dashed line) yields $XX$ operators at one endpoint and $ZZ$ operators at the other, separated by a string of $Y$ operators. (c) The R\'enyi-2 correlator $C^{(2)}\big((XX)_i,(ZZ)_j\big)$ exhibits long-range correlations for $i$ and $j$ along a diagonal.}
\label{fig: ydecohered}
\end{figure*}

We can then define a ``hopping'' operator $S_e$ for each link $e$. 
\begin{align}
S_e=\vcenter{\hbox{\includegraphics[scale=0.21]{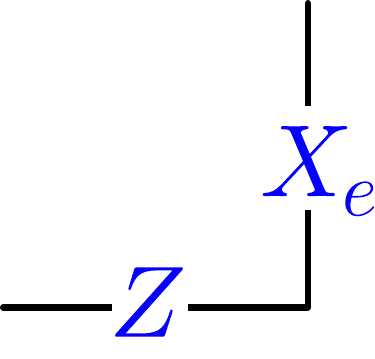}}}, \,\,\quad
    \vcenter{\hbox{\includegraphics[scale=0.21]{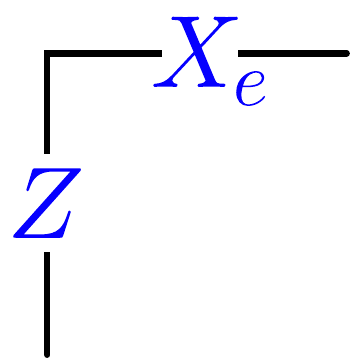}}}.
    \end{align}
The hopping operator $S_e$ creates, annihilates, or moves the fermions between the plaquettes bordered by the edge $e$. We consider the following channel, which incoherently proliferates the fermions:
\begin{equation}
    {\cal N}_f = \prod_e {\cal N}_{f,e}, \quad {\cal N}_{f,e}(\rho)=\frac12(\rho +  S_e\rho S_e).
\end{equation}
For a TC state $\rho_0$, we define $\rho_f={\cal N}_f(\rho_0)$. We refer to this state as the fermion-decohered TC state.

To better understand the decohered state, we employ the fermionization map introduced in \cite{Chen:2017fvr}. This map takes the subspace satisfying $A_vB_{\rm{NE}(v)}=1$, where $\rm{NE}(v)$ is the plaquette to the northeast of the vertex $v$, to the fermion parity even sector of a system with physical fermions on the plaquettes. The initial TC ground state has no $f$ excitations.  
After the channel is applied, one can show that the density matrix in the fermionic Hilbert space takes the following form:
\begin{equation}
    \rho_f = \frac{1}{2^{N_p-1}}{\sum_{\{n_p\}}}'\op{\set{n_p}}\propto \mathbb{1}+P_f,
\end{equation}
where $n_p$ denotes the fermion occupancy at the plaquette $p$, and the sum is over configurations $\{n_p\}$ satisfying $\sum_{p}n_p = 0 \text{ mod }2$.
This gives the maximally mixed state in the ${P_f=1}$ subspace~\footnote{This state can also be thought of as strong-to-weak spontaneous symmetry breaking~\cite{Lessa:2024gcw} of the fermion parity conservation.}. After bosonization, the fermionic density matrix $\rho_f$ is again mapped to the fermion-decohered TC state (hence the same notation).

Interestingly, we find that $\rho_f$ has many other purifications. For example, it can be purified into an Ising TO, or any of the Kitaev's 16-fold ways~\cite{Kitaev:2005hzj}. This is elaborated on in Appendix \ref{app:purify-fermion-TC}.

One can show that $\rho_f$ is not SRE in a bosonic system, but can have a SRE purification, if there are physical fermions in the system. It was recently suggested in Ref.~\cite{Wang:2023uoj} that $\rho_f$ represents an example of an ``intrinsically mixed'' TO. We argue that this is indeed the case in Section~\ref{sec: 1-form symmetry}.

\add{Lastly, we note that the definition of the hopping operators $S_e$ is not unique. In our definition, the fermion is assumed to be the bound state of an $e$ anyon at a vertex $v$ with $m$ at the plaquette to the northeast of $v$. One may choose different conventions for the binding between $e$ and $m$ and then the form of the hopping operators need to be modified accordingly. However, we conjecture that as long as there is a consistent choice for the definition of fermion and hence hopping operators, the channel $\mathcal{N}_f$ all lead to the same mixed-state topological phase. This is motivated by the fact that once a binding of the $e$ and $m$ anyons is specified, one can perform a fermionization duality and employ the logic in Appendix~\ref{app:purify-fermion-TC}.}

\subsubsection{$\Z_2$ toric code with $Y$ noise} \label{sec: Ydecohered TC}

Now, we consider the $\Z_2$ TC in the presence of $Y$ noise. The gauge group is generated by the TC stabilizers and Pauli $Y$ operators. The stabilizer group of the subsystem code is generated by products of $Y$ operators along diagonal paths, as shown in Fig.~\ref{fig: ystabilizers}. This stabilizer group clearly does not have local generators. Thus, the subsystem code is not topological.

Moreover, the TC state with maximum $Y$ decoherence is not locally-correlated, as defined in Section~\ref{sec: generalities}. Therefore, it falls outside of the class of mixed states considered in this work. More specifically, the $Y$-decohered TC state indeed admits a purification into a GGS, and is R\'enyi-1 locally correlated, but it is not R\'enyi-2 locally correlated. 

To see this, we consider a product of $A_vB_{\rm{NE}(v)}$ operators along a diagonal. As shown in Fig.~\ref{fig: vertexproduct}, this produces $XX$ and $ZZ$ operators at the endpoints of the diagonal line with $Y$ operators in between. Since the $Y$ operators are themselves gauge operators, they commute with the decohered state and we have 
\begin{equation}
    (XX)_i(ZZ)_j \rho (ZZ)_j (XX)_i = \rho.
\end{equation}
\add{Here $i$ and $j$ are two well-separated positions.}

On the other hand, $XX$ itself does not commute with the stabilizer group of the decohered state, \add{which implies $(XX)_i \rho (XX)_j$ has orthogonal support as $\rho$.} Therefore we have
\begin{equation}
    \Tr \big[(XX)_i \rho (XX)_j \rho\big]=0,
\end{equation}
and
\begin{equation}
    \add{F\big(\rho, (XX)_i \rho (XX)_j \big)=0.}
\end{equation}
and similarly for $ZZ$. Thus, the R\'enyi-2 and fidelity connected correlator exhibit long-range correlations
\begin{equation}
    C^{(2)}\big((XX)_i,(ZZ)_j\big)=1=C^{F}\big((XX)_i,(ZZ)_j\big),
\end{equation}
with $i$ and $j$ along a diagonal, as in Fig.~\ref{fig: renyiorderparams}.

\begin{figure}[t] 
\centering
\includegraphics[width=.26\textwidth]{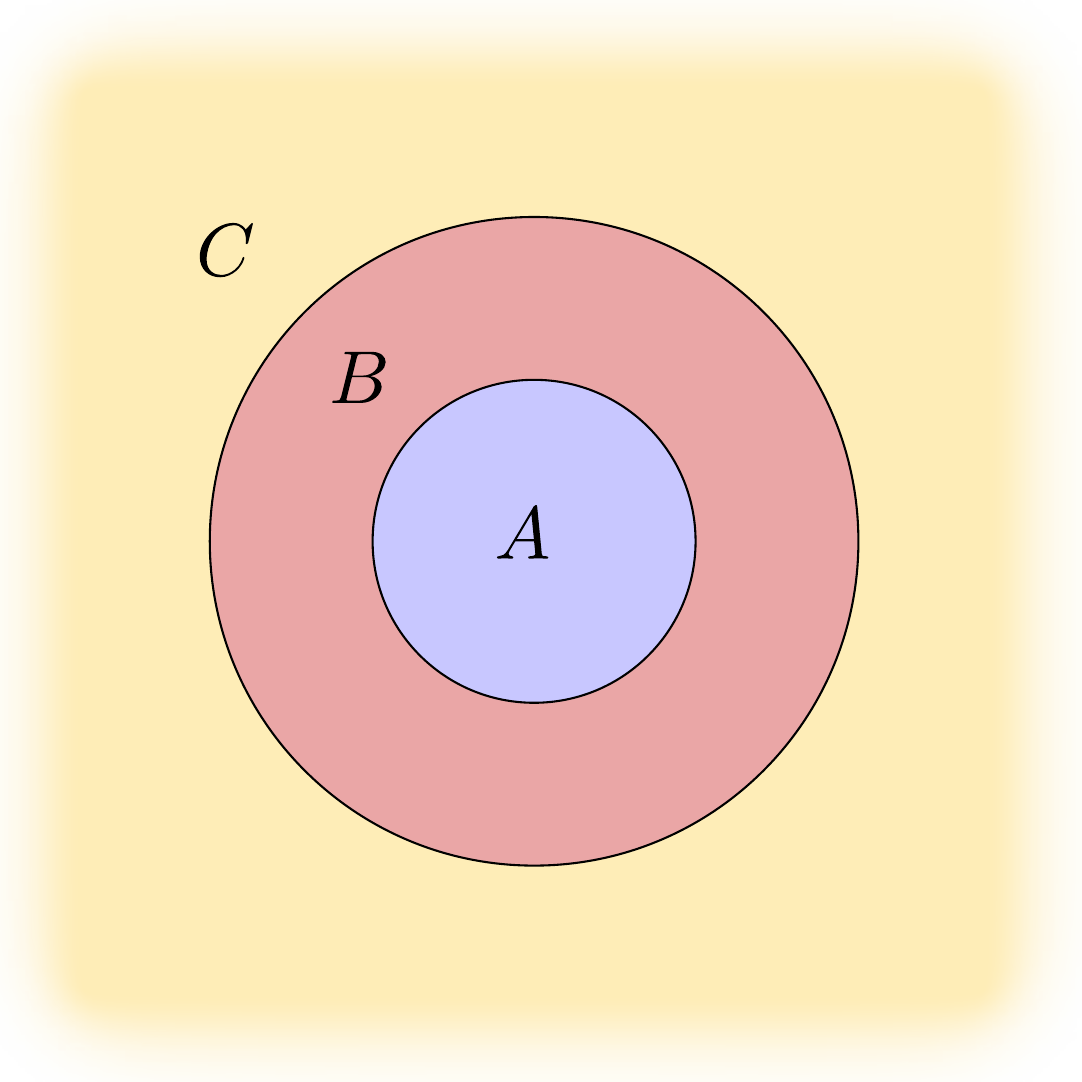}
\caption{Subsystems $A$, $B$, and $C$ for CMI calculation. For any GGS, we expect that the CMI $I(A:C|B)$ decays faster than any power law in the width of the annulus $B$. For the $Y$-decohered TC state, the CMI 
 is non-zero and does not decay with the width of $B$, as argued in Appendix~\ref{app: CMI Ydecohered}.}
\label{fig: ABCgeometry}
\end{figure}

We would also like to point out that, for the TC state decohered by $Y$ noise, the conditional mutual information (CMI) $I(A:C|B)$ for the subsystems $A$, $B$, and $C$ depicted in Fig.~\ref{fig: ABCgeometry}, is non-vanishing in the width of the annulus $B$. This is noteworthy, since the CMI in this geometry is vanishing for short-range correlated GGSs~\cite{Shi2020fusion}. We expect that the CMI vanishes for any mixed state based on a subsystem code that is topological. We compute the CMI for the $Y$-decohered TC state in Appendix~\ref{app: CMI Ydecohered}.

\subsubsection{$\Z_N$ toric code}
For another example, we can consider the $\Z_N$ TC, defined by the Hamiltonian
\begin{align}
    H_\text{TC} = - \sum_v A^{\rm TC}_v - \sum_p B^{\rm TC}_p.
    \end{align}    
The vertex term $A^\text{TC}_v$ and plaquette term $B^\text{TC}_p$ are graphically represented as:
\begin{align}
A_v^{\rm TC} = \vcenter{\hbox{\includegraphics[scale=0.21]{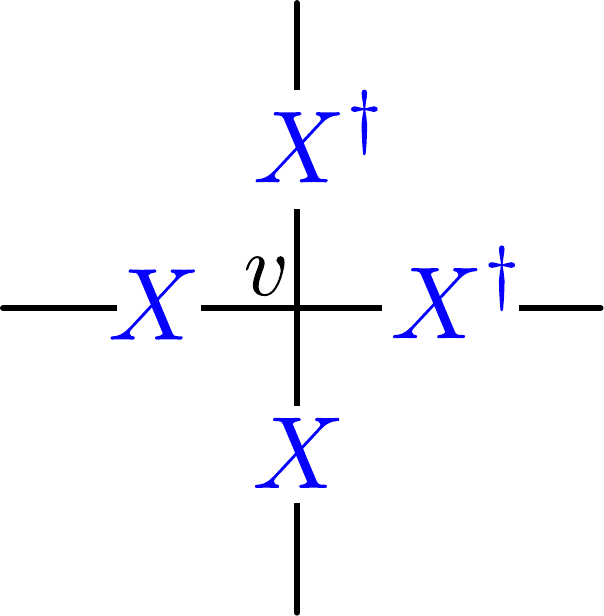}}}, \quad B_p^{\rm TC} = \vcenter{\hbox{\includegraphics[scale=0.21]{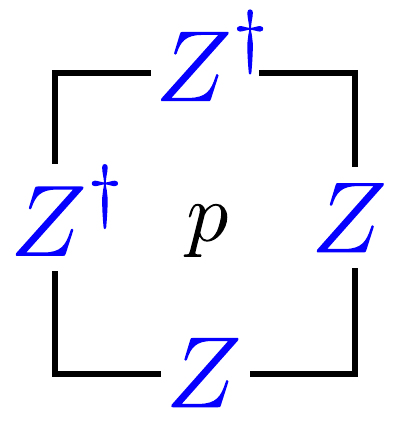}}}.
\end{align}

There is a $\Z_N^{(1)}$ subgroup generated by the $em$ anyons, and $\Z_N^{(-1)}$ generated by $e^{N-1}m$. A useful fact is that for odd $N$, the anyon theory of the $\Z_N$ TC factorizes as $\Z_N^{(1)}\boxtimes \Z_N^{(-1)}$.

Suppose the noise is induced by the following short string operators for $e^{N-1}m$ anyons:
\begin{align}
    \vcenter{\hbox{\includegraphics[scale=0.21]{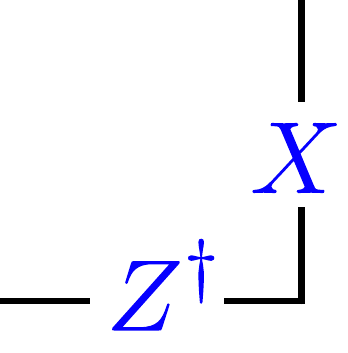}}}, \,\,\quad
    \vcenter{\hbox{\includegraphics[scale=0.21]{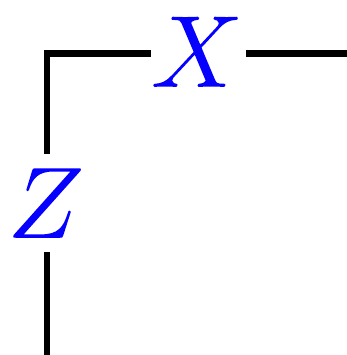}}}.
    \end{align}
   Together with $A_v$ and $B_p$, they generate the gauge group, whose stabilizer group is generated by:
\begin{align} \label{eq: ZN1 stabilizer group construction}
    \mathcal{S}' = \left \langle\: 
    \vcenter{\hbox{\includegraphics[scale=0.21]{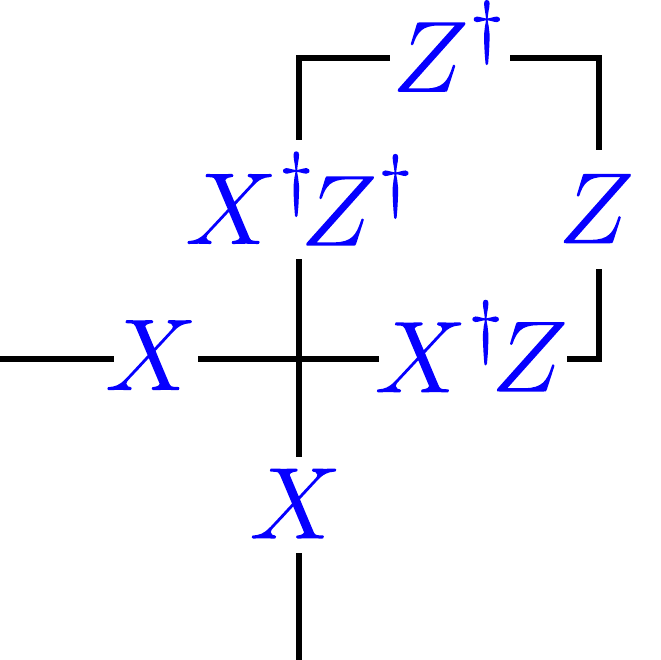}}}\:
    \right \rangle.
\end{align}
Intuitively the local generators are small loops of $em$ anyons.  They can also be defined on non-contractible paths to generate logical operators:
\begin{align} 
    \vcenter{\hbox{\includegraphics[scale=0.21]{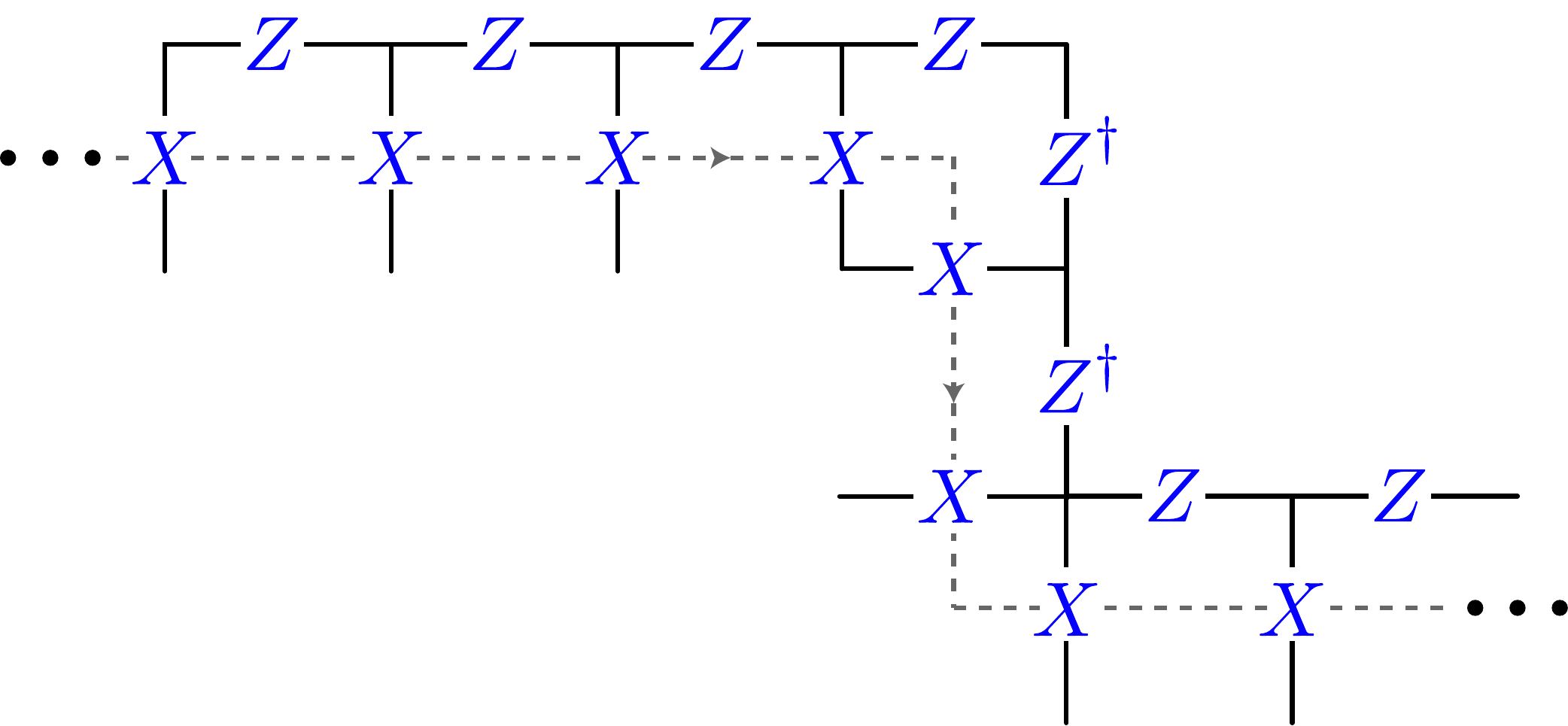}}}
    \label{emlongstring}
\end{align}

On a torus, and for even $N$, ${\cal S}_T$ is generated by the string operators of $e^{N/2}m^{N/2}$ along non-contractible loops.

The anyon theory associated with this topological subsystem code is the $\Z_N^{(1)}$ theory. For odd $N$, the theory is already modular. For even $N$, the transparent center is $\Z_2^{(0)}=\{1,e^{N/2}m^{N/2}\}$.  

\subsubsection{Decohered $\Z_4$ toric code and the symmetry-enriched double semion state}
\label{Z4TC}

We now decohere the $\Z_4$ TC state using noise that proliferates $e^2m^2$ bosons. 
To be explicit, we consider the $\Z_4$ TC 
state with the Krauss operators: 
\begin{equation} \label{eq: Ce terms}
\begin{gathered}
 C_e = \vcenter{\hbox{\includegraphics[scale=.21,trim={0cm 0cm 0cm 0cm},clip]{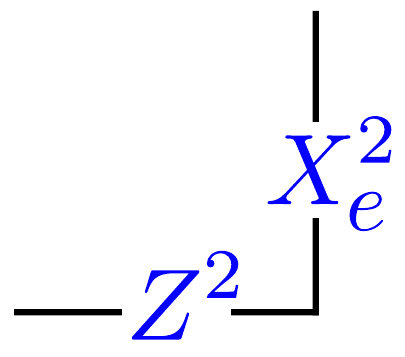}}}, \,\,\quad \vcenter{\hbox{\includegraphics[scale=.21,trim={0cm 0cm 0cm 0cm},clip]{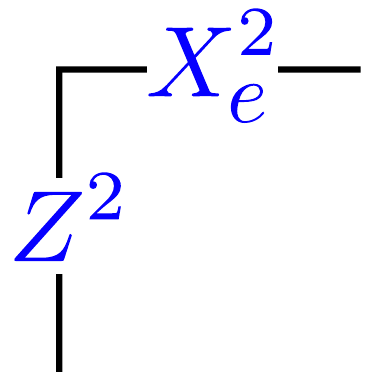}}}.
 \end{gathered}
\end{equation}
They are short string operators that pair create and move $e^2m^2$ anyons. Notice that in the $\Z_4$ TC ground state, $C_e$ satisfies the following constraint at each vertex: 
\begin{equation}
     \prod_{v\in e}C_e=1,
     \label{Ce-prod}
\end{equation}
i.e., it is a product $A_v^{\rm TC}$ and $B_p^{\rm TC}$.
 
Following the analysis in Section~\ref{Pauli}, the stabilizer group $\cal{S}$ of the topological subsystem code is generated by the following operators:
\begin{equation} \label{eq: DS terms}
\begin{gathered}
A_v^{\rm DS} = \vcenter{\hbox{\includegraphics[scale=.21,trim={.5cm 0cm 1.5cm 0cm},clip]{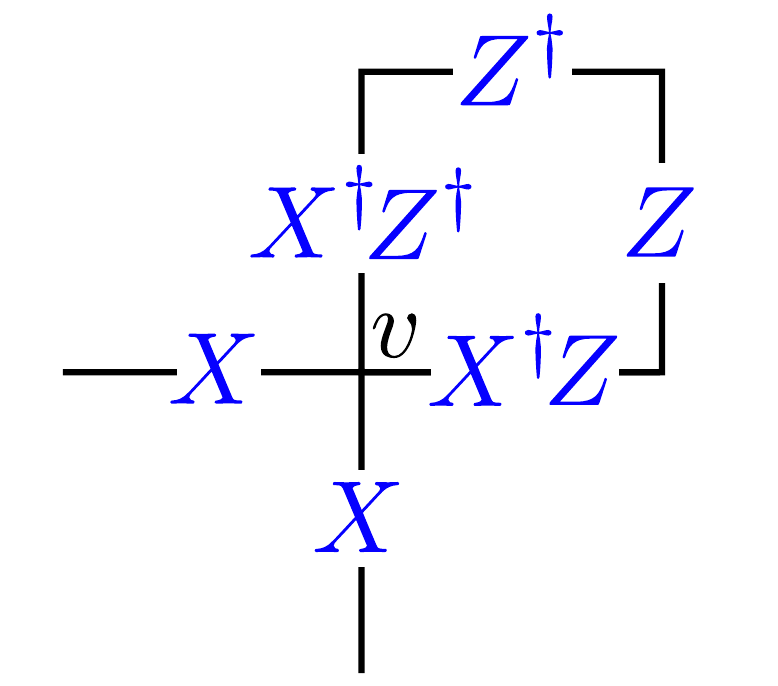}}}, \qquad B_p^{\rm DS} = \vcenter{\hbox{\includegraphics[scale=.21,trim={0cm 0cm 0cm 0cm},clip]{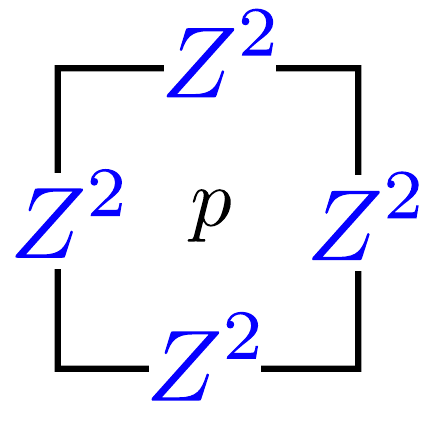}}}. \\
 \end{gathered}
\end{equation}
$A_v^{\rm DS}$ can be viewed as a small $em$ loop and $B_p^{\rm DS}$ as an $e^2$ loop. The mixed state is uniquely determined by the stabilizer group, at least when placed on a sphere -- it is the maximally mixed state in the subspace defined by $A_v^{\rm DS}=B_p^{\rm DS}=1$. Notice that $(A_v^{\rm DS})^2$ can be written as a product of two plaquette operators and four edge operators. 

The logical operators and nonlocal stabilizers are generated by $em$ and $e^3m=e^2\times em$ anyon strings, and can be viewed as $\Z_4^{(1)}\boxtimes \Z_4^{(-1)}$ with the transparent bosons from the two subtheories identified. Further coherently condensing the transparent $e^2m^2$ boson would produce the double semion (DS) theory. 

Alternatively, it proves useful to think of the $\Z_4$ TC state as gauging a $\Z_2$ 0-form symmetry in the DS state. The $\Z_2$ symmetry enriches the DS state in the following way: both the semion $s$ and the anti-semion $s'$ carry ``half charge'' under the $\Z_2$ symmetry. Formally, if we denote the $\Z_2$ charge by $b$, which by definition is a transparent boson, then we have the following fusion rules: $s\times s=s'\times s'=b$. We now claim that the decohered $\Z_4$ TC state can equivalently be represented as a mixture of DS states over all possible configurations of $\Z_2$ defect lines.
 
To see this more explicitly, we consider the following stabilizer Hamiltonian for
the DS state introduced in \cite{Ellison2021}: 
\begin{align}
H_\text{DS}[\mb{s}] = -\sum_v A_v^{\rm DS} - \sum_p B_p^{\rm DS} - \sum_e s_e C_e + \text{h.c.}.
\end{align}
We denote its ground state as $\ket{\psi_{\rm DS}(\mb{s})}$. If all $s_e$ are set to 1, we obtain the translation-invariant DS state as the ground state. Here, we allow $\{s_e=\pm 1\}$ to vary, subject to the constraint given in Eq.~\eqref{Ce-prod}. In other words, the edges with $s_e=-1$ must form contractible closed loops. These are the $\Z_2$ defect loops.

We now define a variant of the double semion stabilizer model, by introducing additional qubits on the plaquettes. The Hamiltonian is modified as shown below:
\begin{equation} \label{eq: DS dynamical domains}
    H_{\rm DS}'=-\sum_v A_v^{\rm TC} X_{\rm{NE}(v)}-\sum_p B_p^{\rm TC} X_p-\sum_e C_e Z_{p_e}Z_{q_e} + \text{h.c.}
\end{equation}
Here, $p_e$ and $q_e$ denote the two plaquettes adjacent to the edge $e$.  

This model has a global $\Z_2$ 0-form symmetry, generated by $\prod_p X_p$. 
Physically, the $C_e Z_{p_e}Z_{q_e}$ pins the $\Z_2$ symmetry defect lines to the domain walls of the plaquette spins. 
Notice that we have $A^{\rm DS}_v=A_v^{\rm TC}B_{{\rm NE}(v)}^{\rm TC}=1$ and $B_p^{\rm DS}=(B_p^{\rm TC})^2=1$. 

Fixing the eigenvalues of the $Z_p$'s, or equivalently choosing a particular domain wall configuration, the Hamiltonian is seen to be exactly equivalent to $H_{\rm DS}[\mb{s}]$ where $s_e=Z_{p_e}Z_{q_e}$. Thus, the ground state wave function can be viewed as a coherent superposition of DS states with varying domain configurations on the plaquettes. If the $\Z_2$ symmetry is absent, one can imagine turning on a Zeeman field to adiabatically connect to the state that satisfies $Z_p=1$ everywhere, which is just the usual DS state.

If the $\Z_2$ 0-form symmetry of the model in Eq.~\eqref{eq: DS dynamical domains} is gauged, we obtain a model in the same phase as the $\Z_4$ TC. Heuristically, the $0$-form symmetry  is gauged by replacing the domain configurations with \add{topological defects}. This implies that the ground states of the $\Z_4$ TC can be viewed as a coherent superposition of (contractible) defect loops in a DS state. By incoherently proliferating the $e^2m^2$ anyons in the TC, the coherent superposition of defect loops is transformed into an equal-weight mixture of defect loops: 
\begin{equation}
    {\sum_{\{s_e\}}}' \op{\psi_{\rm DS}(\mb{s})},
\end{equation}
where ${\sum}'$ indicates that the sum is over $\{s_e\}$ satisfying Eq.~\eqref{Ce-prod}.
In other words, the defect loops of the ket and the bra are bound together, since the decohered state is invariant under conjugation by open $e^2m^2$ string operators, which detect the defect lines. 

Now, we argue that, on a sphere or an infinite plane, the decohered $\Z_4$ TC state can be recovered from a DS state. To see this, we note that the ground state wave function of the model in Eq.~\eqref{eq: DS dynamical domains} can be written as follows:
\begin{equation}
    \ket{\psi_{\rm SET}}=\frac{1}{2^{N_p/2}}\sum_{\{Z_p\}} \ket{\{Z_p\}} \otimes \ket{\psi_{\rm DS}(s_e=Z_{p_e}Z_{q_e})},
\end{equation}
where $\{Z_p\}$ denotes a configuration on the plaquette spins.
Thus, tracing out the plaquette spins, one finds precisely the mixture of DS states with defect loops. Notice that $A^{\rm DS}_v=A_v^{\rm TC}B_{{\rm NE}(v)}^{\rm TC}$ and $B_p^{\rm DS}=(B_p^{\rm TC})^2$ are not affected. In other words, we have found a different purification of the decohered $\Z_4$ TC state, whose TO is described by the DS theory. 

\section{Emergent 1-form symmetries}
\label{sec: 1-form symmetry}

In this section, we discuss a general framework for analyzing the mixed-state TO of Pauli-decohered stabilizer states and mixed states that belong to the same phase. We study, more specifically, the properties of the ``emergent'' symmetries of the mixed states. To get started, we make general statements about symmetries in mixed states. We then introduce 1-form symmetries \cite{Nussinov2009gaugelike, Gaiotto2015, Qi2021higherform}, clarify their connection to anyon theories, and discuss the characterization of mixed-state TOs according to their emergent 1-form symmetries.

\subsection{Strong and weak symmetries}

For a pure state $\ket{\psi}$, a global symmetry is represented by a unitary (or anti-unitary) operator $U$, under which the state is invariant up to a phase: $U\ket{\psi}=e^{i\phi}\ket{\psi}$. We note that, in this work, we only consider unitary symmetries. In a local quantum system, a 0-form symmetry $U$ acts on the entire system. More generally, one can consider symmetry transformations defined on proper subsystems -- for example, on closed 1-dimensional paths, as described in the next section.

To define symmetry in mixed states, we need to distinguish whether the symmetry acts nontrivially on the environment, leading to two distinct notions of global symmetry~\cite{BucaProsen2012, AlbertJiang2014, Albert2018, Lieu2020, deGroot2022}. If the symmetry does not act on the environment, in other words, the system and the environment do not exchange symmetry charges, then the symmetry is said to be ``strong''. By this definition, a mixed state $\rho$ with strong symmetry $U$ can be decomposed into a mixture of pure states all of which have the same total charge under the symmetry. That is, we must have 
\begin{equation}
    U\rho = e^{i\phi}\rho,
\end{equation}
which can be taken as the definition of strong symmetry. 

If on the contrary the symmetry also acts nontrivially on the environment, then the symmetry is called ``weak''. In this case, we only have 
\begin{equation}
    U\rho U^\dag = \rho.
\end{equation}
We note that it has been recently understood that strong and weak symmetries play very different roles in mixed-state SPT orders~\cite{deGroot2022, Ma:2022pvq, Ma:2023rji, Ma:2024kma, Xue:2024bkt, guo2024locally, Hsin2023anomalies}.

\subsection{1-form symmetries of gapped ground states}

A modern view on TO in GGSs is to consider the system's emergent higher-form symmetries~\cite{Hastings:2005xm,  Gaiotto2015, Pace2023} and the non-invertible generalizations~\cite{KongHolographicSym, Cian:2022vjb}. Abelian topological states in (2+1)$d$ are characterized by emergent 1-form symmetries, which are, intuitively, generated by loops of anyon string operators. The TO can then be interpreted as spontaneously broken 1-form symmetry.  

For our purpose, we adopt the following working definition of a 1-form symmetry: for a closed path $\gamma$ (which may be on the lattice or the dual lattice), we associate a unitary operator $W(\gamma)$. In many cases, e.g., in stabilizer models, $W(\gamma)$ is actually a finite-depth local unitary operator supported in the neighborhood of $\gamma$. A GGS $\ket{\psi}$ has the emergent 1-form symmetry if $\ket{\psi}$ is an approximate  eigenstate of $W(\gamma)$ for all contractible $\gamma$:
\begin{equation}
    W(\gamma)\ket{\psi}\simeq e^{i\alpha}\ket{\psi},
\end{equation}
where $\simeq$ means up to $O(L^{-\infty})$ corrections.
Notice that we do not require that the eigenvalues are $1$, although this is the case for all examples considered here. 

For a given GGS, the set of 1-form symmetry operators is naturally endowed with the structure of a group, where the group multiplication is simply the multiplication of the unitary operators. 

We can further associate an anyon theory to a 1-form symmetry. To do so, we first define the notion of a ``breakable'' 1-form symmetry on a state $|\psi\rangle$. Here, ``breakable'' is defined more precisely by considering an open path $\gamma_{ab}$ connecting $a$ and $b$. We let the string operator $W(\gamma_{ab})$ be the truncation of the symmetry operator supported on a large loop containing $\gamma_{ab}$. The open string operator $W(\gamma_{ab})$ is well-defined up to local unitary operators near the end points. We say the 1-form symmetry is breakable if there exists local unitaries $U_a$ and $U_b$, supported near $a$ and $b$, such that $U_aW(\gamma_{ab})U_b\ket{\psi}=e^{i \phi}\ket{\psi}$. In other words, $W(\gamma_{ab})$ only creates local excitations. Note that whether a 1-form symmetry operator is breakable depends in general on the state $\ket{\psi}$.

The anyon theory of a 1-form symmetry $\cal{W}$ for a state $\ket{\psi}$ is defined as $\cal{W}$ modulo the breakable symmetry operators on $\ket{\psi}$. Every GGS has a (possibly trivial) emergent 1-form symmetry group, and the associated anyon theory is invariant throughout the phase.

While we have focused on {emergent} 1-form symmetry, one can also consider \emph{microscopic} (or exact) 1-form symmetry, which are true symmetries of the Hamiltonian. Topological subsystem codes provide many examples of Hamiltonians with exact 1-form symmetry groups~\cite{Ellison:2022web}.

\subsubsection*{Anomaly of 1-form symmetry}

Just as for any global symmetry, 1-form symmetries can exhibit 't Hooft anomalies~\cite{Gaiotto2015, Hsin:2018vcg}. For a finite 1-form symmetry group associated to an anyon theory $\cal{A}$, the anomaly is fully characterized by the exchange statistics $\theta(a)$ for $a\in \cal{A}$, defined in Section~\ref{sec:decoheredSS} for Abelian anyon theories. The 1-form symmetry is non-anomalous if and only if $\theta(a)=1$ for every $a\in {\cal A}$. For pure states, an anomalous symmetry forbids symmetry-preserving short-range entangled states.

As an example, consider the emergent 1-form symmetries in the TC phase. In the original TC model, there are two kinds of loop operators: $W_e(\gamma)$ the product of $Z$ along a direct loop which creates and moves $e$ particles, and $W_m(\gamma^*)$ the product of $X$ along a dual loop $\gamma^*$. For contractible loops, $W_e(\gamma)$ and $W_m(\gamma^*)$ are both written in terms of stabilizers.  
Thus, the ground state of the TC model has $\Z_2\times \Z_2$ 1-form symmetry. In fact, in this case, the $\Z_2\times\Z_2$ 1-form symmetry is exact as the Hamiltonian commutes with the symmetry operators. There is a mixed anomaly between the two $\Z_2$ subgroups (i.e., the braiding statistics between $e$ and $m$ anyons).

If the TC model is tuned away from the fixed-point limit, e.g., by adding a small magnetic field, but still remains in the TC phase, the ground states are no longer eigenstates of $W_e$ and $W_m$. In fact, their expectation values decay exponentially with the length of the loop. However, one can find a new set of loop operators $\tilde{W}_a$, which are emergent 1-form symmetries for the deformed ground state. The string operators $\tilde{W}_a$ can be constructed by conjugating $W_a$'s with a quasi-adiabatic evolution operator~\cite{Hastings:2005xm}. In the generic case, the 1-form symmetry is only emergent for ground states and low-lying excited states.

\subsection{1-form symmetries of mixed states} \label{sec: anyon theories SLDQC}

In this section, we extend the discussion of 1-form symmetries to mixed states. We define a strong 1-form symmetry operator of a mixed state $\rho$ as a loop-like unitary operator $W$, which satisfies $W\rho=e^{i\alpha}\rho$, for some phase factor $e^{i\alpha}$. Similarly, we say a strong 1-form symmetry is breakable, if the open string operator $W(\gamma_{ab})$ satisfies $U_aW(\gamma_{ab})U_b\rho=e^{i\beta}\rho$ for some local unitaries $U_a$ and $U_b$.

Before developing a notion of an \textit{emergent} strong 1-form symmetry for mixed states, let us consider the strong 1-form symmetries of the decohered stabilizer states of Section~\ref{sec:decoheredSS}. As described in Section~\ref{sec:decoheredSS}, topological subsystem codes define a family of decohered stabilizer states, with varying levels of noise. Further, 
as described in Ref.~\cite{Ellison:2022web}, topological subsystem codes are characterized by premodular Abelian anyon theories. The anyon theory of the subsystem code is precisely the strong 1-form symmetry group of the decohered state. 

As a simple example, the subsystem code corresponding to incoherently proliferating $m$ anyons in a $\Z_2$ TC state is characterized by the $\{1,m\}$ anyon theory. In agreement with this is the fact that the strong 1-form symmetry of the decohered state is generated by loops of $m$ string operators. Note that, in this case, the 1-form symmetry is non-anomalous, and the decohered state belongs to the same phase as the maximally mixed state, which has no strong symmetries. As another example, the subsystem code corresponding to incoherently proliferating $e^{-1}m$ in a $\Z_N$ TC is characterized by the anyon theory $\Z_N^{(1)}$. Regardless of the strength of the noise, the mixed state has a strong 1-form symmetry generated by the $e^{-1}m$ string operators.

Now, to move beyond mixed states derived from topological subsystem codes, we consider the effects of QLCs on the strong 1-form symmetries of mixed states. This leads us to a notion of emergent strong 1-form symmetries. 

Suppose that $\rho_1$ and $\rho_2$ are mixed states that can be connected by a QLC $\cal{N}_{21}$, i.e., $\cal{N}_{21}(\rho_1) = \rho_2$. If $\rho_2$ has a strong 1-form symmetry, with an arbitrary symmetry operator represented as $W_2$, such that $W_2\rho_2=\rho_2$, then we have the following chain of equalities:
\begin{eqs} \label{eq: sym rho2 1}
    1=\Tr[\rho_2] = \Tr[W_2 \rho_2] = \Tr[W_2 \cal{N}_{21}(\rho_1)].
\end{eqs}
If we further purify the channel in terms of a $\mathrm{polylog}(L)$-depth local circuit $V_{21}$, we obtain:
\begin{eqs} \label{eq: sym rho2 2}
    1=\Tr[W_2 V^\dagger \rho_1 \otimes \op{0} V] = \Tr[V W_2 V^\dagger \rho_1 \otimes \op{0}],
\end{eqs}
where $\op{0}$ is a many-body product state in an ancillary Hilbert space. 

This implies that $\rho_1 \otimes \op{0}$ has a strong 1-form symmetry represented by $V W_2 V^\dagger$.\footnote{It follows from the simple lemma that if $U$ is unitary and $|\Tr [U\rho]|=1$, then $U$ is a strong symmetry of $\rho$. To prove this, we expand $\rho$ in its eigenbasis: $\rho=\sum_n p_n \op{\psi_n}$. Then, $\Tr [U\rho]=\sum_n p_n \braket{\psi_n|U|\psi_n}$. Since $U$ is unitary, $|\braket{\psi_n |U|\psi_n}|\leq 1$, thus $\Tr [U\rho] \leq \sum_n p_n=1$. The equality is reached when $\braket{\psi_n|U|\psi_n}=e^{i\alpha}$ for some $\alpha$ independent of $n$, which implies $U\ket{\psi_n}=e^{i\alpha}\ket{\psi_n}$, so $U$ is a strong symmetry of $\rho$.} Because $V$ is a QLUC, $VW_2V^\dagger$ remains a 1-form symmetry operator, with exactly the same group structure and anomaly. Thus, every strong 1-form symmetry of $\rho_2$ corresponds to one for $\rho_1\otimes \op{0}$. Note that, in determining the strong 1-form symmetries of a state, we allow ourselves to freely append ancillas. Therefore, the strong 1-form symmetries of $\rho_1 \otimes \op{0}$ are, by definition, the same as those of $\rho_1$. This gives us
\begin{eqs} \label{eq: 1-form sub condition}
    \cal{W}_2 \subset \cal{W}_1,
\end{eqs}
where $\cal{W}_1$ and $\cal{W}_2$ are the strong 1-form symmetries of $\rho_1$ and $\rho_2$, and $\subset$ denotes a subgroup. In this sense, the strong 1-form symmetry of $\rho_2$ is emergent for $\rho_1$. 

As an example, the reasoning here can be used to define an emergent 1-form symmetry for the decohered $\Z_2$ TC state when the noise strength is small -- in that case, one can find an explicit QLC (the ``recovery'' channel) that maps the decohered $\Z_2$ TC state back to the pure one~\cite{Sang:2023rsp}. Thus, once purifying the channel, one finds strong 1-form symmetry operators for the decohered $\Z_2$ TC state (tensored with ancilla in a product state). 

If $\rho_1$ and $\rho_2$ are two-way connected by QLCs, then their strong 1-form symmetries must be equivalent. However, it is important to note  that this does not imply that the anyon theories of $\rho_1$ and $\rho_2$ are equivalent. What is missing is the ``breakability'' condition. In what follows, we study the implications of Eq.~\eqref{eq: 1-form sub condition} for the associated anyon theories.

To get started, we make the following observation:
\begin{itemize}
    \item If a symmetry operator $W_2$ is unbreakable on $\rho_2$ and corresponds to a transparent boson, then the symmetry $VW_2V^\dagger$ may be breakable on $\rho_1$.
\end{itemize}
Physically, this means that a QLC is capable of turning a trivial anyon into a (nontrivial) transparent boson. This is exhibited by the decohered $\Z_4$ TC example in Section~\ref{Z4TC}. In that case, there is a QLC that maps the pure DS state to the decohered $\Z_4$ TC. The pure DS state does not have any transparent bosons, while the decohered $\Z_4$ TC does have one. More explicitly, the 1-form symmetry corresponding to $e^2m^2$ is unbreakable for the decohered $\Z_4$ TC, while it is breakable for the DS state.  The mechanism behind this phenomenon can be intuitively understood as follows: the breakable 1-form symmetry operator in $\rho_1$ terminates on certain local operators, which are traced out by the quantum channel, making the 1-form symmetry unbreakable. 

This observation implies that the anyon theories $\cal{A}_1$ and $\cal{A}_2$, derived from $\cal{W}_1$ and $\cal{W}_2$, must satisfy
\begin{eqs} \label{eq: anyon subtheories}
    \cal{A}_2/\cal{B}_2 \subset \cal{A}_1,
\end{eqs}
where $\subset$ denotes a subtheory. Here, $\cal{A}_2/\cal{B}_2$ is the anyon theory obtained by condensing some subgroup $\cal{B}_2$ of the transparent bosons of $\cal{A}_2$. The subgroup $\cal{B}_2$ is necessary to include in Eq.~\eqref{eq: anyon subtheories}, since unbreakable 1-form symmetries of $\rho_2$ may correspond to breakable 1-form symmetries of $\rho_1$.

To gain intuition for Eq.~\eqref{eq: anyon subtheories}, let us consider a few examples. As a trivial example, the maximally-mixed state can be prepared from any other mixed state by applying a depolarizing noise channel. Since the maximally-mixed state does not have any strong 1-form symmetries, the expression in Eq.~\eqref{eq: anyon subtheories} is trivially satisfied with $\cal{A}_2 = \id$. 

As a second example, the fermion-deochered TC state in Section~\ref{fermion-TC} can be prepared from a pure state TC by a QLC. The expression in Eq.~\eqref{eq: anyon subtheories} simply tells us that the anyon theory of the fermion-decohered TC state (i.e., $\Z_2^{(1)}$) is a subtheory of the TC anyon theory (i.e., the subtheory generated by $em$). 

As a final example, the decohered $\Z_4$ TC state can be prepared from a pure DS state with a QLC. In this case, $\cal{B}_2$ must be nontrivial for the expression to hold. $\cal{B}_2$ can be taken to be the subgroup of transparent bosons generated by $e^2m^2$ for the decohered $\Z_4$ TC state.

We now consider two mixed states $\rho_1$ and $\rho_2$ that are two-way connected by QLCs. According to Eq.~\eqref{eq: anyon subtheories}, there exists subgroups of transparent bosons $\cal{B}_1$ and $\cal{B}_2$ such that
\begin{eqs} \label{eq: two-way connected anyons}
    \cal{A}_2/\cal{B}_2 \subset \cal{A}_1, \quad \cal{A}_1/\cal{B}_1 \subset \cal{A}_2,
\end{eqs}
where $\cal{A}_1$ and $\cal{A}_2$ are the anyon theories of $\rho_1$ and $\rho_2$. 

These conditions impose strong constraints on the anyon theories $\cal{A}_1$ and $\cal{A}_2$. 
Let us discuss three consequences of the conditions in Eq.~\eqref{eq: two-way connected anyons}:
\begin{enumerate}
    \item Let $\cal{T}_i$  be the full subgroup of transparent bosons of $\cal{A}_i$, and define $\cal{A}_i^{\rm min} = \cal{A}_i/\cal{T}_i$.  We prove in Appendix~\ref{app: Amin} that Eq.~\eqref{eq: two-way connected anyons} implies that
\begin{eqs} \label{eq: equality of mins}
    \cal{A}_1^{\rm min} = \cal{A}_2^{\rm min},
\end{eqs}
This means that $\cal{A}_i^{\rm min}$ is invariant under QLCs and thus, can be used to distinguish between mixed-state TOs. 

\item If $\cal{A}_1$ or $\cal{A}_2$ is modular, one can show that Eq.~\eqref{eq: two-way connected anyons} implies $\cal{A}_1=\cal{A}_2$. This is a special case of a more general theorem proven in Section~\ref{sec: general intrinsic}.

\item We prove in Appendix~\ref{app: single generator} that if $\cal{A}_1$ and $\cal{A}_2$ both have a single generator, Eq.~\eqref{eq: two-way connected anyons} also implies $\cal{A}_1=\cal{A}_2$.
\end{enumerate}

Our discussion so far leads to a partial classification of Abelian mixed-state TOs in terms of the minimal anyon theory $\cal{A}_i^{\rm min}$ -- that is, two mixed states with different minimal anyon theories must belong to different mixed-state phases. This classification is consistent with the result proven in \cite{Coser2019classificationof}, that $\Z_M$ and $\Z_N$ TCs belong to different mixed-state phases when $M \neq N$. 

Another implication of our classification result is that the fermion-decohered TC discussed in Section~\ref{fermion-TC} is a mixed-state TO distinct from any ground state TO in a bosonic system. Therefore, in this sense, it is an example of an ``intrinsically'' mixed-state TO, as proposed in Ref.~\cite{Wang:2023uoj}. More generally, any mixed-state TO characterized by a premodular anyon theory is an intrinsically mixed-state TO, excluding cases where the anyon theory decomposes as $\C \boxtimes \cal{T}$, for a modular theory $\C$ and a theory of transparent bosons $\cal{T}$.  

Based on these results, we conjecture that the conditions in Eq.~\eqref{eq: two-way connected anyons} are sufficient to prove that $\cal{A}_1 = \cal{A}_2$, for arbitrary Abelian premodular anyon theories. This holds for all of the examples considered in this text, and we are currently unaware of any counterexamples. 

\subsubsection*{Weak 1-form symmetry}

We briefly comment on the weak 1-form symmetries of mixed states, focusing on those obtained by decohering stabilizer states with Pauli noise. For these examples, we have $W\rho W^\dag = \rho$, for every 1-form symmetry operator $W$ of the pure stabilizer state. The decohered TC state, for example, for any amount of bit-flip noise, has weak 1-form symmetries generated by the $e$ and $m$ string operators.

More generally, the weak 1-form symmetries of a Pauli-decohered stabilizer state correspond to the ``fluxes'' of a topological subsystem code, using the language of Refs.~\cite{Bombin2014structure, Ellison:2022web}. Loosely speaking, the fluxes of a topological subsystem code are created by string operators that commute with all of the stabilizers along their length. They may or may not commute with the gauge operators outside of the stabilizer group. Therefore, the string operators along closed paths, in particular, commute with all of the stabilizers. Since the maximally-decohered state is a sum of stabilizers, the closed flux string operators commute with the mixed state. Hence, they yield weak 1-form symmetries.

We note that weak 0-form symmetries are important to our understanding of symmetry-protected mixed states~\cite{Ma:2022pvq, Ma:2023rji} and strong-weak spontaneous symmetry breaking states. Weak 0-form symmetries also feature in our construction of topologically ordered mixed states in Section~\ref{sec: classical gauging}. However, it is unclear whether weak 1-form symmetries play a larger role in the classification and characterization of mixed-state topological orders. We leave this to future investigations. 

\section{Mixed states with generalized 1-form symmetries}
\label{sec: stringnet}

We now go beyond Pauli stabilizer states and consider mixed-state TOs built from models that support non-Abelian anyons. In other words, they exhibit generalized, non-invertible 1-form symmetries. As a first example, we construct a mixed state by decohering an Ising string-net model. The resulting mixed state is characterized by an anyon theory that is non-Abelian and thus, falls outside of the purview of the previous sections. We study the locally indistinguishable states obtained by creating non-Abelian anyons and by defining the system on a torus.

We then generalize the construction of the decohered Ising string-net model to $G$-graded string-net models. 
Subsequently, we further generalize the construction to mixed states built by ``classically gauging'' the weak symmetry of a bosonic symmetry-enriched topological (SET) order. Finally, we give the most general construction, in which we construct mixed states from decohering Walker-Wang models. We conclude this section by discussing a general algebraic characterization of mixed-state TOs in terms of premodular anyon theories and the equivalence relations induced by QLCs.

\subsection{Example: decohered Ising string-net model} \label{sec: decohered Ising}

We begin by briefly summarizing the relevant details of Ising string-net models. For more complete expositions, we recommend Refs.~\cite{Levin:2004mi, Lin:2020bak}. We will also give more details for the general case in Section~\ref{sec: G-graded stringnet}. String-net models are exactly-solvable, commuting-projector Hamiltonians with topologically ordered ground states. In their most general form, they can realize any (2+1)$d$ TO that admits a gapped boundary (known as a quantum double). Some of the examples considered in the previous section, such as the $\Z_N$ TC, can be viewed as special cases of string-net models. Below, we focus on the so-called Ising string-net model to illustrate the more general construction of mixed states.

\subsubsection*{Pure state wave function}

First, we briefly review the ground state wave function of the Ising string-net model. For convenience, the model is defined on a honeycomb lattice. On each edge of the lattice there is a 3-dimensional Hilbert space, with an orthonormal basis labeled as $\ket{1}, \ket{\sigma}$ and $\ket{\psi}$. They will be referred to as ``string types'', and graphically represented as
\begin{align}
    1:\vcenter{\hbox{\includegraphics[scale=1.0]{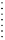}}} \quad\quad \sigma&:\vcenter{\hbox{\includegraphics[scale=1.0]{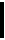}}} \quad\quad \psi:\vcenter{\hbox{\includegraphics[scale=1.0]{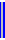}}}\, .
\end{align}
We further define an operator $\mu^z_e$ for each edge $e$, such that $\mu^z\ket{\sigma}=-\ket{\sigma}$ and $\mu^z\ket{1/\psi}=\ket{1/\psi}$. $\mu^z$ defines a ``$\Z_2$ grading'' on the Hilbert space.

On each vertex where three strings meet, we impose the following branching constraints: (1) a $\sigma$ string can never terminate on a vertex, and (2) a $\psi$ string can terminate on a vertex only if a $\sigma$ string passes through the vertex. String configurations that satisfy these constraints are called Ising string-net states. For later use, for each vertex $v$ we define a projector $A_v$ that is 1 on states that satisfy the branching constraint on $v$, and 0 otherwise. The product $\prod_v A_v$ projects to the space of string-net states.
Graphically, the string-net states have loops of $\sigma$ strings and $\psi$ strings that either form loops or terminate on the $\sigma$ lines.  An example of a string-net state in a honeycomb lattice is shown in Fig. \ref{fig: Isingconfig} (ignoring the degrees of freedom on the hexagons).

The ground state wave function $\ket{\Psi}$ is a superposition of all string-net states. For a given string-net state $X$, the amplitude in the (un-normalized) wave function is given by 
\begin{equation}
    \braket{X|\Psi}=2^{N_\sigma/2} f(X),
\end{equation}
where $N_\sigma$ is the number of $\sigma$ loops in $X$, and $f(X)$ is $0$ or $\pm 1$.  We refer the readers to \cite{HeinrichPRB2016} for the explicit expression of $f(X)$.  

The wave function $\ket{\Psi}$ is the ground state of the following Hamiltonian:
\begin{equation}
    H_{\rm Ising}=-\sum_v A_v - \sum_p B_p^+,
    \label{eq: Ising SN Hamiltonian}
\end{equation}
Here, $p$ sums over all plaquettes, and $B^+_p$ is defined as 
\begin{equation}
    B_p^\pm=\frac14(1+B_p^\psi\pm\sqrt{2}B_p^\sigma).
\end{equation}
For the definitions of the $B_p^\psi$ and $B_p^\sigma$ operators we refer the readers to Ref.~\cite{ChengSET2017} -- intuitively, they fuse a $\psi$ or $\sigma$ string into the plaquette $p$, respectively. For now, it suffices to notice the following: $(B_p^\psi)^2=1$ and $\left(B^+_p\right)^2=B^+_p$. The operator $B_p^\psi$ does not change the $\Z_2$ grading, determined by $\mu^z_e$, but $B_p^\sigma$ flips the $\Z_2$ grading on all 6 edges of the hexagon $p$. Thus, the operator $B_p^\sigma$ anti-commutes with $\mu_e^z$ for $e\in \partial p$. 

We now give two alternative representations of the ground state, which turn out to be useful later. We denote a configuration of $\sigma$ loops by $\set{\sigma}$ and define a projector $P(\set{\sigma})$, which annihilates any string-net state whose $\sigma$ loops are different from $\set{\sigma}$. With this, we can define 
\begin{equation}
    \ket{\set{\sigma}} = P(\set{\sigma})\ket{\Psi}.
\end{equation}
The ground state wave function of the Ising string-net model can be written as 
\begin{equation}
     \ket{\Psi}= \sum_{\set{\sigma}}\ket{\set{\sigma}}.
\end{equation}
Note that, if there are no $\sigma$ loops at all, then the state is just a quantum superposition of all closed $\psi$ loops, i.e., a $\Z_2$ TC state. The $\sigma$ loops amount to inserting certain topological defect loops into the TC.
 
Lastly, since the Hamiltonian is a sum of commuting projectors, the ground state density matrix (on a sphere) can be written as 
\begin{equation}
    \rho_0=\op{\Psi}=\prod_p B^+_p \prod_v A_v.
    \label{Isingrho0}
\end{equation}

\subsubsection*{Decohered density matrix}
We now couple the system to the following decoherence channel: 
\begin{equation} \label{eq: Ising decoherence}
    \cal{N}=\prod_e \cal{N}_e, \quad \: {\cal N}_e(\rho)=\frac12(\rho+\mu_e^z\rho\mu_e^z).
\end{equation}
One can see that $\cal{N}(\rho_0)$ takes the following form:
\begin{equation}
   \cal{N}(\rho_0) \propto \sum_{\set{\sigma}} \op{\set{\sigma}}.
    \label{rhoIsing}
\end{equation}
The $\sigma$ loops only proliferate probabilistically in the ensemble. Note that the $\psi$ lines can still fluctuate coherently if the $\sigma$ loops are fixed.

Using the representation of the ground state given in Eq.~\eqref{Isingrho0}, we see that, alternatively,
\begin{equation}
\cal{N}(\rho_0) \propto\Big(\sum_{\{s_p=\pm\}}\delta_{\prod_p s_p,1}\prod_p B_p^{s_p}\Big)\prod_v A_v,
\end{equation}
Physically, $B_p^-$ projects to the state with a $\psi\bar{\psi}$ anyon in the plaquette $p$. This excitation can be measured by the $B_p^\sigma$ operator, due to the following relation:
\begin{equation}
    B_p^\sigma B_p^\pm = \pm \sqrt{2} B_p^\pm, \quad B_p^\psi B_p^\pm =  B_p^\pm.
\end{equation}

Assuming that the model is placed on a sphere, then there can only be an even number of $\psi\bar{\psi}$ anyons. In other words, the state with an odd number of $\psi\bar{\psi}$'s must be 0. Thus the $\delta_{\prod_p s_p,1}$ factor in the sum can be dropped, and we find
\begin{equation}
\cal{N}(\rho_0) \propto\prod_p \frac{1+B_p^\psi}{4}\prod_v A_v.
\end{equation}

\subsubsection*{Purification to the SET state}

We now show that the decohered doubled Ising mixed state defined in Eq.~\eqref{rhoIsing} can be purified into a symmetry-enriched $\Z_2$ TC state, where the $\Z_2$ symmetry permutes $e$ and $m$ anyons, at least when the underlying manifold is a sphere.

\begin{figure}[t] 
\centering
\includegraphics[width=.45\textwidth]{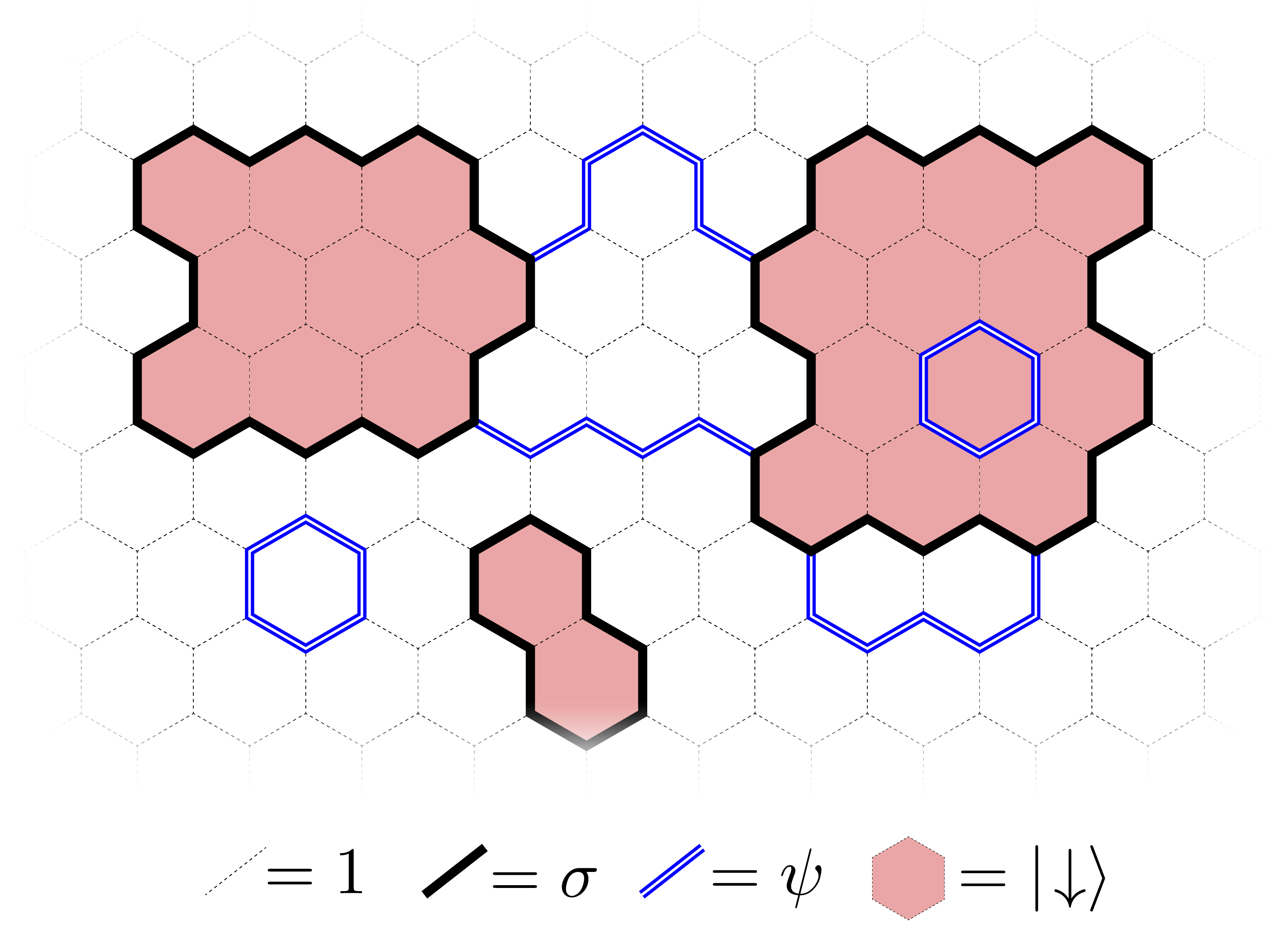}
\caption{A configuration of the symmetry-enriched TC with a nonzero amplitude. The symmetry-enriched TC in Ref.~\cite{HeinrichPRB2016} and \cite{ChengSET2017} is defined on a honeycomb lattice with three basis states $\ket{1}$, $\ket{\sigma}$, and $\ket{\psi}$ on the edges and an Ising spin on each plaquette. The ground state wave function has nonzero amplitudes for configurations for which the $\sigma$ lines appear on the domain walls of the plaquette spins, and the $\psi$ lines either form loops or terminate on a $\sigma$ line. Note that, ignoring the plaquette spins, the edge degrees of freedom give a string-net state for the Ising string-net state.}
\label{fig: Isingconfig}
\end{figure}
 
First, we review the symmetry-enriched TC state, which can be constructed using a variation of the string-net model as described in \cite{HeinrichPRB2016} and \cite{ChengSET2017}.
Starting from the Ising string-net model, we add Ising spins to each hexagon, whose Pauli operators are denoted by $\tau^\alpha$, for $\alpha = x,y,z$. Then we impose the constraint that the $\Z_2$ grading $\mu_e^z$ on an edge $e$ must be equal to $\tau_p^z\tau_q^z$, where $p$ and $q$ are the two adjacent hexagons. Namely, the $\sigma$ loops are bound to domain walls of $\tau$ spins (see Fig.~\ref{fig: Isingconfig} for an illustration). The resulting model has a global $\Z_2$ symmetry generated by $\prod_p \tau_p^x$.  

Within the space of string-net states, the Hamiltonian for the $\Z_2$ TC state takes the following form:
\begin{equation}
     H_{\rm SET}=-\sum_p \frac14(B_p^1+B_p^\psi+\tau_p^x B_p^\sigma)-\sum_e P_e,
\end{equation}
where the edge projector is $P_e=\frac12(1+\tau_{p_e}^z\mu_e^z\tau_{q_e}^z)$. The ground state density matrix is given by 
\begin{equation}
     \rho_{\rm SET}=\prod_p \frac{1+B_p^\psi+\tau_p^xB_p^\sigma}{4}\prod_v A_v\prod_e P_e.
\end{equation}
Alternatively, the SET state is
\begin{equation}
     \ket{\Psi_{\rm SET}}=\sum_{\{\tau^z\}} \ket{\{\tau^z\}}\otimes \ket{\set{\partial\tau}}.
 \end{equation}
Here $\partial\tau$ denotes the domain wall configuration of the plaquette spins.
The ground state is in the same phase as the $\Z_2$ TC, without imposing the $\Z_2$ 0-form symmetry~\cite{ChengSET2017}. This can be seen by simply polarizing the plaquette spins, which results in a coherent fluctuation $\psi$ lines without any $\sigma$ lines. We note that the $\Z_2$ SET order is related to the doubled Ising TO by gauging the $\Z_2$ symmetry.

To see that the state $\ket{\Psi_{\rm SET}}$ is a purification of the mixed state $\rho$ in Eq.~\eqref{rhoIsing}, we just need to trace out the plaquette spins. 
This gives us
\begin{equation}
    \rho\propto\sum_{\{\sigma\}}\op{\set{\sigma}}.
\end{equation}
Here, the sum is over all contractible $\sigma$ loop configurations. On a sphere, this is identical to decohered doubled Ising state.

\subsubsection*{Anyons and string operators}

For the Ising string-net model, there are nine anyon types labeled by $a\bar{b}$ where $a,b\in \{1,\sigma,\psi\}$. For brevity, we write $a\bar{1}$ as just $a$, and similarly $1\bar{a}$ as just $\bar{a}$. Furthermore, we write the trivial anyon as $I = 1\bar{1}$. These anyons are created and moved by string operators $W_{a\bar{b}}(\gamma)$, where $\gamma$ is a path on the lattice. When $\gamma$ is closed, the string operator keeps the ground state invariant. When $\gamma$ is open, $W_{a\bar{b}}(\gamma)\ket{\Psi}$ has two excitations created at the end points of $\gamma$, and away from the end points the state is locally indistinguishable from $\ket{\Psi}$.

We focus on those string operators that act ``diagonally'' in the basis $\ket{\set{\sigma}}$. 
More precisely, the closed string operators keep each of the $\ket{\set{\sigma}}$ states invariant (up to an overall factor). Only the following anyon types satisfy this requirement: $I,  \psi,\bar{\psi}, \psi\bar{\psi}$ and $\sigma\bar{\sigma}$. These string operators share a common feature, that is, they do not change the $\Z_2$ grading on the edges, thus remain well-defined in the presence of strong decoherence. 

The $\psi\bar{\psi}$ string is most straightforward to write down. Choose a path $\gamma$ on the dual lattice, then 
\begin{equation}
    W_{\psi\bar{\psi}}(\gamma)=\prod_{e\perp \gamma} \mu_e^z,
    \label{eq:Wboson}
\end{equation}
Here $e\perp \gamma$ means that the edge $e$ intersects $\gamma$. Note that here $\gamma$ does not have to be closed.  For the density matrix $\rho$, it is easy to see that 
\begin{equation}
    W_{\psi\bar{\psi}}(\gamma)\rho W_{\psi\bar{\psi}}(\gamma)=\rho,
    \label{Wboson}
\end{equation}
for any closed or open path $\gamma$. This is consistent with the picture that $\psi\bar{\psi}$ has proliferated incoherently, therefore $W_{\psi\bar{\psi}}$ does not create excitations. However, note that in contrast to coherent condensation, $\psi \bar{\psi}$ is not identified with the trivial anyon. In other words, $W_{\psi\bar{\psi}}$ is not ``strongly'' breakable.

To write down the other string operators, especially the one for $\sigma\bar{\sigma}$, we need to introduce the following ``local'' graphical representation for string operators~\cite{Levin:2004mi}. A string operator $W_{\bm{a}}$ is represented by a directed string acting along an open or closed path on the lattice. Graphically, we draw a string lying on top of the graph state to represent the string operator. Its action on a given basis state is defined using the following rule to resolve each overcrossing:
\begin{equation}
\AddFig{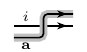}
     =\sum_{jst} \Omega^j_{\bm{a},ist}
\AddFig{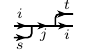}
\end{equation}
On the left-hand side, $\bm{a}$ labels the anyon of the string operator, and $i$ labels the string type on the edge of the lattice. On the right-hand side, we have introduced artificial \add{strings} that are to be ``fused'' into the lattice (see Ref.~\cite{Levin:2004mi}).
In other words, once all crossings are resolved, the string diagrams can then be reduced to a superposition of string-net states using the diagrammatic rules of the Ising fusion category.

In this representation, the $\psi$/$\bar{\psi}$ strings are given by
\begin{align}
    \AddFig{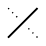} &= \AddFig{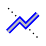}\\
\AddFig{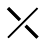} &= \pm i \AddFig{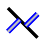}\\
\AddFig{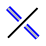} &= -\AddFig{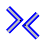}
\end{align}
In these diagrams, the string operator on the left-hand side has positive slope, while the edge of the lattice has negative slope.
Note that $\psi$ and $\bar{\psi}$ differ by a $\psi\bar{\psi}$ string. So when conjugating the density matrix, $\psi$ and $\bar{\psi}$ have identical effect, even though they are distinct string operators.

Now we turn to the string operator for the non-Abelian anyon $\sigma\bar{\sigma}$:
\begin{align}
    \AddFig{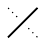} &= \AddFig{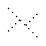} + \AddFig{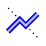}\\
    \AddFig{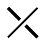} &= \AddFig{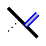} + \AddFig{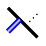}\\
    \AddFig{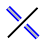} &= - \AddFig{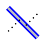} + \AddFig{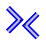}
    \label{eq:string for sigma-bar-sigma}
\end{align}
A few remarks are in order. First, the string operator does not affect the $\sigma$ loop configurations. It only changes the state of the $\psi$ lines within a ``sector'' of a given configuration of $\sigma$ loops. This is necessary in order for the string operator to act nicely on the density matrix. Second, a noticeable feature of the rules is that if we ignore the diagram for crossing on $\sigma$ edge, the rules to resolve crossings essentially decompose into two sets: one is that we only use the first diagram on the right hand side, which will be refereed to as the $m$-type diagram, and the other is to use the second diagram, referred to as the $e$-type diagram. The types of diagrams are interchanged whenever there is a $\sigma$ string.

Having defined the string operators, we note that they satisfy the following algebra when acting on the ground state or the mixed state:
\begin{equation}
    W_{\gamma}(a)W_\gamma(b)=\sum_{c}N_{ab}^c W_\gamma(c),
\end{equation}
Here $N_{ab}^c$ is the fusion rule of anyons: namely, fusing $a$ and $b$ can produce a $c$ anyon if $N_{ab}^c\neq 0$. The most important fusion rules is
\begin{equation}
    \sigma\bar{\sigma}\times\sigma\bar{\sigma}=I+\psi+\bar{\psi}+\psi\bar{\psi}.
\end{equation}
This means that the quantum dimension of $\sigma\bar{\sigma}$ is $2$. The string operators considered here keep fixed $\sigma$ line configurations $\ket{\set{\sigma}}$ invariant. Since the ground state $\ket{\Psi}$ is invariant under closed string operators up to an overall factor, the same must be true for $\ket{\set{\sigma}}$ and the proportionality constant is independent of $\set{\sigma}$. We normalize them so that 
\begin{equation}
    W_\gamma(a) \ket{\set{\sigma}}=d_a \ket{\set{\sigma}},
\end{equation}
where $d_a$ is the quantum dimension of the anyon $a$.
It then follows that 
\begin{equation}
    W_\gamma(a) \rho = d_a \rho.
    \label{exact-sym-SN}
\end{equation}

These string operators can thus be viewed as generalizations of 1-form symmetry to non-Abelian TOs~\cite{Shao:2023gho}. The anyon theory formed by $I,\psi, \bar{\psi}, \psi\bar{\psi}$ and $\sigma\bar{\sigma}$ can still be endowed with the structure of fusion and braiding inherited from the parent 
doubled Ising theory, thus forming a braided fusion category. The only difference is that the braiding is not modular. In particular, $\psi\bar{\psi}$ braids trivially with every other anyon. This is consistent with the $\psi\bar{\psi}$ boson being incoherently proliferated, but \emph{not} coherently proliferated (or ``condensed''). Such a braided fusion category without modularity is a premodular category.

\subsubsection*{Non-Abelian local indistinguishability} \label{sec: nonabelian anyons}

In the pure state, when applying $W_a(\gamma)$ along an open path $\gamma$ to $\ket{\Psi}$, a pair of anyons $a$ and $\bar{a}$ are created at the end points of the path $\gamma$. We refer to the anyons as excitations even though the Hamiltonian is not necessary. The state $W_a(\gamma)\ket{\Psi}$ is locally indistinguishable from $\ket{\Psi}$ except at the end points of $\gamma$. Because of the fusion rule, $\sigma\bar{\sigma}$ is a non-Abelian anyon with quantum dimension 2.
This implies that the space of locally indistinguishable states with four $\sigma\bar{\sigma}$ anyons is four dimensional, with a basis labeled by the fusion channels of any two of the anyons. In other words, the four $\sigma \bar{\sigma}$ anyons encode two qubits.
 
\begin{figure}
   \centering
   \includegraphics[width=0.46\columnwidth]{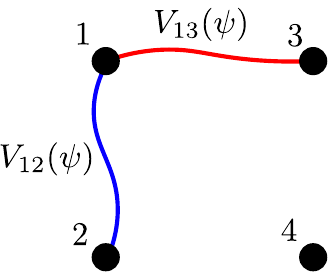}
   \caption{The string operators $V_{12}(\psi)$ and $V_{13}(\psi)$ for four $\sigma \bar{\sigma}$ anyons at points 1, 2, 3, and 4. The string operators $V_{12}(\bar{\psi})$ and $V_{12}(\bar{\psi})$ are analogous.}
   \label{fig:nadeg}
\end{figure} 

Let us study this space in more detail. For definiteness, we label the four $\sigma\bar{\sigma}$ anyons as $1,2,3,4$. We denote by $V_{ij}(a)$ the open string operator of $a$ connecting $i$-th and $j$-th $\sigma\bar{\sigma}$. One can label the basis for the four-dimensional space by the eigenvalues of the string operators $V_{12}(\psi)$ and $V_{12}(\bar{\psi})$. The conjugate operators are $V_{13}(\psi)$ and $V_{13}(\bar{\psi})$. See Fig. \ref{fig:nadeg} for an illustration of the string operators. The logical Pauli operators can be defined as
 \begin{eqs}
     X_1&=V_{12}(\psi), \quad \! X_2=V_{12}(\bar{\psi}), \\
     Z_1&=V_{13}(\psi), \quad Z_2=V_{13}(\bar{\psi}).
 \end{eqs}
The coherent subspace should have fixed eigenvalue under any nonlocal stabilizers, which include $V_{ij}(\psi\bar{\psi})$. To illustrate, let us assume that $V_{ij}(\psi\bar{\psi})=1$. Observe that 
\begin{eqs}
     V_{12}(\psi\bar{\psi}) &= V_{12}(\psi) V_{12}(\bar{\psi}) =X_1X_2, \\ 
     V_{13}(\psi\bar{\psi}) &= V_{13}(\psi)V_{13}(\bar{\psi}) = Z_1Z_2.
\end{eqs}
Thus, the coherent subspace is satisfies the constraints $X_1X_2=1$ and $Z_1Z_2=1$, which is one-dimensional space. In other words, the non-Abelian degeneracy becomes classical: the 4-dimensional Hilbert space is decohered into 4 classical states (i.e. isolated extremal points).

\subsubsection*{Local indistinguishability on a torus}

If the Ising string-net model is defined on a torus, then there are nine locally-indistinguishable ground states. A basis for the ground state space is given by $\{\ket{a}\}$, where $a\in \{1,\psi, \sigma\} \times \{1, \bar{\psi}, \bar{\sigma}\}$. We choose the states $\ket{a}$ to be eigenstates of the closed string operators $W_y(b)$ along the non-contractible path in the $y$ direction. More specifically, we can choose the states such that
\begin{equation} \label{eq: ground state properties}
    W_y(b)\ket{a}=\frac{S_{ab}}{S_{Ia}}\ket{a}, \quad W_x(b)\ket{a}=\sum_{c}N_{ab}^c \ket{c},
\end{equation}
where $W_x(b)$ is the string operator wrapped around the non-contractible path in the $x$ direction.

Before adding noise to the Ising string-net model, it is instructive to first consider {coherently} condensing the $\psi \bar{\psi}$ anyons. The effect of condensation is that the ground states are projected onto the mutual $+1$ eigenspace of the $\psi \bar{\psi}$ string operators, reflecting the fact that the $\psi \bar{\psi}$ anyons can be freely created and annihilated. Note that, according to Eq.~\eqref{eq: ground state properties}, the ground states $\ket{a}$ labeled by anyons that have nontrivial braiding relations with $\psi \bar{\psi}$ are annihilated by the projector. Writing the image of the ground state $|a \rangle$ under the projection as $|a \rangle_{\psi \bar{\psi}}$, we are left with the following three nontrivial states:
\begin{equation} \label{eq: Ising degeneracy 1}
    \ket{I}_{\psi \bar{\psi}}, \quad \ket{\sigma\bar{\sigma}}_{\psi \bar{\psi}}, \quad \ket{\psi }_{\psi \bar{\psi}}.
\end{equation}
Here, the states $\ket{\psi \bar{\psi}}_{\psi \bar{\psi}}$ and $\ket{\bar{\psi}}_{\psi \bar{\psi}}$ have become identified with $\ket{I}_{\psi \bar{\psi}}$ and $\ket{\psi }_{\psi \bar{\psi}}$, respectively.

One might naively conclude that the ground state subspace is three-dimensional. However, there is a fourth basis state, which can be understood as follows. 
Because of the non-Abelian fusion rule $\sigma\bar{\sigma}\times \psi\bar{\psi}=\sigma\bar{\sigma}$, it is possible to create an excited state on the torus with a single $\psi\bar{\psi}$, as long as there is a $\sigma\bar{\sigma}$ anyon flux threading the torus. We refer to this state as $\ket{\sigma \bar{\sigma}^{(\psi\bar{\psi})}}$. After condensation, this state is locally indistinguishable from the states in Eq.~\eqref{eq: Ising degeneracy 1}, which have an even parity of $\psi \bar{\psi}$ anyons. Thus, the ground state space becomes 4-dimensional. Note that this agrees with the fact that condensing $\psi \bar{\psi}$ produces a model with the same TO as the $\Z_2$ TC.   In fact, we can make the following identifications:
\begin{align}
\label{eq: identify Ising and TC states}
    \ket{I}_{\psi\bar{\psi}}&\sim \ket{1}_{\rm TC}\\
    \ket{\psi}_{\psi\bar{\psi}}&\sim \ket{\psi}_{\rm TC}\\
    \ket{\sigma\bar{\sigma}}_{\psi\bar{\psi}}&\sim \frac{1}{\sqrt{2}}(\ket{e}_{\rm TC}+\ket{m}_{\rm TC})\\
     \ket{\sigma\bar{\sigma}^{(\psi\bar{\psi})}}&\sim \frac{1}{\sqrt{2}}(\ket{e}_{\rm TC}-\ket{m}_{\rm TC}).
\end{align}
Here $\ket{a}_{\rm TC}$ denotes the $\Z_2$ TC ground states.\footnote{Note that the state $\ket{\sigma\bar{\sigma}^{(\psi\bar{\psi})}}$ is identified with a state that is odd under the $\Z_2$ symmetry that exchanges $e$ and $m$. This is because the $\psi\bar{\psi}$ anyon is the $\Z_2$ gauge charge obtained from gauging the symmetry-enriched TC. In accordance, the state $\ket{\sigma\bar{\sigma}^{(\psi\bar{\psi})}}$ has an odd number of $\psi\bar{\psi}$ anyons.}

We now study the locally indistinguishable states obtained by decohering the Ising string-net model. 

Given any ground state $\rho=\ket{\Psi}\bra{\Psi}$, we apply the noise channel $\cal{N}$ in Eq.~\eqref{eq: Ising decoherence} to obtain the mixed state $\cal{N}(\rho)$. Since the noise incoherently proliferates $\psi \bar{\psi}$ anyons, the mixed state is invariant under conjugation by the string operators $W_x(\psi \bar{\psi})$ and $W_y(\psi \bar{\psi})$:
\begin{eqs}
    \cal{N}(\rho) &= W_x(\psi \bar{\psi}) \cal{N}(\rho) W_x(\psi \bar{\psi}), \\
    \cal{N}(\rho) &= W_y(\psi \bar{\psi}) \cal{N}(\rho) W_y(\psi \bar{\psi}).
\end{eqs}
This means that the state $\cal{N}(\rho)$ preserves the $W_x(\psi \bar{\psi})$ and $W_y(\psi \bar{\psi})$ eigenspaces. Thus, it is block diagonalized in the eigenbasis of $W_x(\psi \bar{\psi})$ and $W_y(\psi \bar{\psi})$. Therefore, the coherent subspaces are labeled by the eigenvalues of the $\psi \bar{\psi}$ string operators. Drawing analogy to the examples in Section~\ref{Pauli}, the string operators $W_x(\psi \bar{\psi})$ and $W_y(\psi \bar{\psi})$ play the same role as the nonlocal stabilizers of the subsystem code.

Each block of $\cal{N}(\rho)$ can be expressed as a mixture of states, each of which is obtained by decohering ground states within an eigenspace of $W_x(\psi \bar{\psi})$ and $W_y(\psi \bar{\psi})$. In other words, under decoherence, the eigenspaces of these string operators yield extremal points of the manifold of locally indistinguishable states. Therefore, there is a continuum of extremal points, with connected components labeled by the eigenvalues of $W_x(\psi \bar{\psi})$ and $W_y(\psi \bar{\psi})$. To make the discussion more explicit, the eigenspace with $W_x(\psi \bar{\psi})=W_y(\psi \bar{\psi})=1$ is given by
\begin{itemize}
\item $W_x(\psi \bar{\psi})=1$, $W_y(\psi \bar{\psi})=1$:
    \begin{equation} \label{eq: Ising degeneracy}
      \text{span}\Big\{\frac{1}{\sqrt{2}}(\ket{I}+\ket{{\psi\bar{\psi}}}), \quad \ket{\sigma\bar{\sigma}}, \quad \frac{1}{\sqrt{2}}(\ket{\psi }+\ket{\bar{\psi}}) \Big\}.
    \end{equation}
\end{itemize}

$W_{\psi\bar{\psi}}^x$ and $W_{\psi\bar{\psi}}^y$ play the same role as the nonlocal stabilizers. 

One might notice that the conclusion so far, is very similar to the case of coherently condensing $\psi\bar{\psi}$. It is thus natural to ask whether the state $|\sigma \bar{\sigma}^{(\psi \bar{\psi})} \rangle$ with a single $\psi\bar{\psi}$ plays any role here. Intuitively, since the $\psi\bar{\psi}$ anyons have been incoherently proliferated, at $p=1/2$ one can not locally detect whether there is a $\psi\bar{\psi}$ anyon. Therefore, we have an exponentially large number of locally-indistinguishable states. However,  this phenomenon already shows up in the much simpler decohered TC example as discussed in Section~\ref{sec: decohered TC}, and only happens at $p=1/2$. Therefore, we do not include the state in the locally indistinguishable space. 

For completeness, we list the coherent subspaces for the other eigenvalues of $W_x(\psi\bar{\psi})$ and $W_y({\psi\bar{\psi}})$.
\begin{itemize}
\item $W_x(\psi \bar{\psi})=-1$, $W_y(\psi \bar{\psi})=1$:
    \begin{equation}
   \!\!\!\!\!\!\!\!\!\!\!\!\!\!\!\!\!\!\!\!\!\!\!\!\text{span}\Big\{\frac{1}{\sqrt{2}}(\ket{I}-\ket{{\psi\bar{\psi}}}),  \quad \frac{1}{\sqrt{2}}(\ket{\psi }-\ket{\bar{\psi}})\Big\}
    \end{equation}
\item $W_x(\psi \bar{\psi})=1$, $W_y(\psi \bar{\psi})=-1$: 
    \begin{equation}
   \!\!\!\!\!\!\!\!\!\!\!\!\!\!\!\!\!\text{span}\Big\{\frac{1}{\sqrt{2}}(\ket{\sigma }+\ket{\sigma \bar{\psi}}), \quad \frac{1}{\sqrt{2}}(\ket{\bar{\sigma} }+\ket{\psi\bar{\sigma}})\Big\}
    \end{equation}
\item $W_x(\psi \bar{\psi})=-1$, $W_y(\psi \bar{\psi})=-1$: 
    \begin{equation}
    \text{span}\Big\{\frac{1}{\sqrt{2}}(\ket{\sigma }-\ket{\sigma \bar{\psi}}), \quad \frac{1}{\sqrt{2}}(\ket{\bar{\sigma} }-\ket{\psi\bar{\sigma}})\Big\}.
    \end{equation}
\end{itemize}
Notice that each of these subspaces is 2-dimensional. Intuitively, in the pure-state case, these correspond to inserting \add{$\Z_2$} defect lines along non-contractible paths in the $\Z_2$ TC. 

\subsubsection*{\add{Modular transformations on torus}}
For ground state TOs defined on a torus, the locally-indistinguishable ground states transform nontrivially under the modular transformations of the torus. The universal part of the transformation, known as the modular data, is related to the braiding statistics of the anyons. 
In a lattice model, the modular transformations can be implemented by coordinate transformations~\cite{Zhang:2011jd}. For example, the $S$ transformation corresponds to a $\pi/2$ rotation, {which swaps the $x$ and $y$ axes,} and the $T$ transformation can be implemented by a shear deformation, or a Dehn twist. The modular matrices can then be found by computing the matrix representations of the corresponding coordinate transformations in the ground-state subspace and removing non-universal contributions.

It is natural to ask how the modular transformations act on the locally-indistinguishable states in mixed-state TOs, such as the decohered doubled Ising TO. To this end, we study the expectation values of the modular transformations on states inside the coherent space. Here we consider both ordinary linear-in-$\rho$ expectation value, as well as the non-linear expectation values. In both cases, we find that up to a normalization factor, the expectation values are equal to the pure state expectation values, up to overall normalization factors. Details of the calculations can be found in Appendix \ref{app: modular data}.

In the coherent space defined in Eq.~\eqref{eq: Ising degeneracy}, we find that the S and T transformations are represented by
\begin{equation} \label{eq: decohered Ising S and T main text}
    S=\frac12
    \begin{pmatrix}
        1 & \sqrt{2} & 1\\
        \sqrt{2} & 0 & -\sqrt{2}\\
        1 & -\sqrt{2} & 1
    \end{pmatrix}, \quad
    T=\begin{pmatrix}
        1 & 0 & 0\\
        0 & 1 & 0\\
        0 & 0 & -1
    \end{pmatrix}.
\end{equation}
Curiously, these modular matrices do not correspond to any anyon theory: the S matrix is identical to that of a (chiral) Ising TO, but the T matrix differs in the second diagonal entry from any of the modular anyon theories with the same S matrix.

\subsection{$G$-graded string-net models} \label{sec: G-graded stringnet}

We now discuss generalizations of the construction of the decohered doubled Ising state in the previous section. We begin by further introducing details about string-net models. For a general string-net model, the input data is a unitary fusion category $\C$.\footnote{We assume that $\C$ has no fusion multiplicity, so the fusion coefficient is either 0 or 1.} On each edge of the lattice, one has a local Hilbert space with an orthonormal basis labeled by simple objects (also referred to as ``string types'') of $\C$. Each vertex of the lattice corresponds to a fusion of the three strings on the edges emanating from the vertex. A string-net state is a basis state (i.e., a configuration of string labels on the edges) that satisfies the fusion rules of $\C$ at all vertices. 

The commuting-projector Hamiltonian is given by 
\begin{equation}
    H=-\sum_v A_v-\sum_p B_p,
\end{equation}
which is a generalization of Eq.~\eqref{eq: Ising SN Hamiltonian}. Here $A_v$ is a vertex projector that enforces the fusion rule at each vertex $v$. The plaquette term $B_p$ takes the following form
\begin{equation}
    B_p=\frac{1}{\mathcal{D}^2}\sum_a d_a B_p^a,
\end{equation}
where $B_p^a$ pictorially fuses a loop of $a$ string type to a plaquette $p$. Furthermore, $d_a$ is the quantum dimension of $a$ and $\cal{D}$ is the total quantum dimension. For details, we refer to Ref.~\cite{Lin:2020bak}.

The ground state wave function of the string-net model is a certain coherent superposition of all string-net states that satisfies $B_p=1$ (and $B_p^a=d_a$). The relative amplitudes of the string-net states are determined by a set of local rules, which involve further categorical data (e.g. $F$-symbols) of $\C$.
The TO in the ground state is described by the so-called Drinfeld center $\cZ(\C)$. 

We can further assume that the input category is $G$-graded, where $G$ is taken to be a finite group. A $G$-graded fusion category $\C$ has the following decomposition:
\begin{equation}
    \C=\bigoplus_{\mb{g}\in G}\C_\mb{g},
\end{equation}
such that $\C_\mb{g}\times \C_\mb{h}\subset \C_\mb{gh}$. In other words, fusion rules in $\C$ respect the $G$-grading:
\begin{equation}
    a_\mb{g}\times b_\mb{h}=\sum_{c_{\mb{gh}}\in \C_\mb{gh}}N_{a_\mb{g}b_\mb{h}}^{c_{\mb{gh}}} c_\mb{gh}.
\end{equation}
We say that $\C$ is a $G$-extension of $\C_1$, the identity component.  

\def\ginv{\overline{\mb{g}}}

A particularly useful fact is that the TO $\cZ(\C)$ can be obtained from $\cZ(\C_1)$ by gauging a $G$ symmetry. This means that $\cZ(\C)$ contains a subcategory of (possibly non-Abelian) bosons isomorphic to $\Rep(G)$, the category of finite-dimensional linear representations of $G$. Physically, $\Rep(G)$ describes the $G$ charges of the gauge theory. Condensing $\Rep(G)$ in $\cZ(\C)$ again yields $\cZ(\C_1)$.
Based on this relation, Refs.~\cite{HeinrichPRB2016} and \cite{ChengSET2017} constructed exactly solvable models for SET phases whose underlying topological order is given by $\cZ(\C_1)$, and such that gauging the $G$ symmetry yields thte TO $\cZ(\C)$.

As an example, the Ising fusion category is $\Z_2=\{1,\mb{g}\}$ graded, with  $\C_1=\{1,\psi\}$ and $\C_{\mb{g}}=\{\sigma\}$. The Drinfeld center $\cZ(\cal{C})$ is the doubled Ising TO, which is related to the $\Z_2$ TC (the Drinfeld center $\cZ(\C_1)$), by gauging the $\Z_2$ symmetry that permutes the $e$ and $m$ anyons.

We now use the $G$-graded string-net models to produce examples of mixed states with mixed-state TO, similar to the construction with the Ising string-net model. 
We start from the ground state $\rho_0=\ket{\Psi}\!\bra{\Psi}$ of the string-net model for $\C$. Because of the $G$-graded structure of $\C$, the $G$-grading labels on the edges should form a $G$-defect network. For a given $G$-defect network $\{\mb{g}\}$, we define a projector $P(\{\mb{g}\})$, and let
\begin{equation}
    \ket{\Psi_{\{\mb{g}\}}}= P(\{\mb{g}\})\ket{\Psi}.
    \label{eq: Psi_g}
\end{equation}
The fixed-point density matrix is given by 
\begin{equation}
    \rho_{G}\propto \sum_{\{\mb{g}\}}\ket{\Psi_{\{\mb{g}\}}}\!\bra{\Psi_{\{\mb{g}\}}} .
    \label{rhoGgraded}
\end{equation}
For the Ising fusion category, the projector $P(\set{\mb{g}})$ is a projector onto a fixed configuration of $\sigma$ loops, and the state $\rho_{G}$ is a mixture of these configurations. 

The mixed state $\rho_G$ can be obtained from $\rho_0$ using the QLC $\cal{N}$ defined below -- thus, $\rho_G$ has a purification into a GGS. The QLC $\cal{N}$ is
defined as:
\begin{equation}
    {\cal N}=\prod_e {\cal N}_e, \quad
    {\cal N}_e(\rho)=\frac{1}{|G|}\sum_{\mb{g}\in G}T_e^\mb{g} \rho T_e^{\mb{g}}.
    \label{eq: G dephasing}
\end{equation}
Here, the operator $T^{\mb{g}}$ (suppressing the edge label) is given by
\begin{equation} \label{eq: Tg def}
\begin{split}
     T^{\mb{g}} \ket{a_\mb{h}}=\delta_{\mb{g,h}}\ket{a_\mb{h}} .
\end{split}
\end{equation}
The string-net ground state has the following density matrix:
\begin{equation}
    \rho_0=\prod_v A_v \prod_p B_p.
\end{equation}
After applying the channel, it becomes
\begin{equation}
    \cal{N}(\rho)\propto\prod_v A_v \prod_p \Big(\sum_{a\in \C_1} d_a B_p^a\Big),
    \label{eqn: rho_G}
\end{equation}
which is an equivalent representation of $\rho_G$.

Physically, as will be explained in more details later in Section \ref{sec: classical gauging}, these $T^{\mb{g}}$ operators create gauge charges, which are labeled by irreducible representations of $G$. $\rho_G$ is thus obtained from $\rho_0$ by incoherently proliferating the gauge charges. Note that, when $G$ is a non-Abelian group, the gauge charges may be non-Abelian.   

We also note that, in the special case when the input category is just the group algebra ${\rm Vec}_G$ (i.e. $\C_\mb{g}=\{\mb{g}\}$), the string-net model is equivalent to the Kitaev's quantum double model~\cite{Kitaev:1997wr}. In this case, the density matrix in Eq.~\eqref{rhoGgraded} describes a classical ensemble of $G$ gauge fields, i.e., a classical $G$ gauge theory~\cite{Liu:2023opg}. We use this observation in the next section to generalize the construction beyond $G$-graded string-net models.

We now discuss another way to prepare the state $\rho_G$, generalizing the observations about the decohered doubled Ising model in Section~\ref{sec: decohered Ising}. 

In particular, we show that $\rho_G$ can always be purified into a symmetry-enriched $\cZ(\C_1)$ state, meaning that $\rho_G$ can be prepared by tracing out certain degrees of freedom in the SET state. 
Intuitively, the state $\ket{\Psi_{\set{\mb{g}}}}$ can be viewed as the ground state of the $\C_1$ string-net model but with insertions of topological defects. 
$\rho_G$ is then a classical mixture of ground states with defects, similar to the decohered doubled Ising state in Section~\ref{sec: decohered Ising}. 

To see the purifications explicitly, we 
follow Refs.~\cite{ChengSET2017, HeinrichPRB2016} and begin by introducing plaquette $G$ spins with an orthonormal basis $\ket{\mb{g}}$, for $\mb{g}\in G$. The string-net Hamiltonian is then modified as follows. For each edge, we introduce a projector that aligns the $G$-grading of the edge with the domain wall of the adjacent plaquette spins. The plaquette term is also modified to
 \begin{equation}
     B_p=\frac{1}{\cal{D}^2}\sum_{a_\mb{g}\in \C}d_{a_\mb{g}} L_p^{\mb{g}}B_{p}^{a_\mb{g}}=\frac{1}{|G|}\sum_\mb{g}L_p^{\mb{g}}B_{p}^{\mb{g}},
 \end{equation}
 where $L_p^{\mb{g}}$ acts as left multiplication: $L_p^{\mb{g}}\ket{\mb{g}_p}=\ket{\mb{g}\mb{g}_p}$. 
We have also defined
\begin{equation} \label{eq: Bpg}
    B_p^{\mb{g}}=\frac{1}{\cal{D}_1^2}\sum_{a_\mb{g}\in \C_\mb{g}} d_{a_\mb{g}} B_p^{a_\mb{g}}.
\end{equation}
The operator $B_p$ fluctuates the plaquette spins and changes the grading on the edges to abide by the edge projectors.
We note that the operators $B_p^{\mb{g}}$ obey the following relations:
\begin{equation} \label{eq: Bpg definition}
\begin{split}
    B_p^{\mb{g}}B_p^{\mb{h}}&=B_p^{\mb{gh}},\\
    B_p^{\ginv}&=(B_p^{\mb{g}})^\dag\\
    [B_p^\mb{g}, B^\mb{h}_{p'}]&=0, \text{ for }p\neq p',
\end{split}
\end{equation}
with $\ginv$ defined as
\begin{equation}
    \bar{\mb{g}}\equiv \mb{g}^{-1}.
\end{equation}
These properties of $B_p^{\mb{g}}$ essentially follow from the properties of plaquette operators of string-net models, proven in Appendix D of Ref.~\cite{Lin:2020bak}. The Hamiltonian is invariant under $U_\mb{g}=\prod_p R_p^\mb{g}$, where $R_p^\mb{g}\ket{\mb{g}_p}=\ket{\mb{g}_p\mb{g}}$.

When the system is defined on a sphere, the ground state wave function can be written as 
\begin{equation}
     \ket{\Psi_{\rm SET}}=\frac{1}{\sqrt{|G|}}\sum_{\{\mb{g}_p\}}\ket{\{\mb{g}_p\}}\otimes \ket{\Psi_{\{\partial\mb{g}_p\}}},
     \label{eq: SET wavefunction}
\end{equation}
where $\ket{\Psi_{\{\mb{g}\}}}$ is defined in Eq.~\eqref{eq: Psi_g}, and $\{\partial\mb{g}_p\}$ refers to the $G$ defects defined by the $\{\mb{g}_p\}$ domain walls. After tracing out the plaquette spins, we are left with the state $\rho_G$. Therefore, $\ket{\Psi_{\rm SET}}$ is a purification of $\rho_G$, as claimed.

The fact that the mixed state $\rho_G$ can be purified into an SET state suggests a potential generalization of the construction. In particular, we can build an analog of $\rho_G$ starting from a SET mixed state, as opposed to a pure state. This allows for additional possibilities, due to the fact that mixed states can be enriched by weak symmetries, which do not suffer from t'~Hooft anomalies~\cite{Ma:2022pvq}. As a consequence, there are SET mixed states with anomalous $G$ symmetries that do not have any pure-state counterpart with an onsite $G$ symmetry. In the next section, we construct mixed-state TOs from $G$ SET states, which may have an t'~Hooft anomaly, by ``classically gauging'' the symmetry. 

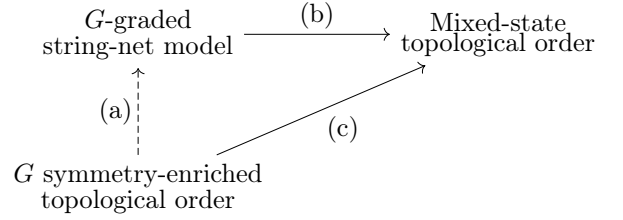
\begin{figure}
    \centering
    \[\begin{tikzcd}
	{\substack{{\text{\normalsize \text{$G$-graded}}} \\ \text{\normalsize string-net model}}} && {\substack{\text{\normalsize Mixed-state}\\ \text{\normalsize topological order}}} \\
	\\
	{\substack{\text{\normalsize \text{$G$ symmetry-enriched}}\\ \text{\normalsize topological order}}}
	\arrow["{\text{\normalsize (b)}}", from=1-1, to=1-3]
	\arrow["{\text{\normalsize (a)}}", dashed, from=3-1, to=1-1]
	\arrow["{\text{\normalsize (c)}}"', from=3-1, to=1-3]
\end{tikzcd}\]
    \caption{Commutative diagram for the two constructions of mixed states in Sections~\ref{sec: G-graded stringnet} and \ref{sec: classical gauging}. The diagram is commutative when the $G$ SET order admits a gapped boundary and $G$ is non-anomalous.  (a) The construction in Section~\ref{sec: G-graded stringnet} can be interpreted as starting with a non-anomalous $G$ SET state with gapped boundaries, such as those from Refs.~\cite{ChengSET2017, HeinrichPRB2016}, and (quantum) gauging the $G$ symmetry. We have used a dashed line, as a reminder that this is only possible if the $G$ symmetry is non-anomalous. This results in the TO of a $G$-graded string-net model. (b)  Starting from a $G$-graded string-net model, we add noise to transform coherent fluctuations of the $G$-gradings into incoherent fluctuations. Equivalently, the process can be viewed as incoherent proliferation of anyons in the $\Rep(G)$ subtheory. (c) Starting from an arbitrary SET state (which may be anomalous or not admit a gapped boundary), the global symmetry can be classically gauged to obtain a mixed state characterized by the $G$ symmetrization of the SET order.}
    \label{fig: commutative diagram}
\end{figure}

\subsection{Classically gauging SET orders} \label{sec: classical gauging}

It has been well-established by now that gauging (generalized) global symmetries is a powerful tool for relating certain pure state TOs to one another. For example, gauging a finite 0-form symmetry in an SPT state gives a topological gauge theory, while gauging a finite 0-form symmetry in an SET state leads to a new TO with a gauge structure~\cite{SET, TeoSET2015}. Here, we interpret the decohered string-net states discussed in the previous section in terms of ``classically gauging'' weak symmetries. We then argue that this procedure can be applied to more general SET states beyond the string-net constructions, e.g., chiral TOs. \add{We summarize the abstract classification of SET states in Section~\ref{sec: weak SET classification}.}

First, we describe what we mean by classical gauging at a conceptual level. To draw a comparison, we start by reviewing the standard (quantum-mechanical) gauging. Suppose we have a pure many-body state $\ket{\psi}$, invariant under a (strong) finite-group symmetry $G$. We additionally assume that the symmetry is implemented by finite-depth local-unitary circuits. The first step in gauging is to couple the system to background gauge fields of the $G$ symmetry. 
In other words, symmetry defects associated with a flat background gauge field are inserted by acting with the symmetry in a (generally disconnected) region $R$ and flipping the gauge fields on the boundary of $R$. 
Thus, for each background gauge field, schematically denoted by $A$, we have a state $\ket{\psi(A)}$.\footnote{Note that, by this construction, the gauge fields have trivial holonomy. One could also consider adding defects along nontrivial cycles to map to sectors with nontrivial holonomies.} 

Next, in gauging the symmetry of the pure many-body state, one forms a coherent superposition of the states with defects, summing over all possible defect configurations:
\begin{equation}
    \ket{\psi_G}\propto\sum_{A}e^{i\alpha(A)}\ket{\psi(A)}.
\end{equation}
Here, the weight $e^{i\alpha(A)}$ is a phase factor that depends locally on $A$, known as the ``local counterterm'' in the field theory literature. In order for this superposition to make sense as a gauge-invariant state, $e^{i\alpha(A)}\ket{\psi(A)}$ must be single-valued over the space of $A$. If this can not be achieved with any choice of the local counterterm $\alpha(A)$, then we say there is an obstruction to gauging. In other words, the symmetry has an 't Hooft anomaly. This step is equivalent to imposing Gauss's law strongly.

Now, we modify this procedure to define how to classically gauge a weak symmetry group ${G}$ of a mixed state $\rho$. The first step of the procedure remains essentially the same: for each background gauge field $A$ (or equivalently, insertion of symmetry defects), one can canonically define a state $\rho(A)$. The next step is now considerably more straightforward: we simply form a classical mixture
\begin{equation} \label{eq: classical gauging rho}
    \rho_G\propto \sum_A \rho(A).
\end{equation}
Notice that there is no issue with anomalous phase factors in this case, since the Gauss's laws are only imposed weakly. That is, the mixed state is only required to be invariant under the conjugation by local gauge transformations.

\subsubsection{Classical gauging in  string-net states}

Let us spell out classical gauging more concretely by reproducing the mixed state $\rho_G$ from the previous section, this time through classical gauging. We start with a weak SET state $\rho_{\rm wSET}$ derived from the SET state in Eq.~\eqref{eq: SET wavefunction}:
\begin{equation}
    \rho_{\rm wSET}\propto \sum_{\{\mb{g}_p\}}\ket{\{\mb{g}_p\}}\!\bra{\{\mb{g}_p\}}\otimes \rho_{\rm SN}(\{\partial\mb{g}_p\}),
    \label{eq: wSET2}
\end{equation}
where $\rho_{\rm SN}(\{\partial\mb{g}_p\})$ is the $\C_1$ string-net state with defects
\begin{equation}
    \rho_{\rm SN}(\{\partial\mb{g}_p\}) = \ket{\Psi_{\{\partial\mb{g}_p\}}}\!\bra{\Psi_{\{\partial\mb{g}_p\}}}.
\end{equation}
For notations in these expressions we refer the readers to discussions around Eq. \eqref{eq: SET wavefunction}.
Note that the weak SET state $\rho_{\rm wSET}$ is obtained from $\ket{\Psi_{\rm SET}}$ in Eq.~\eqref{eq: SET wavefunction} by dephasing the $G$ spins. This reduces the strong $G$ symmetry to a weak $G$ symmetry.\footnote{Of course, any strong symmetry is also a weak symmetry, but we dephase the $G$ spins here to simplify the discussion.}

Now, we follow the steps above to classically gauge the weak $G$ symmetry. The first step is to add gauge fields, which is accomplished by adding a $G$ spin to each edge of the lattice.\footnote{ As the ``matter'' spins live on the plaquettes, gauge fields should be defined on the edges of the dual lattice. Since the edges of the dual lattice are in one-to-one correspondence with the edge of the original lattice, we do not need to make the distinction.} We then insert a configuration of topological defects corresponding to $\{\mb{h}_p\}$ to obtain
\begin{eqs}
    &\rho_{\rm wSET}(\{\mb{h}_p\}) \propto  \\
    &\,\,\,\,\,\,\sum_{\{\mb{g}_p\}}\op{\{\mb{h}_p\mb{g}_p\}}\otimes \rho_{\rm SN}(\{\partial\mb{g}_p\}) \otimes \rho_{\rm GF}(\{\partial\mb{h}_p\}),
    \label{eq: wSETR}
\end{eqs}
where $\rho_{\rm GF}(\{\partial\mb{h}_p\})=\op{\set{\partial\mb{h}_p}}$ denotes the state on the gauge field degrees of freedom. 

The last step is to form a classical mixture of the states $\rho_{\rm wSET}(\{\mb{h}_p\})$ over all of the topological defect configurations. The classically gauged state is
\begin{eqs}
   \sum_{\{\mb{h}_p\}} \sum_{\{\mb{g}_p\}}\op{\{\mb{h}_p\mb{g}_p\}}\otimes \rho_{\rm SN}(\{\partial\mb{g}_p\}) \otimes \rho_{\rm GF}(\{\partial\mb{h}_p\})
\end{eqs}
Notice that this classical mixture is weakly invariant under local gauge transformations.
It can be further simplified by copying the $G$ grading of the string-net state to the gauge field degrees of freedom and redefining the variables. Doing so, we arrive at
\begin{eqs}
   \sum_{\{\mb{k}_p\}} \op{\{\mb{k}_p\}} \otimes \rho_{\rm GF}(\{\partial\mb{k}_p\}) \otimes \sum_{\{\mb{g}_p\}} \rho_{\rm SN}(\{\partial\mb{g}_p\})
\end{eqs}

Finally, we trace out both the $G$ spins on the plaquettes and the gauge fields to find
\begin{eqs}
    \rho_G \propto \sum_{\{\mb{g}_p\}} \rho_{\rm SN}(\{\partial\mb{g}_p\}).
\end{eqs}
Therefore, we have constructed $\rho_G$ by classically gauging a weak SET state. 
In this case, the result is actually equivalent to tracing out the plaquette spins directly. We also note that it is straightforward to generalize $\rho_G$ to arbitrary closed surfaces, where the gauge fields may be topologically nontrivial.

Let us make the strong 1-form symmetry of $\rho_G$ explicit, from the perspective of classical gauging.
First, we show that $\rho_G$ has $\Rep(G)$ as a generalized 1-form symmetry. Intuitively, this stems from the proliferation of closed $G$ defects. More precisely, we define the string operators for the $\Rep(G)$ subtheory as follows. Denote by $\mb{g}_e$ the $G$-grading of the (oriented) edge $e$. For a closed, oriented path $\gamma^*$ on the dual lattice starting at a plaquette $p_0$, following Ref.~\cite{Kitaev:1997wr}, we define 
\begin{equation}
    W_\mb{h}(\gamma^*)\ket{\set{a_\mb{g}}}=\delta_{\mathcal{P}\prod_{e\in \gamma^*}\mb{g}_e, \mb{h}}\ket{\set{a_\mb{g}}}.
    \label{eq:def Wh}
\end{equation}
Here, the ``path-ordered'' product $\mathcal{P}\prod_{e\in \gamma^*}$ means that the $\mb{g}_e$'s are multiplied in the order of the edges along the oriented path starting at $p_0$. One can also think of $W_\mb{h}(\gamma^*)$ as the generalization of a Wilson loop in a $G$ lattice gauge theory. For $G=\Z_2$ in the Ising string-net case, this is the $W_{\psi\bar{\psi}}$ operator defined in Eq.~\eqref{eq:Wboson}. Since the $G$ network in each allowed string-net state is closed, we have
\begin{equation}
    W_\mb{h}(\gamma^*)\rho_G=\begin{cases}
        \rho_G, & \text{if } \mb{h}=1\\
        0, & \text{otherwise}.
    \end{cases}
    \label{eq:Wh}
\end{equation}

Naively, these string operators are labeled by group elements, but similar to lattice gauge theory, they should be organized into operators labeled by irreps $\pi \in \Rep(G)$. For one, this removes the dependence on a choice of base point for the path-ordered product. Explicitly, we write the string operator on a closed path as
\begin{equation}
    W_\pi(\gamma^*)=\sum_{\mb{h}\in G} \chi^*_\pi(\mb{h})W_\mb{h}(\gamma^*),
    \label{eq:WhC}
\end{equation}
where $\chi_\pi(\mb{h})$ is the character of the representation.
From Eq.~\eqref{eq:Wh}, it follows that
\begin{equation}
    W_\pi(\gamma^*)\rho_G = \chi_\pi(1)\rho_G=\dim \pi\cdot \rho_G.
\end{equation}

We can also define a string operator along an open path $\gamma^*_{pp'}$ from a plaquette $p$ to $p'$. Generalizing Eq.~\eqref{eq:WhC}, we define a set of open string operators as follows~\cite{Cong:2017ffh}:
\begin{equation}
    W^{\alpha,\alpha'}_\pi(\gamma^*_{pp'}) = \sum_{\mb{h}\in G}\pi^{-1}(\mb{h})_{\alpha\alpha'}W_\mb{h}(\gamma^*_{pp'}).
\end{equation}
Here, $W_\mb{h}(\gamma_{pp'}^*)$ is the straightforward generalization 
 of Eq. \eqref{eq:def Wh} to the open path $\gamma_{pp'}^*$, where the product is taken along the path from $p$ to $p'$. $\alpha,\alpha'=1,\dots, \dim\pi$ label an orthonormal basis for the irrep, and $\pi$ denotes a unitary matrix representation with a matrix element $\pi(\mb{h})_{\alpha\alpha'}$. 
 
 It is instructive to note that, if $\gamma^*_{pp'}$ connects neighboring plaquettes, then the operator $W_\mb{h}(\gamma^*_{pp'})$ is precisely $T_e^{\mb{h}}$ in Eq.~\eqref{eq: Tg def}, for the edge $e$ bordering the two plaquettes. Acting on a string-net ground state, $W_{\pi}^{\alpha,\alpha'}(\gamma^*_{pp'})$ creates a pair of anyons carrying gauge charges $\pi$ and $\pi^*$, and $\alpha,\alpha'$ represent local degrees of freedom.

We expect that the Rep($G$) anyons have been incoherently proliferated in $\rho_G$. To this point, we observe that $\rho_G$ satisfies:
\begin{equation}
    \sum_{\alpha,\alpha'}W^{\alpha,\alpha'}_\pi(\gamma^*_{pp'}) \rho_G[W^{\alpha,\alpha'}_\pi(\gamma^*_{pp'})]^\dag=\dim\pi\cdot \rho_G.
    \label{eq:general incoherent proliferation}
\end{equation}

This can be derived as follows. From the definition of $W^{\alpha,\alpha'}_\pi(\gamma^*_{pp'})$, we have (omitting $\gamma_{pp'}^*$ for brevity):
\begin{multline}
    \sum_{\alpha,\alpha'} W_\pi^{\alpha,\alpha'}\rho_G (W_\pi^{\alpha,\alpha'})^\dag \\
    = \sum_{\alpha,\alpha'}\sum_{\mb{h},\mb{h}'}{\pi}^{-1}(\mb{h})_{\alpha\alpha'}{\pi^{-1}}(\mb{h}')^*_{\alpha\alpha'} W_\mb{h}\rho_G W_{\mb{h}'}.
\end{multline}
Since $\rho_G$ is a convex sum of states with $G$ defects, and $W_\mb{h}(\gamma^*)$ is a projector that enforces the product of all $G$ lines crossing $\gamma_{pp'}^*$ to be $\mb{h}$, this reduces to
\begin{align} \label{eq: irrep step}
    \sum_{\alpha,\alpha'}\sum_{\mb{h}}|{\pi}^{-1}(\mb{h})_{\alpha\alpha'}|^2 W_\mb{h}\rho_G W_{\mb{h}}.
\end{align}
Then, the fact that the irrep is unitary gives us
\begin{align}
   \sum_{\alpha,\alpha'}|\pi_{\alpha\alpha'}(\mb{h})|^2=\dim \pi. 
\end{align}
Plugging this into Eq.~\eqref{eq: irrep step}, we find
\begin{eqs}
    \sum_{\alpha,\alpha'}W_\pi^{\alpha,\alpha'}\rho_G (W_\pi^{\alpha,\alpha'})^\dag
    &=\dim \pi \sum_\mb{h}W_\mb{h}\rho_G W_\mb{h} \\
    &= \dim \pi \cdot\rho_G.
\end{eqs}

To gain intuition for this expression, note that, when $\pi$ is one-dimensional, $\alpha$ and $\alpha'$ can be suppressed, so $W_\pi$ becomes a unitary operator. In this case, Eq.~\eqref{eq:general incoherent proliferation} reduces to $W_\pi(\gamma^*_{pp'})\rho_G W^\dag_{\pi}(\gamma^*_{pp'})=\rho_G$, which describes the incoherent proliferation of the Abelian anyon $\pi$. We thus propose that Eq.~\eqref{eq:general incoherent proliferation} describes incoherent proliferation of (possibly non-Abelian) $\Rep(G)$ anyons.

Next we consider the string operators of anyons that are already present in $\cZ(\C_1)$. To this end, it is more convenient to use the previous ``decohering out'' construction starting from the $\cZ(\C)$ string-net ground state. Since the $\cZ(\C)$ TO can be obtained from gauging a $G$ symmetry in a $\cZ(\C_1)$ TO, the anyons in $\cZ(\C)$ can be labeled by their $G$ fluxes (conjugacy classes of $G$). In particular, string operators for anyons carrying the trivial flux, denoted by the subtheory $\cZ(\C)_1$, have the feature that they do not change the $G$ grading on the edge labels. As a result, these string operators are well-defined for each $\ket{\Psi_{\{\mb{g}\}}}$, and therefore, they become strong 1-form symmetries of $\rho_G$ (in the sense of Eq.~\eqref{exact-sym-SN}). We should note that $\cZ(\C)_1$ contains the $\Rep(G)$ anyons as well.

Mathematically,  $\cZ(\C)_1$ is the $G$ symmetrization (also known as ``equivariantization'') of $\cZ(\C_1)$~\cite{SET}. To understand the effect of symmetrization, it is instructive to consider the case when $G$ permutes anyons. To be more concrete, suppose a set of anyon labels are permuted into each other under $G$, i.e., they form an orbit under the $G$ action. When $G$ symmetry defects are present, the anyon string operators must change type when crossing a defect that permutes the anyon types. Thus, with the proliferation of $G$ defects, the string operator for an anyon in the orbit is no longer a strong 1-form symmetry. However, their direct sum remains so. A concrete example was discussed in Section~\ref{sec: decohered Ising}, where $\cZ(\C_1)$ is the $\Z_2$ TO, and the $\Z_2$ symmetry permutes $e$ and $m$ (so they form an orbit). Indeed, as discussed below Eq.~\eqref{eq:string for sigma-bar-sigma}, the $\sigma\bar{\sigma}$ string is precisely such a superposition of $e$ and $m$ string operators. More generally, each orbit of anyons under $G$ becomes a set of new anyons, carrying different representations under the stabilizer group of the orbit.
A more complete description of the ``symmetrization'' can be found in Ref.~\cite{SET}, and we provide some 
more examples in Appendix \ref{premodular-cat}. However, we should note that the extra non-Abelian degeneracy due to the symmetrization is classical in nature, as exemplified in Section~\ref{sec: nonabelian anyons}. 

\subsubsection{Weak SET states with anomalous symmetries}

Having seen an example of classically gauging a weak SET state, we now generalize to the case where the $G$ symmetry may have an anomaly. Indeed,  the $G$ symmetry is non-anomalous in the previous example. This is because, by construction, the symmetry can be (quantum-mechanically) gauged to obtain a $G$-graded string-net model. 

To make the generaliztion explicit, let us re-examine the data of a $G$-graded fusion category. Recall that the data of a fusion category consists of the label set (i.e. string types), fusion rules and the $F$ symbols. The definition requires that the $F$ symbols satisfy the pentagon identity, which is crucial for obtaining a GGS that is a self-consistent superposition of string-net states. However, in our context,  we only need a classical mixture of states with defects. Therefore, we can relax the pentagon equation to hold up for a phase factor $O_4$ that depends only on the $G$-grading of the external lines. More formally, $O_4$ is a 4-cocycle of $G$, and the group cohomology class $[O_4]$ fully characterizes the 't~Hooft anomaly of $G$ for the SET order~\cite{ChengSET2017}.

A direct consequence of having a nontrivial $O_4$ is that the $B_p^{\mb{g}}$ operators from adjacent plaquettes (see Eq.~\eqref{eq: Bpg}) no longer commute. This means that we can no longer write down an exactly-solvable parent Hamiltonian, nor its ground state wave function. The extra phase factor due to $O_4$, however, only depends on the $G$-gradings. Therefore, when acting on a state with a fixed grading, the $B_p^{\mb{g}}$ operators only fail to commute by a phase factor~\cite{Lin:2020bak}. This can be exploited to build a weak SET state with an anomalous $G$ symmetry, where the symmetry is only anomalous in the sense of quantum-mechanical gauging. 

We proceed by directly constructing a $\mb{g}$ defect network by applying plaquette operators. More concretely, we define the following mixed state with string-net degrees of freedom: 
\begin{equation}
    \rho_{\rm SN}(\{\partial\mb{g}_p\})=\Big(\prod_p B_p^{\mb{g}_p}\Big)\ket{\Psi_0}\!\bra{\Psi_0}\Big(\prod_p B_p^{\mb{g}_p}\Big)^\dag,
    \label{eq:rho SN 2}
\end{equation}
for arbitrary choices of $\set{\mb{g}_p}$. Here, $\ket{\Psi_0}$ is in the $B_p^1=1$ subspace and has grading $1$ on each edge (or equivalently, the ground state of the string-net model with $\C_1$ as input). Because $B_p^\mb{g}$'s only fail to commute up to a phase when acting on $\ket{\Psi_0}$, there is no issue with the order in which the $B_p^\mb{g}$'s are multiplied in the definition of $\rho_{\rm SN}(\{\partial\mb{g}_p\})$. When $O_4=1$, Eq. \eqref{eq:rho SN 2} is nothing but $\op{\Psi_{\set{\mb{g}}}}$~\footnote{\add{More generally, the 't Hooft anomaly vanishes when $O_4=d\omega_3$, where $\omega_3$ is a group 3-cochain. In this case, one can redefine the F symbols of the $G$-graded fusion category as $F^{a_\mb{g}b_\mb{h}c_\mb{k}}\rightarrow F^{a_\mb{g}b_\mb{h}c_\mb{k}}\omega_3(\mb{g,h,k})$ so that $O_4=1$. This redefinition can be implemented by a depth-1 unitary circuit on the state.}}.

Crucially, $\rho_{\rm SN}(\{\partial \mb{g}_p\})$ only depends on the domain wall configuration $\{\partial\mb{g}_p\}$. To see this, we show that for all $\mb{g}$,
\begin{equation}
    \prod_p B_p^{\mb{g}}\ket{\Psi_0}\propto\ket{\Psi_0}.
\end{equation}
First of all, the operator $\prod_p B_p^\mb{g}$ does not change the grading on the edges. Moreover, it commutes with all of the $B_p^1$ operators, by Eq.~\eqref{eq: Bpg definition}. Therefore, it commutes with the $\C_1$ string-net Hamiltonian. This implies that, on a sphere, the ground state $\ket{\Psi_0}$ must be invariant up to an overall factor. 
Given that $B_p^{\ginv} = (B_p^{\mb{g}})^\dagger$ and $B_p^{\ginv}B_p^{\mb{g}}=B_p^{\mb{1}}$, the operator $B_p^{\mb{g}}$ acts unitarily in the $B_p^{\mb{1}}=1$ subspace. Hence, the ground state $\ket{\Psi_0}$ is invariant under $\prod_p B_p^{\mb{g}}$, up to a phase, as claimed.

Inspired by the state in Eq.~\eqref{eq: SET wavefunction}, we define a weak SET state as follows:
\begin{equation}
    \rho_{\rm wSET}\propto \sum_{\{\mb{g}_p\}}\ket{\{\mb{g}_p\}}\!\bra{\{\mb{g}_p\}}\otimes \rho_{\rm SN}(\{\partial\mb{g}_p\}).
    \label{eq: wSET}
\end{equation}
In contrast to Eq.~\eqref{eq: SET wavefunction}, the anomaly $[O_4]$ need not vanish. 
Classical gauging now amounts to replacing the domain walls with defects. In our case, we can simply trace out the plaquette spins in Eq.~\eqref{eq: wSET} to yield: 
\begin{equation}
    \rho_G\propto \sum_{\set{\mb{g}_p}} \rho_{\rm SN}(\set{\partial \mb{g}_p}),
\end{equation}
which is a direct generalization of Eq.~\eqref{rhoGgraded}.

For completeness, we argue that $\rho_{\rm wSET}$ belongs to the same mixed-state TO as the ground state $\rho_0 = \op{\Psi_0}$ of the $\C_1$ string-net model (ignoring the symmetry). We first show that $\rho_{\rm wSET}$ can be obtained from $\rho_0$ with a QLC. This can be achieved by the following channel:
\begin{eqs}
    \cal{N} = \prod_p \cal{N}_p, 
\end{eqs}
with $\cal{N}_p$ defined as
\begin{eqs}
    \cal{N}_p(\rho) =\frac{1}{|G|}\sum_{\mb{g}\in G} L_p^\mb{g}B_p^\mb{g} \Pi_1 \rho \Pi_1  ( L_p^\mb{g}B_p^\mb{g})^\dag + \Pi^\perp_1 \rho \Pi^\perp_1.
\end{eqs}
Here, $\Pi_1$ is a projector onto the $B_p^{\mb{1}}=1$ subspace in the vicinity of $p$ and $\Pi_1^\perp$ is the projector onto its orthogonal complement.
Notice that $B_p^{\mb{g}}$ acts unitarily, because it is in the $B_p^{\mb{1}}=1$ subspace. Applying this channel to $\rho_0\otimes \ket{\set{\mb{g}_p=1}}\! \bra{\set{\mb{g}_p=1}}$ yields Eq.~\eqref{eq: wSET}.

Next we need to find a channel mapping $\rho_{\rm wSET}$ to $\rho_{0}$. This can be constructed as follows. We first apply a unitary with the spin at $p$ as the control. If the spin at $p$ is $\mb{g}_p$, we apply the $B_p^{\ginv_p}$ operator. This way, all symmetry defects are removed, and the plaquette spins and the string-net degrees of freedom are disentangled. At this point, the plaquette spins are in the maximally mixed states and one can trace out the spins to recover the state $\rho_0$.

Lastly, we discuss the strong 1-form symmetry of the state $\rho_G$. The construction of the $\Rep(G)$ anyon string operators described in Eq.~\eqref{eq:WhC} still applies. While there is no parent $\cZ(\C)$ string-net state (since $\C$ cannot be a fusion category), $\rho_G$ can still be interpreted as classically gauging an (possibly anomalous) $G$ symmetry in the $\cZ(\C_1)$ string-net state. Therefore, we expect that the strong generalized 1-form symmetry is still described by the $G$-symmetrization of the $\cZ(\C_1)$ TO.~\footnote{As discussed in Appendix \ref{premodular-cat}, the presence of an nontrivial 't Hooft anomaly means that the premodular anyon theory does not have a minimal modular extension. Whereas, when the anomaly vanishes, a minimal modular extension is the $\cZ(\C)$ TO.}

\subsubsection{Classifying weak SET orders} \label{sec: weak SET classification}

We now consider the effects of classical gauging at the level of the anyon theory, for a general bosonic SET state with a finite unitary symmetry, which may be anomalous. To get started, we review the classification of pure-state bosonic SET phases. Let us consider a $G$-symmetric GGS, whose TO is given by a modular category $\C$. Denote by $\mathcal{A}$ the group of Abelian anyons in $\C$. The symmetry ${G}$ can then enrich the TO in three ways~\cite{SET, TarantinoSET2016, TeoSET2015}: 
\begin{enumerate}
   \item  There is a group homomorphism $\varphi$ from ${G}$ to the group of auto-equivalence maps $\mathrm{Aut}(\C)$,
\begin{equation}
   \varphi\, : \, {G} \to \mathrm{Aut}(\C).
\end{equation}
Here, ${\rm Aut}(\C)$ consists of all the permutations of anyon types that keep the fusion and braiding properties invariant.\footnote{Note that more precisely, ${\rm Aut}(\C)$ is the group of braided tensor auto-equivalences of $\C$ and there can be nontrivial elements which do not permute any anyons. However, such examples are only known to occur for very complicated $\C$, and for simplicity, we do not consider them here.} Basically, the map $\varphi$ tells us how ${G}$ transformations permute the anyon types. The map $\varphi$ is uniquely associated with a group cohomology class $[O_3]\in \H^3_\varphi(G, \cal{A})$.
\item The anyons of $\cal{C}$ may carry fractionalized quantum numbers under ${G}$. However, given a $\varphi$, there is a possible obstruction to symmetry fractionalization, given by the class $[O_3]$. When $[O_3]$ vanishes, distinct symmetry fractionalization classes form a torsor over $\H^2_{\varphi}({G},\mathcal{A})$. \add{That is, different possible patterns of
symmetry fractionalization can be related to each other by
elements of $\H^2_{\varphi}({G},\mathcal{A})$.}
\item Once $\varphi$ and the symmetry fractionalization of anyons are known, we then need to specify the fusion and braiding properties of ${G}$ symmetry defects. 
In particular, given $\varphi$ and the symmetry fractionalization of anyons, the global symmetry has an 't Hooft anomaly $[O_4]$ valued in $\H^4({G}, \U)$~\cite{SETanomaly, SET, BarkeshliAnomaly1, BarkeshliAnomaly2}. When the class $[O_4]$ vanishes, distinct equivalence classes form a torsor over $\H^3({G},\U)$, up to further identifications~\cite{aasen2022torsorial, ChengPRR2020}.  Physically, an element of $\H^3(G, \U)$ means stacking with a bosonic $G$ SPT state.
\end{enumerate}

It is useful to interpret this data in terms of an SET state with fluctuating symmetry domain walls. 
Each domain wall is associated with an anyon permutation action given by $\varphi$. The obstruction $[O_3]$ means that defect fusion may fail to be associative~\cite{FidkowskiPRB2017}. If $[O_3]$ vanishes, then the defect fusions can be made associative with appropriate decorations of Abelian anyons on the junctions. Inequivalent patterns of decorations are classified by a torsor over $\H^2_\varphi({G}, \cal{A})$. 

Lastly, once we have well-defined defect fusions, including decorations on tri-junctions, there may be a Berry phase in the space of states with defects, which gives the $\H^4$ anomaly. From this interpretation, it is clear that with a non-trivial $\H^3_\varphi({G}, \cal{A})$ class the map $\varphi$ does not make sense in a pure (2+1)$d$ system. The class $[O_4]$ gives the 't Hooft anomaly of the ${G}$ symmetry.

Before discussing classically gauging the symmetry of an SET, we review the effects of gauging the strong symmetry of a pure state at the level of the modular category. For a finite $G$, gauging an SET state results in a TO, whose anyon theory corresponds to a modular category denoted by $\C_G$. It can be constructed from $\C$ as follows:
\begin{enumerate}
    \item First, as already discussed in Section~\ref{sec: G-graded stringnet}, gauge invariance under $G$ means that the anyon theory $\C$ must be symmetrized to form a new premodular category, denoted by $(\C_G)_1$, which has Rep$(G)$ as its transparent center. The premodular category $(\C_G)_1$ resulting from symmetrization contains all information about the symmetry action on anyons~\cite{LanPRB2016}. Importantly, given $\varphi$, this step of symmetrizing can be done if and only if the $[O_3]$ class vanishes. 
    
    \item $(\C_G)_1$ is only a subcategory of $\C_G$.  The modular category $\C_G$ is in fact a \emph{modular extension} of $(\C_G)_1$.  Physically, the anyons of $\C_G$ that are not in $(\C_G)_1$ are $G$ flux anyons and braid nontrivially with $G$ gauge charges in $\Rep(G)\subset (\C_G)_1$, thus restoring modularity. See Appendix \ref{premodular-cat} for more on modular extensions.  
\end{enumerate}
 
Classically gauging the weak symmetry $G$ means that a classical mixture of SET states with arbitrary $G$ defects is formed. 
The precise structure of the density matrix is determined by imposing Gauss's law weakly. In other words, the density matrix must be invariant under conjugation by local gauge transformations, which, intuitively speaking, deform the $G$ defects locally. This condition then requires that the obstruction $[O_3]$ vanishes, otherwise defect configurations which only differ locally can have different total anyon charge, and such configurations can not be related by a local gauge transformation. We also notice that there is no longer an anomaly valued in $\H^4(G,\U)$, since the classical mixture has no coherent phase information. It is worth noting that the results here are similar to the classification of ``average SET orders'' proposed in Ref.~\cite{Ma:2023rji}.\footnote{We note however that average SET orders are defined for disordered ensembles of Hamiltonians with topologically ordered ground states. While they can be formally thought of as density matrices with a preferred basis, the equivalence relations are completely different. In particular, Ref.~\cite{Ma:2023rji} considers the possibility of ``Anderson localization'' of Abelian anyons, which trivializes both the $\H^3_\varphi(G, \cal{A})$ obstruction and the $\H^2_\varphi(G, \cal{A})$ fractionalization classes when $G$ is an average symmetry.}

The analysis of the topological order in the classically gauged state proceeds similarly following the discussions in the $G$-graded string-net example, so we will be brief here. Observe that the arbitrary insertions of the $G$ defects imply that only $G$-invariant string operators are well-defined as strong 1-form symmetry. This is to say that the $\C$ anyon theory must be symmetrized to form the premodular $(\C_G)_1$ category. Another way to see this when there is no 't Hooft anomaly is to note that the anyon string operators in $(\C_G)_1$ do not alter the $G$ defects. Since they give rise to emergent 1-form symmetry in the gauged SET pure state, they must form strong 1-form symmetry for the classically gauged mixed state. We note that the description in terms of premodular anyon theory automatically excludes those symmetries with nontrivial $\H^3$ obstruction classes.

In the other direction, as discussed in Appendix \ref{premodular-cat}, every premodular anyon theory whose transparent center is purely bosonic can be viewed as symmetrization of a certain modular anyon theory. Thus the classical gauging construction can realize mixed TOs with such premodular anyon theories as strong 1-form symmetry.

Lastly, we comment that, if the symmetry is non-anomalous, then one can first quantum-mechanically gauge the symmetry to get a new TO, which contains $\Rep(G)$ as a subcategory (the gauge charges). Now, a QLC can be applied to proliferate the $\Rep(G)$ anyons. When the SET state is non-chiral, or more generally, can be realized by a string-net model, then this is precisely the construction in Section~\ref{sec: G-graded stringnet}.

\subsection{Walker-Wang models} \label{sec: walkerwang}

\begin{figure}[t] 
\centering
\includegraphics[width=.4\textwidth]{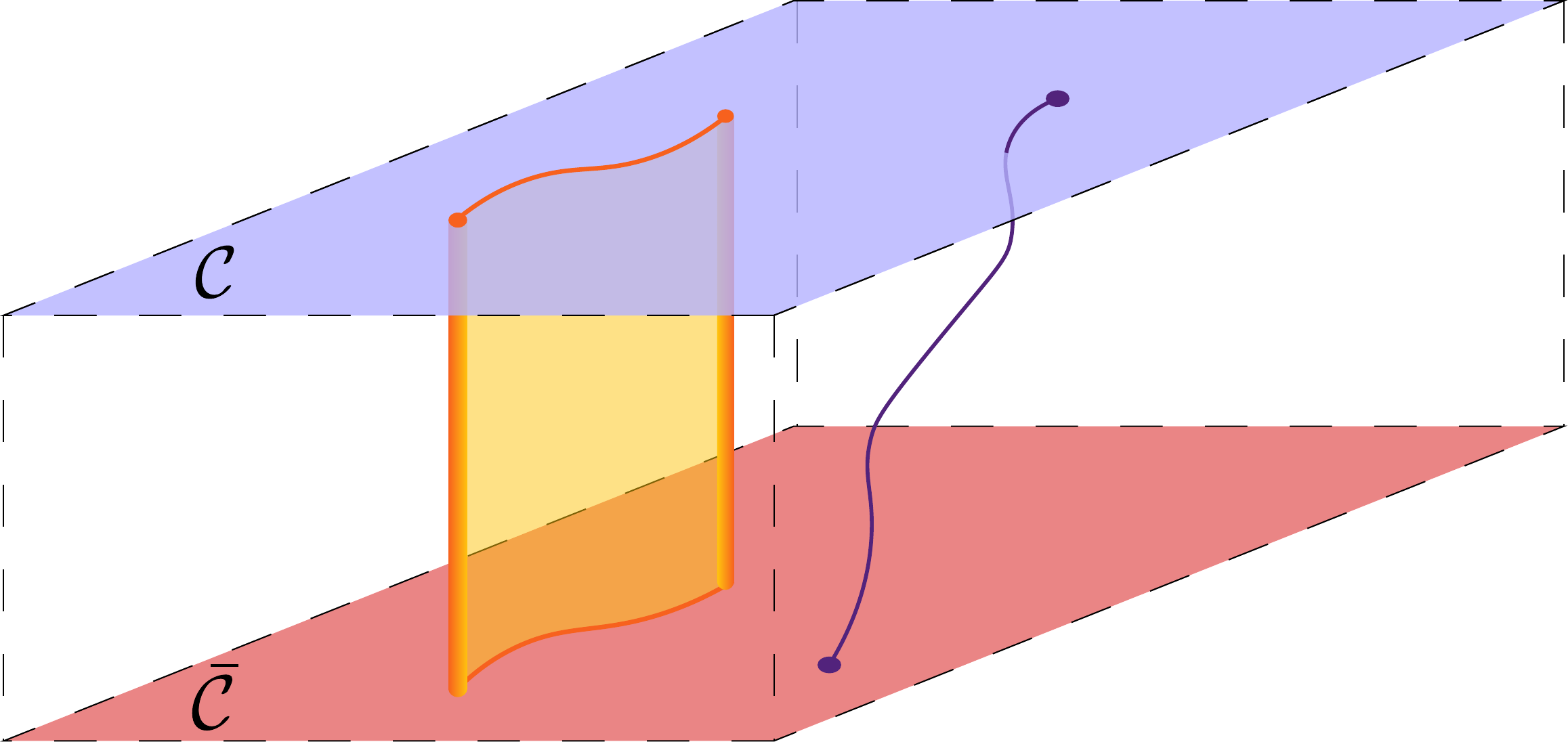}
\caption{Anyons of a WW model. The input of a WW model is a premodular category $\cal{C}$. For a quasi-2D slab, the anyons in $\cal{C}$ are hosted on the top surface, while the bottom surface hosts the conjugate anyon theory $\bar{\cal{C}}$. The transparent anyons in $\cal{C}$ (purple) are bulk excitations and are shared by both surfaces. The flux anyons (orange) are created by a quasi-1D membrane operator that terminates on the two surfaces.}
\label{fig: WW}
\end{figure}

The mixed states in the previous two sections, constructed from $G$-graded string-net models or through classically gauging bosonic SETs, have the property that all of transparent anyons are bosons. There are, of course, premodular categories with transparent fermions. To realize mixed states whose anyon theories are arbitrary premodular categories, we employ Walker-Wang (WW) models~\cite{WW}. We show that for any premodular category $\C$, we can decohere the corresponding WW model to construct a mixed TO with $\C$ as the strong 1-form symmetry.

WW models take as an input any premodular category $\C$, and produce a (3+1)$d$ commuting-projector Hamiltonian. 
The topological order in the ground state of the WW model can be understood from the mathematical structure of the premodular category. As reviewed in Appendix \ref{premodular-cat}, each premodular category has a transparent center of anyons, all of which braid trivially with every other anyon. This transparent center is uniquely associated with a finite group $G$, such that it can be identified as (bosonic or fermionic) gauge charges of $G$. 
The bulk of the Walker-Wang model turns out to be a (possibly twisted) $G$ gauge theory.
The anyon theory $\C$ is realized on the surface of the model, with the transparent center identified as the gauge charges in the bulk. 
When the input category $\C$ is modular with a trivial center, the resulting TO is trivial (invertible). 

We make a slab of the 3D model, which has both top and bottom surfaces. When viewed as a quasi-2D system, the anyon types can be divided into three groups. 
\begin{enumerate}
    \item  The first are those that are confined to the top and bottom surfaces. We choose, as a convention, that the top surface hosts the anyon theory $\C$, while the bottom has the conjugate theory $\overline{\C}$.
    \item  The transparent center $\mathcal{T}$ are gauge charges that are mobile within the 3D bulk. Thus, the top and bottom surfaces share the same transparent center $\cal{T}$. 
    \item The last group consists of anyons that braid nontrivially with $\cal{T}$, which are descendants of flux loop excitations in the 3D bulk created by membrane operators -- hence, we call them flux-like anyons.  The loop-like excitations can condense on the surfaces (which should actually be taken as the definition of the boundary condition), so the flux-like anyons can be created with a membrane operator stretching from the top surface to the bottom surface, as in Fig.~\ref{fig: WW}. In the quasi-2D slab, these  membrane operators become string operators for the flux-like anyons.
\end{enumerate}
Together, the slab realizes the quantum double (or Drinfeld center) of $\C$. The three types of excitations are illustrated in Fig.~\ref{fig: WW}.

Next, imagine cutting the slab through the middle to form two thin slabs, then trace out the lower slab. Here, we assume that the slab is thick enough compared to the width of support of the string operators of the surface anyons (including the transparent center). Since the WW Hamiltonian is a fixed-point model with zero correlation length, we expect all string operators of the surface anyons have finite-width support (see \cite{Keyserlingk2013} for examples). So in the upper-half slab these string operators are still well-defined, and not affected by tracing out the lower half. As a result, they remain (generalized) 1-form symmetry of the state. On the other hand, string operators of the flux anyons must have support in the lower half, so they are no longer strong symmetries. The resulting state thus has $\C$ as its generalized 1-form symmetry. 

\subsection{Phase equivalence for general premodular categories} \label{sec: general intrinsic}

We have seen through a number of examples that decohered TO states can be assigned premodular anyon theories. The anyon string operators generate (non-invertible and invertible) strong 1-form symmetries of the state. It is natural to conjecture that this holds for any mixed-state TO. We now discuss how the anyon theory is affected by the application of QLCs, based on observations made in the previous sections and generalizations of the Abelian case in Section~\ref{sec: anyon theories SLDQC}.

 One way to construct mixed-state TOs is to start from a pure-state TO, whose anyon theory is a modular category $\C$, and apply a QLC to ``decohere out'' a subcategory $\cal{D}$ of anyons. In other words, one forms a classical mixture of excited states, obtained from exciting anyons in the subcategory. One expects that once the density of excitations is higher than some threshold, the decohered state is in a new phase. The remaining 1-form symmetry in this decohered phase is the ``commutant'' subcategory $\cal{D}'$ of the subcategory.\footnote{We thank Roger Mong for discussions on this construction.} That is, $\cal{D}'$ contains the anyons that braid trivially with those in $\cal{D}$. Here, trivial braiding between an anyon $a$ and $b$ means the $S$ matrix element $S_{ab}/S_{11}=d_ad_b$, where $d_{a/b}$ are the quantum dimensions. The same construction can obviously be applied to a mixed state to ``decohere out'' a subcategory of anyons.

Another way that a QLC can affect the anyon theory is to classically gauge a finite symmetry. \add{That is, if a mixed state $\rho$ with anyon theory $\C$ contains a bosonic transparent subcategory $\cal{B}$, it can be obtained from a mixed state with anyon theory $\C/\cal{B}$ by applying a QLC. Here, $\C/\cal{B}$ is the premodular category obtained by condensing $\cal{B}$ in $\C$.} 
As reviewed in Appendix \ref{premodular-cat}, in this case, there always exists a finite group $G$, such that $\cal{B}$ is isomorphic to the category $\Rep(G)$ of finite-dimensional linear representations. Mathematically, it means that $\C$ can be recovered by symmetrizing the $G$ symmetry in $\C/\mathcal{B}$. Thus, one can start from a mixed-state with anyon theory $\C/\cal{B}$, and classically gauge a $G$ symmetry to obtain a mixed state with anyon theory $\C$.

Suppose that $\rho_1$ and $\rho_2$ are mixed states that can be connected by a QLC $\cal{N}_{21}$, such that $\cal{N}_{21}(\rho_1) = \rho_2$. Suppose further that $\cal{N}_{21}$ can be purified into a unitary $V$ acting on $\rho_1\otimes \op{0}$.
By our assumption, both $\rho_1$ and $\rho_2$ can be associated to  premodular categories, $\mathcal{C}_1$ and $\C_2$, respectively. A straightforward generalization of the argument in Section~\ref{sec: 1-form symmetry} shows that if $\rho_2$ has a strong generalized 1-form symmetry operator $W$, then $\rho_1\otimes \op{0}$ has a strong 1-form symmetry given by $VWV^\dag$. However, if $W$ represents a transparent boson, it may become breakable in $\rho_1$, in light of the examples in Section~\ref{sec: decohered Ising} and \ref{sec: G-graded stringnet}.

Thus, the premodular anyon theories $\C_i$ for the mixed state $\rho_i$ should be related by the following: there exists a transparent, bosonic subcategory $\mathcal{B}_2$, such that 
\begin{equation}
     \C_2/\mathcal{B}_2\subset \C_1.
     \label{C12-relation}
\end{equation}
Here, $\subset$ means that $\C_2/\mathcal{B}_2$ is a subcategory of $\C_1$. Physically, the relation describes two effects of a QLC on the anyon theory: it can ``decohere out'' a subcategory, leaving its commutant, or classically gauge a symmetry. 

If $\rho_1$ and $\rho_2$ are two-way connected by QLCs, then there exists  bosonic transparent subcategories $\cal{B}_i\subset \C_i$ such that
\begin{equation}
    \C_2/\mathcal{B}_2\subset \C_1, \quad \C_1/\mathcal{B}_1\subset \C_2.
    \label{eqn:general-equiv}
\end{equation}

Using this argument, one can show that the decohered doubled Ising state and the $\Z_2$ TC do not belong to the same phase. Indeed, if there is a QLC to transform the decohered doubled Ising state to the $\Z_2$ TC state, then one can purify the $\Z_2$ TC state into a doubled Ising state. This is clearly impossible, as the any purification of the $\Z_2$ TC needs to have all of its $\Z_2\times\Z_2$ 1-form symmetries, which are absent in the doubled Ising TO.
 
One immediate conclusion is that, if $\C_1$ and $\C_2$ contain no bosonic transparent subcategory (meaning they are modular up to transparent fermions), then Eq.~\eqref{eqn:general-equiv} implies $\C_1=\C_2$. If only one of them, say $\C_1$, is modular, then we find
\begin{equation}
    \C_1\subset \C_2, \quad \C_2/\mathcal{B}_2\subset \C_1.
\end{equation}
The first relation implies $\C_2=\C_1\boxtimes \C_1'$, where $\C_1'$ is the commutant of $\C_1$ (see Appendix \ref{premodular-cat}). However, then the second relation implies $\C_1\boxtimes (\C_1'/\mathcal{B}_2)\subset \C_1$, which implies that $\C_1'/\mathcal{B}_2$ is trivial, so $\C_1'$ is a bosonic transparent subcategory itself, which can be freely removed from $\C_2$. We thus conclude that $\C_1=\C_2$.

Now suppose $\cal{T}_b$ is the maximal bosonic transparent subcategory in $\C$. We define $\C^{\rm min}=\C/\cal{T}_b$ as the premodular category obtained from condensing $\cal{T}_b$. By definition, the transparent center of $\C^{\rm min}$ is either trivial, in which case $\C^{\rm min}$ is modular, or given by $\Z_2^{(1)}$, in which case $\C^{\rm min}$ is super-modular~\cite{Bruillard_2017}. Super-modular categories describe fermionic TOs for GGSs. Similar to the Abelian case, we conjecture for general premodular categories that if $\C_1$ and $\C_2$ satisfy Eq.~\eqref{eqn:general-equiv}, they must have the same $\C^{\rm min}$. 

The discussion so far closely parallels that of (invertible) strong 1-form symmetry for Abelian TOs in Section~\ref{sec: anyon theories SLDQC}. However, unlike Abelian anyon theories, a non-Abelian super-modular category does not necessarily factorize into the product of a modular category and $\Z_2^{(1)}$. If the category does not factorize, we refer to it as ``intrinsically fermionic''. Theories that are intrinsically fermionic are not captured by decohering Pauli stabilizer models or the constructions in Sections~\ref{sec: G-graded stringnet} and \ref{sec: classical gauging}.

The simplest nontrivial example of an intrinsically fermionic anyon theory is the so-called SO(3)$_3$ category~\cite{LanPRB2016}. This anyon theory has four anyon types $1,f,s,sf$, where $f$ is the transparent fermion, and the non-Abelian anyon $s$ satisfies the following fusion rule:
\begin{equation}
    s\times s = 1 + s + sf.
\end{equation}
From this, the quantum dimension of $s$ is $d_s=1+\sqrt{2}$. Furthermore, the self statistics of $s$ is $\theta(s)=i$. A minimal modular extension of SO(3)$_3$ is the SU(2)$_6$ theory. The fermion $f$ is identified as the spin-3 particle, and $s$ as the spin-$1$. There are infinitely many intrinsically fermionic premodular categories -- for example, the integer-spin subcategory of the $\mathrm{SU}(2)_{4m+2}$ modular categories for all $m \geq 0$. We can thus construct a mixed-state TO by proliferating the emergent spin-$(2m+1)$ fermion in a SU(2)$_{4m+2}$ TO or using the WW construction.  

Finally, we conjecture that the strong 1-form symmetries, described by premodular categories, provide a full classification of locally-correlated mixed-state TOs. We have provided evidence that two mixed states that are two-way connected by QLCs have the same premodular anyon theories. The converse is a difficult problem, and remains an open question even in the case of ground state TOs (although, it is widely believed to be true).

\section{Conclusion and discussion} \label{sec: discussion}

We have proposed a classification of mixed-state TOs according to their strong (generalized) 1-form symmetries. Notably, this includes mixed-state TOs that are intrinsically mixed, as first suggested in Ref.~\cite{Wang:2023uoj}. These are characterized by premodular anyon theories with transparent anyons, i.e., at least one anyon has trivial braiding relations. We established strong constraints on the anyon theories exhibited by mixed states belonging to the same mixed-state TO, and we proved that the minimal anyon theory $\cal{A}^{\rm{min}}$ is an invariant of the phase. We conjecture that, more generally, the anyon theory itself characterizes the mixed-state TO but leave the proof in the general case as an open question. 

Furthermore, we constructed a wide variety of examples of fixed-point mixed states. Firstly, we established that topological subsystem codes provide a natural framework for studying stabilizer states under Pauli noise. This led to examples of mixed states characterized by arbitrary Abelian premodular anyon theories. We then constructed examples of mixed states characterized by non-Abelian anyon theories by leveraging $G$-graded string-net models. We subsequently generalized the construction by classically gauging SET states, which is allowed to have an anomalous $G$ symmetry. Lastly, we showed that mixed states characterized by arbitrary premodular anyon theories can be constructed from a slab of WW model with depolarizing noise on the lower half of the system.

Our work has left a number of open questions and avenues for future work. For one, a fundamental aspect of our approach is specifying a space of mixed states with which to define mixed-state TOs. Our current definition of ``locally-correlated mixed states'' restricts to the class of states that can be purified into a GGS. While the definition includes many interesting classes of examples, e.g., all decohered topological states, one can easily imagine mixed states that do not fit into this definition. It is thus highly desirable to have a definition of mixed-state TO that does not make any reference to the purification or Hamiltonians. 

One potential source of inspiration in this direction comes from the entanglement bootstrap program~\cite{Shi2020fusion,Shi2020Verlinde}, which has seen success in characterizing pure-state TOs without any reference to a parent Hamiltonian. Hence, it may be fruitful to extend the entanglement bootstrap program to mixed states. In this case, the class of mixed states might be determined by imposing conditions on the entanglement or the conditional mutual information (CMI) in particular, similar to the axioms employed in entanglement bootstrap.

Indeed, it is widely expected that, for ground state TOs, the CMI vanishes for the geometry given in Fig.~\ref{fig: ABCgeometry}. One can also argue that this is the case for all examples in Section~\ref{sec:decohered ss examples} associated with topological subsystem codes, but not for the $Y$-decohered TC state, whose subsystem code is not topological. Thus, we expect that CMI should play an important role in charting the class of mixed states~\cite{Sang:2023rsp, Sang:2024vkl}.

In fact, even relaxing the R\'enyi-2 short-range correlated assumption in the definition of locally-correlated mixed states already opens up possibilities for new types of mixed states. For example, in a recent work~\cite{Lessa:2024wcw} a topologically nontrivial mixed state in (1+1)$d$ was discovered, with long-range R\'enyi-2 correlations.

For ground state TO, certain data of the TO can be extracted from the ground state wave function. For instance, the entanglement entropy contains a subleading correction lower bounded (and in many cases, equal to) by the logarithm of the total quantum dimension~\cite{Kim:2023ydi}.  Generalizations of these results to mixed-state TO are worth investigating. In fact, the ``topological correction'' to the logarithmic negativity has been studied in the $\Z_2$ TC model under bit-flip channel~\cite{Bao:2023zry}. In an upcoming work we will compute topological negativity in other decohered stabilizer models, which provide further evidence for the conjectured classification.

TOs in ground states in (2+1)$d$ can be associated with a 3$d$ topological quantum field theory (TQFT). For TOs in mixed states, while one does not expect there is a full 3$d$ TQFT description, it is possible that a weaker notion of TQFTs without assuming full spacetime symmetry still makes sense. For example, we have shown that one can still define state space on closed oriented manifolds (in terms of coherent spaces), as well as the modular data (i.e. the actions of the mapping class group). Further structures required by the TQFTs at the level of state spaces are interesting to investigate, such as states associated with punctures, closely related to anyon excitations. Such a mathematical structure may be considered as a ``mixed-state TQFT''~\cite{Zini:2021lte}, providing an alternative characterization of mixed state TOs.

Lastly, a natural avenue for further work is the generalizations to TOs in three spatial dimensions. In the case of ground states, a complete understanding of (3+1)$d$ TOs that admit a topological field theory description has been achieved recently~\cite{Lan3DTO1, Lan3DTO2, JF2020}. We expect that the methods developed in this work can also shed light on the classification of mixed-state TOs in (3+1)$d$, which may include intrinsically mixed TOs whose pure state counterparts are anomalous. An example of this kind is the $\Z_2$ gauge theory with both fermionic charge and fermionic loops, as described in Refs.~\cite{JF2020, Chen:2021xks, Fidkowski:2021unr}. It can be realized by a (4+1)$d$ generalization of the WW-type construction discussed in Section~\ref{sec: walkerwang}, which we briefly outline: 
 first, a generalization of the WW model by Ref. \cite{Fidkowski:2021unr} realizes a (4+1)$d$ invertible state with the anomalous (3+1)$d$ TO on the boundary. Then we make a slab of the model, and trace out one of the two boundaries. Now, the anomalous TO is realized in (3+1)$d$ as a mixed state.

\vspace{0.2in}
\noindent{\it Acknowledgements -- } T.D.E. is grateful to Yimu Bao, Ruihua Fan, Tim Hsieh, Zhu-Xi Luo, and Shengqi Sang for inspiring discussions about mixed-state phases of matter. T.D.E. also thanks Fiona Burnell and Kyle Kawagoe for helpful conversations about non-Abelian boson condensation. M.C. acknowledges Ruochen Ma for extensive discussions on mixed state phases and collaborations on related projects, as well as Shawn X. Cui,
Yifan Wang and Zhenghan Wang for enlightening conversations. We are especially grateful to Roger Mong for insightful comments on the general ``decohering out'' construction. We thank Abhinav Prem and Ramanjit Sohal for communicating about their unpublished work.  M.C. acknowledges support from NSF under award number DMR-1846109.

\vspace{0.2in}
\noindent{\it Note added -- } While preparing the manuscript we became aware of a closely related preprint with overlapping results~\cite{Sohal:2024qvq}, and two upcoming works~\cite{Lessa_unpub, Luo_unpub} on related topics.

\appendix

\section{Purifications of the fermion-decohered toric code}
\label{app:purify-fermion-TC}

In this Appendix, we show that the fermion-decohered TC state in Section~\ref{fermion-TC} can be purified into any state that is the bosonization of a fermionic ground state with total even fermion parity. 

First, we briefly recall the notation used in the (2+1)$d$ bosonization map of Ref.~\cite{Chen:2017fvr}, defined on a square lattice. We associate a complex fermion to each plaquette, which can be represented as a pair of Majorana operators $\gamma_p, \gamma_p'$. We take all of the edges of the lattice to be oriented. We then define the fermion parity operator $B_p=(-1)^{n_p}=i\gamma_p\gamma_p'$ at a plaquette $p$ and the hopping operator $S_e=i\gamma_{L(e)}\gamma_{R(e)}'$, which transfers fermion parity between neighoring plaquettes. Here, $L(e)$ ($R(e)$) denotes the plaquette to the left (right) of the oriented edge $e$.

With this notation, we consider the following two channels
\begin{equation}
\begin{split}
    {\cal N}_1 = \prod_e {\cal N}_{1,e}, \quad {\cal N}_{1,e}(\rho)=\frac12(\rho +  S_e\rho S_e),\\
    {\cal N}_2 = \prod_e {\cal N}_{2,p}, \quad {\cal N}_{2,p}(\rho)=\frac12(\rho +  B_p\rho B_p),
\end{split}
\end{equation}
and their composition $\mathcal{N}=\mathcal{N}_1\circ \mathcal{N}_2$. We claim that for any fermionic pure state $\ket{\psi}$ with even fermion parity $\prod_{p}B_p=1$, we have 
\begin{equation} \label{eq: fermionic fermion-decohered}
    \mathcal{N}(\op{\psi})\propto 1+\prod_p B_p.
\end{equation}

To prove this, first notice that $\mathcal{N}_2$ is a fully dephasing channel in the $n_p$ basis. Thus, after applying $\mathcal{N}_2$, the density matrix becomes diagonal in the $\ket{\{n_p\}}$ basis. Next, notice that $\mathcal{N}_1(\rho)$ is a fixed point of $\mathcal{N}_1$. By applying products of $S_e$'s, all different basis states $\ket{\{n_p\}}$ can be connected with each other, with matrix element always equal to $\pm 1$. Therefore, the fixed point of $\mathcal{N}_1$ must be an equal-weight mixture of all $\op{\set{n_p}}$ states.

To identify purifications of the fermion-decohered TC state, we can bosonize the construction. This is because the bosonization of the state in Eq.~\eqref{eq: fermionic fermion-decohered} is precisely the fermion-decohered TC state. Moreover, both $S_e$ and $B_p$ are mapped to local Pauli operators, so the channels $\cal{N}_1$ and $\cal{N}_2$ are local after bosonization. We can then choose $\ket{\psi}$ to be any ground state of a fermionic Hamiltonian $H_f$, which can always be expressed in terms of $S_e$ and $B_p$. In the spin representation, $\ket{\psi}$ is the ground state of the bosonized Hamiltonian (see Ref.~\cite{Chen:2017fvr}). Finally, the fermion-decohered TC state is obtained by applying the bosonized channels $\cal{N}_1$ and $\cal{N}_2$. Thus, if we purify the channels $\cal{N}_1$ and $\cal{N}_2$ we have a purification of the fermion-decohered TC state.

\section{R\'enyi correlations of Pauli-decohered stabilizer states} \label{app: Renyi2 topological codes}

Here, we argue that, when the subsystem code associated to a Pauli-decohered stablizer state is topological, then the associated mixed state under maximal decoherence is R\'enyi-1 and -2 locally correlated, as defined in Section~\ref{sec: generalities}. We start by considering the R\'enyi-1 correlations. 

We let $\rho$ be a mixed state defined by a topological subsystem code, where the state is maximally mixed in the gauge subsystem. To derive a contradiction, we assume that there exists operators $M_i$ and $M_j$ localized at the sites $i$ and $j$, such that the R\'enyi-1 correlations of $M_i$ and $M_j$ in the state $\rho$ do not vanish in the separation between $i$ and $j$

To make this assumption more explicit, let us choose both a set of local generators for the gauge group and a set of local generators for the stabilizer group on the infinite plane (or a sphere). Then, by the topological property, there exists a finite distance $\ell$, such that the support of every gauge generator and stabilizer generator can be contained in a box of dimensions $\ell \times \ell$. We assume more explicitly that, for $|i-j| \gg \ell$, there exists operators $M_i$ and $M_j$ localized at $i$ and $j$ such that 
\begin{align} \label{eq: renyi1 app}
    \Tr[M_iM_j \rho] - \Tr[M_i \rho] \Tr[M_j \rho] \neq 0.
\end{align}
Note that for simplicity, we assume that $M_i$ and $M_j$ have bounded support. 

To make progress, let us decompose $M_i$ and $M_j$ into Pauli operators:
\begin{align}
    M_i = \sum_{P_i} C_{P_i} P_i, \quad  M_j = \sum_{P_j} C_{P_j} P_j.
\end{align}
Here, the $P_{i}$ and $P_j$ operators are Pauli operators localized near $i$ and $j$, and $C_{P_i}$, $C_{P_j}$ are complex coefficients.  
The correlator in Eq.~\eqref{eq: renyi1 app} can be rewritten using the decomposition of $M_i$ and $M_j$ to give
\begin{align} \label{eq: renyi1 app 2}
    \sum_{P_i,P_j} \tilde{C}_{P_iP_j}\left( \Tr[P_iP_j \rho] - \Tr[P_i \rho] \Tr[P_j \rho] \right) \neq 0,
\end{align}
for some coefficient $\tilde{C}_{P_iP_j}$.  

Now, we consider different possibilities for $P_iP_j$. First, if $P_iP_j$ fails to commute with at least one stabilizer, then the summand is zero, since $\rho$ is a projector onto the code space. If $P_iP_j$ is a gauge operator, not belonging to the stabilizer group, then the summand also vanishes. This is because, due to the decoherence, the expectation value of gauge operators outside of the stabilizer group is zero. Therefore, the summand can only be nonzero if $P_iP_j$ is a stabilizer and $P_i$, $P_j$ are not stabilizers. 

This leads us to a contradiction, since it is not possible for $P_iP_j$ to be a stabilizer, while $P_i$ and $P_j$ are not stabilizers. Suppose $P_i$ and $P_j$ are gauge operators. Then, they must fail to commute with at least on gauge generator, since they are not in the stabilizer group. Given that $P_iP_j$ is a stabilizer, they must fail to commute with the same gauge generators. This is not possible, because $i$ and $j$ are well separated relative $\ell$, the maximum linear size of a gauge generator. Similarly, if $P_i$ and $P_j$ fail to commute with a stabilizer, then they must fail to commute with the same stabilizer. Again, this cannot happen due to the fact that each stabilizer generator can be contained in a box of linear size $\ell$. Therefore, $\rho$ must have vanishing R\'enyi-1 correlations.

An analogous calculation shows that $\rho$ must also have vanishing R\'enyi-2 correlations. To argue that this is the case, the summand in Eq.~\eqref{eq: renyi1 app 2} is replaced by the R\'enyi-2 correlator
\begin{align}
   \left( \Tr[P_iP_j \rho (P_iP_j)^\dagger \rho] - \Tr[P_i \rho P_i^\dagger \rho] \Tr[P_j \rho P_j^\dagger \rho] \right) / \Tr[\rho^2].
\end{align}
This is only nonzero if $P_iP_j$ is a gauge operator and $P_i$, $P_j$ are not gauge operators. This is not possible, however, because that implies that $P_i$ and $P_j$ fail to commute with the same stabilizer generator, even though they are well separated relative to $\ell$. We conclude that the R\'enyi-2 correlations must also vanish.

Thus, if $\rho$ corresponds to a topological subsystem code and is obtained by decohering a stabilizer state, then it admits a purification into a GGS and has vanishing R\'enyi-1 and -2 correlations. In other words, $\rho$ is a locally-correlated mixed state.

\section{Conditional mutual information of the $Y$-decohered TC state} \label{app: CMI Ydecohered}

We prove here that the CMI $I(A:C|B)$ is non-vanishing in the $Y$-decohered TC state $\rho_Y$, for the subsystems $A$, $B$, and $C$ shown in Fig.~\ref{fig: ABCgeometry}, as claimed in Section~\ref{sec: Ydecohered TC}. We begin by recalling a formula for the CMI, applicable to stabilizer states (which may be mixed).  

The CMI can be expressed in terms of entanglement entropies as
\begin{align}
    I(A:C|B) = S(AB) + S(BC) - S(B) - S(ABC).
\end{align}
For a stabilizer state, the entanglement entropy for an arbitrary subsystem $A$ is
\begin{align}
    S(A) = n_A - k_A,
\end{align}
where $n_A$ is the number of qubits in the subsystem $A$ and $k_A$ is the dimension of the subgroup of stabilizers whose support can be entirely contained within $A$. Substituting the formula for the entanglement entropy into the expression for the CMI, we find
\begin{align}
    I(A:C|B) = k_{ABC} + k_B - k_{AB} -k_{BC}.
\end{align}
Notice that the dependencies on the number of qubits cancel. 

With this formula, the calculation of $I(A:C|B)$ for the $Y$-decohered TC state is straightforward. For simplicity, we assume that $\rho_{Y}$ is defined on a torus with dimensions $L \times L$. Since the stabilizers of $\rho_Y$ are generated by products of Pauli $Y$ operators along the diagonals (see Fig.~\ref{fig: ystabilizers}), there are no stabilizers supported entirely on $B$ or $AB$. This means that $k_B=k_{AB}=0$.  
The formula for the CMI in the state $\rho_Y$ reduces to
\begin{align}
    I_{\rho_Y}(A:C|B) = k_{ABC} -k_{BC}.
\end{align}

There are certainly stabilizers supported within $ABC$ that are not contained within $BC$ -- namely, the stabilizers that pass through the subsystem $A$. Therefore, the CMI is nonzero. Furthermore, the number of stabilizers that pass through $A$ depends only on the volume and geometry of $A$. Importantly, it is independent of the separation between $A$ and $C$. Thus, as claimed, the CMI is non-vanishing in the width of the subsystem $B$.

\section{Minimal anyon theory as a topological invariant} \label{app: Amin}

In this appendix, we show that Eq.~\eqref{eq: equality of mins} follows from Eq.~\eqref{eq: two-way connected anyons}. We start by constructing an injective map from $\cal{A}_1^{\rm min}$ to $\cal{A}_2^{\rm min}$. First, the anyons of $\cal{A}_1^{\rm min}$ can be lifted to anyons in $\cal{A}_1 / \cal{B}_1$. This is only ambiguous up to fusing with transparent bosons in $\cal{A}_1 / \cal{B}_1$. We denote this injective map by 
\begin{eqs}
    f_1: \cal{A}_i^{\rm min} \to \cal{A}_1 / \cal{B}_1.
\end{eqs} 
According to Eq.~\eqref{eq: two-way connected anyons}, $\cal{A}_1 / \cal{B}_1$ is a subtheory of $\cal{A}_2$, so we can define an injective map
\begin{eqs}
    g_1: \cal{A}_1 / \cal{B}_1 \to \cal{A}_2.
\end{eqs}
Finally, we define a non-injective map from $\cal{A}_2$ to $\cal{A}_2^{\rm min}$ by condensing all of the transparent bosons
\begin{eqs}
    h_1: \cal{A}_2 \to \cal{A}_2^{\rm min}.
\end{eqs}

Although $h_1$ is non-injective in general, when restricted to image of $g_1 \circ f_1$, it is injective. This follows from the fact that the anyons in the image of $f_1$ must correspond to distinct anyons in $\cal{A}_1^{\rm min}$ after condensing all of the transparent bosons. Therefore, they differ from one another in $\cal{A}_1/\cal{B}_2$ by more than fusing with transparent bosons. This means that the anyons in the image of $g_1 \circ f_1$ also differ by more than fusing with transparent bosons, and hence, after condensing the transparent bosons in $\cal{A}_2$, they must correspond to distinct elements of $\cal{A}_2^{\rm min}$. 

Similar to the construction of $h_1 \circ g_1 \circ f_1$, we can define an injective map from $\cal{A}_2^{\rm min}$ to $\cal{A}_1^{\rm min}$. Since we have injective maps between $\cal{A}_1^{\rm min}$ and $\cal{A}_2^{\rm min}$, the orders of the anyon theories must be the same, i.e., $|\cal{A}_1^{\rm min}| = |\cal{A}_2^{\rm min}|$.
 
Next, we argue that the anyons in $\cal{A}_1^{\rm min}$ and $\cal{A}_2^{\rm min}$ have the same exchange statistics and fusion rules. The fact that the exchange statistics are the same follows immediately from the observation that $f_1$, $g_1$, and $h_1$ preserve the statistics of the anyons. The maps also preserve the braiding relations. 

To see that the fusion rules of $\cal{A}_1^{\rm min}$ and $\cal{A}_2^{\rm min}$ are the same, we note that there are two possibilities for $\cal{A}_i^{\rm min}$.
\begin{enumerate}
    \item All of the transparent anyons in $\cal{A}_i$ are bosons, in which case, $\cal{A}_i^{\rm min}$ is modular.
    \item $\cal{A}_i$ has a transparent fermion, and thus, $\cal{A}_i^{\rm min}$ takes the form $\cal{A}_i^{\rm min} = \cal{C}_i \boxtimes \Z_2^{(1)}$, for some modular theory $\cal{C}_i$~\cite{FermionLSM}. 
\end{enumerate}
Because the braiding and statistics are preserved, if $\cal{A}_1^{\rm min}$ has a transparent fermion, then so too does $\cal{A}_2^{\rm min}$. Hence, in either case, the map $h_1 \circ g_1 \circ f_1$ maps the modular part of $\cal{A}_1^{\rm min}$ to the modular part of $\cal{A}_2^{\rm min}$, implying that the modular factors have the same exchange statistics and braiding relations. Finally, according to the Verlinde formula, which relates the statistics and braiding of the anyons to their fusion rules, the modular parts of $\cal{A}_1^{\rm min}$ and $\cal{A}_2^{\rm min}$ must have the same fusion rules. We then conclude that $\cal{A}_1^{\rm min}$ must be equal to $\cal{A}_2^{\rm min}$. 

\section{Phase equivalence of Abelian single-generator anyon theories} \label{app: single generator}

In Section~\ref{sec: anyon theories SLDQC}, we showed that if mixed states $\rho_1$ and $\rho_2$ are two-way connected by QLCs, then their anyon theories $\cal{A}_1$ and $\cal{A}_2$ satisfy
\begin{eqs} \label{eq: app anyon subtheories}
    \cal{A}_2/\cal{B}_2 \subset \cal{A}_1, \quad \cal{A}_1/\cal{B}_1 \subset \cal{A}_2,
\end{eqs}
for some transparent boson subgroups $\cal{B}_1$ and $\cal{B}_2$. Here, we consider the implications of these constraints when both $\cal{A}_1$ and $\cal{A}_2$  admit a single generator, i.e., we take the anyon theories to be
\begin{eqs} \label{eq: single generator A1 A2}
    \cal{A}_1 = \Z_{N}^{(q)}, \quad \cal{A}_2 = \Z_{M}^{(r)},
\end{eqs}
which are defined in Section~\ref{sec: anyon theories of subsystem codes}.

We start by noting that, in general, the anyon theories $\Z_{N}^{(q)}$ and $\Z_{M}^{(r)}$ can be factorized into subtheories whose orders are powers of primes -- analogous to the fundamental theorem of finite Abelian groups. To make this explicit, we write $N$ and $M$ as products of primes 
\begin{align}
    N = \prod_{p}p^{n_p}, \quad M = \prod_{p} p^{m_p}.
\end{align}
The anyon theories $\cal{A}_1$ and $\cal{A}_2$ then factorize as
\begin{eqs} \label{eq: ZN decomposition}
    \Z_{N}^{(q)} = \text{\raisebox{-1.3pt}{$\substack{\mathlarger{\mathlarger{\mathlarger{\mathlarger{\boxtimes}}}} \\ {\scriptstyle p}}$}} \, \Z_{p^{n_p}}^{(q_p)}, \quad
    \Z_{M}^{(r)} = \text{\raisebox{-1.3pt}{$\substack{\mathlarger{\mathlarger{\mathlarger{\mathlarger{\boxtimes}}}} \\ {\scriptstyle p}}$}} \, \Z_{p^{m_p}}^{(r_p)},
\end{eqs}
for the $q_p$ and $r_p$ specified below. The generator of the factor associated to the prime $p$ is $[N/p^{n_p}]$ and $[M/p^{m_p}]$, respectively. It can be checked using the formula for the braiding relations in Eq.~\eqref{eq: ZN braiding} that the generators for different prime factors have trivial braiding relations with each other, meaning that the factors are indeed independent. 

To determine $q_p$ and $r_p$, we compute the statistics of the generators. Using the formula for the exchange statistics in Eq.~\eqref{eq: ZN exchange}, we find:
\begin{align}
    \theta([N/p^{n_p}]) &= \exp\left\{\frac{2 \pi i}{p^{n_p}} \frac{qN}{p^{n_p}} \right\}, \\
    \theta([M/p^{m_p}]) &= \exp\left\{\frac{2 \pi i}{p^{m_p}} \frac{rM}{p^{m_p}} \right\}.
\end{align}
This implies that $q_p$ and $r_p$ are
\begin{eqs}
    q_p = \frac{qN}{p^{n_p}}, \quad r_p = \frac{rM}{p^{m_p}}.
\end{eqs}
Notice that if $q$ is divisible by $p^{n_p}$, then $q_p = 0 \text{ mod }p^{n_p}$. This would mean that there is a factor in the decomposition of $\Z_N^{(q)}$ in Eq.~\eqref{eq: ZN decomposition} that is purely composed of transparent bosons. Since QLCs can add and remove factors of transparent bosons, we restrict ourselves to the case where, respectively, $q$ and $r$ are not divisible by $p^{n_p}$ and $p^{m_p}$, for any prime $p$ and $n_p,m_p \neq 0$. This, of course, excludes $q=0$ and $r=0$ from our consideration.

With this condition on $q$ and $r$, we prove that Eq.~\eqref{eq: app anyon subtheories} implies that $\cal{A}_1=\cal{A}_2$. We begin by arguing that if $n_p \neq 0$, then $m_p \neq 0$. That is, for each prime subgroup of $\Z_N^{(q)}$ there is a corresponding subgroup of $\Z_M^{(r)}$. This follows immediately from the fact that $\cal{A}_1/\cal{B}_1$ is a subgroup of $\cal{A}_2$. Suppose $n_p \neq 0$, for some prime $p$, then there is a $\Z_{p^{k}}$ subgroup of $\cal{A}_1/\cal{B}_1$, for some $k \in \{1, 2, \ldots p^{n_p}-1 \}$, which must be a subgroup of $\Z_M^{(r)}$. This is only possible if $m_p \neq 0$. Likewise, the condition that $\cal{A}_2/\cal{B}_2 \subset \cal{A}_1$ tells us that if $m_p \neq 0$, then $n_p \neq 0$.  

We now focus only on the factor associated to $p$ and relate $q_p$ and $r_p$. We use the notion of a minimal anyon theory, discussed in Section~\ref{sec: anyon theories SLDQC} and Appendix~\ref{app: Amin}. The anyon theories $\cal{A}_1^{\rm min}$ and $\cal{A}_2^{\rm min}$ admit a prime factorization, as in Eq.~\eqref{eq: ZN decomposition}. Moreover, the subgroups of $\cal{A}_1^{\rm min}$ and $\cal{A}_2^{\rm min}$ associated to the prime $p$ must be generated by images of the generators of $\Z_{p^{n_p}}^{(q_p)}$ and $\Z_{p^{m_p}}^{(r_p)}$ after condensing transparent bosons. Then, since $\cal{A}_1^{\rm{min}} = \cal{A}_2^{\rm{min}}$, the generators of $\Z_{p^{n_p}}^{(q_p)}$ and $\Z_{p^{m_p}}^{(r_p)}$ must have the same exchange statistics. This gives us
\begin{eqs}
    e^{2 \pi i q_p/p^{n_p}} = e^{2 \pi i r_p/ p^{m_p}},  
\end{eqs}
which implies
\begin{eqs}
    q_p = r_p p^{m_p-n_p} \, \rm{mod}\, p^{m_p}.
\end{eqs}
Notice that, if $n_p = m_p$, then the expression above gives $q_p=r_p$, and we have that the factors are the same. Thus, we only need to consider the case where, without loss of generality, $m_p > n_p$.

Assuming that $m_p > n_p$, we again consider the condition $\cal{A}_1/\cal{B}_1 \subset \cal{A}_2$. In this case, the subgroup of  $\cal{A}_1/\cal{B}_1$ associated to $p$ must be a proper subtheory of $\Z_{p^{m_p}}^{(r_p)}$, which is generated by an anyon with the same statistics as the generator of $\Z_{p^{n_p}}^{(q_p)}$. Since the generator of $\Z_{p^{n_p}}^{(q_p)}$ has the same statistics as the generator of $\Z_{p^{m_p}}^{(r_p)}$, we see that $\Z_{p^{m_p}}^{(r_p)}$ must have an anyon with the property that it generates a proper subgroup of $\Z_{p^{m_p}}^{(r_p)}$ and has the exchange statistics of the generator. Let us write the generator of $\Z_{p^{m_p}}^{(r_p)}$ as $a$. Then there must be an anyon $a^{p^k}$ with $0<k\leq m_p$ that has the same exchange statistics as $a$. Note that $k$ must be greater than $0$ so that it generates a proper subgroup. The condition that $\theta(a) = \theta(a^{p^k})$ gives\footnote{Here, we have used the general property of Abelian anyons that $\theta(a^m)=\theta(a)^{m^2}$, for any anyon $a$ and integer $m$ \cite{Kitaev:2005hzj, Ellison2021}.}
\begin{eqs}
    e^{2\pi i r_p/ p^{m_p}} = e^{2\pi i r_p p^{2k}/ p^{m_p}},
\end{eqs}
which implies
\begin{eqs} \label{eq: number theory single generator 1}
    r_p = r_p p^{2k} \, \text{mod } p^{m_p}.
\end{eqs}

We now argue that there is no solution to this equation (other than $r_p=0$, which we have ruled out). The condition in Eq.~\eqref{eq: number theory single generator 1} can be re-expressed as
\begin{eqs} \label{eq: number theory single generator 2}
    r_p (p^{2k} - 1) = \ell p^{m_p},
\end{eqs}
for some integer $\ell$. The right-hand side is divisible by $p^{m_p}$, so the left-hand side must also be divisible by $p^{m_p}$. However, the factor $(p^{2k} - 1)$ is not divisible by any positive power of $p$.\footnote{To see this, one can derive a contradiction by assuming that $p^{2k}-1 = \ell p$, for some integer $\ell$. Rearranging, we find $p^{2k}-\ell p=p(p^{2k-1}-\ell) =1$. The left-hand side is divisible by $p$, but the right-hand side is not. Hence, $p^{2k}-1$ is not divisible by $p$. 
} 
Therefore, according to Eq.~\eqref{eq: number theory single generator 2}, $r_p$ must be divisible by $p^{m_p}$. However, $r_p$ satisfies $0<r_p < p^{m_p}$, so it also cannot be divisible by $p^{m_p}$.

This leads us to conclude that the anyon theories $\cal{A}_1$ and $\cal{A}_2$ in Eq.~\eqref{eq: single generator A1 A2} can only satisfy the conditions in Eq.~\eqref{eq: app anyon subtheories} if $n_p=m_p$ and $q_p=r_p$, for every prime $p$. This implies that $\cal{A}_1=\cal{A}_2$.

\section{Coordinate transformations and modular data}
\label{app: modular data}
 
In this appendix, we compute the expectation values of the modular transformations inside a coherent space of the decohered Ising string-net model. 
In particular, we consider the Ising string-net model on a torus, and choose an arbitrary ground state with $W_x(\psi\bar{\psi})= W_y(\psi\bar{\psi}) =1$. This subspace is spanned by the states in Eq.~\eqref{eq: Ising degeneracy}. The decohered state $\cal{N}(\op{\Psi})$ belongs to the coherent subspace labeled by $W_x(\psi\bar{\psi})= W_y(\psi\bar{\psi}) =1$.

Let $R$ be a modular transformation operator. First, we consider the standard expectation value of $R$ for a state within the coherent space with $W(\psi\bar{\psi})=1$. We can compute 
\begin{equation}
    \Tr [R\cal{N}(\rho)]=\Tr [\cal{N}^*(R)\op{\Psi}]=\braket{\Psi|\cal{N}^*(R)|\Psi},
\end{equation}
where $R$ is an arbitrary coordinate transformation.
We then find, for $\cal{N}$ in Eq.~\eqref{eq: Ising decoherence}, 
\begin{equation}
\begin{split}
    \cal{N}^*(R)&=\frac{1}{2^{N_e}}\sum_{\mb{e}}\prod_{e\in \mb{e}}\mu^z_e R \prod_{e\in \mb{e}}\mu^z_e\\
    &= \frac{1}{2^{N_e}}\sum_{\mb{e}}\prod_{e\in \mb{e}}\mu^z_e \prod_{e'\in R(\mb{e})}\mu^z_{e'} R,
\end{split}
\end{equation}
where the sums are over collections of edges $\mb{e}$.
If $\mb{e}\neq R(\mb{e})$, then the expectation value is 0 since $\mu_z^e$ excites plaquette terms. Thus, we only need to consider $\mb{e}$ such that $\mb{e}=R(\mb{e})$. The result is 
\begin{equation}
    \Tr [R\cal{N}(\rho)]=\frac{N_{R}}{2^{N_e}} \braket{\Psi|R|\Psi}.
\end{equation}
Here, $N_R$ counts the total number of $R$-invariant subsets of edges $\mb{e} = R(\mb{e})$.
Therefore, we conclude that up to an overall constant, the result is identical to the pure state expectation value. 

We note that another way to define modular matrices is to consider the ``distance'' between the mixed state $\rho=\cal{N}(\op{\psi})$ and the transformed state $R\rho R^\dag$. For simplicity, we consider the Hilbert-Schmidt (or ``Renyi-2'') distance. For two states $\rho$ and $\sigma$, we define the ``Renyi-2'' distance as
\begin{equation}
    \frac{\Tr \rho \sigma}{\sqrt{\Tr \rho^2}\sqrt{\Tr\sigma^2}}.
\end{equation}
For $\sigma=R\rho R^\dag$, since $R$ is unitary we have $\Tr\sigma^2=\Tr \rho^2$. For $\rho=\cal{N}(\op{\Psi})$, the numerator is given by
\begin{equation}
\begin{split}
    \Tr(\rho R\rho R^\dag)&= \Tr \big[\cal{N}(\op{\Psi})\cal{N}(\op{\Psi'})\big]\\
    &=\Tr \big[\op{\Psi}(\cal{N}^*\circ\cal{N})(\op{\Psi'})\big]\\
    &=\Tr \big[\op{\Psi}\cal{N}(\op{\Psi'})\big],
\end{split}
\end{equation}
where $\ket{\Psi'}=R\ket{\Psi}$. We can then expand $\cal{N}(\op{\Psi'})$ into a convex sum of $\ket{\Psi'}$ with various plaquette excitations. The only term that contributes is $\op{\Psi'}$, which should have a weight $\frac{1}{2^{N_p-1}}$. Here, we have used the fact that $W(\psi\bar{\psi})=1$ on $\ket{\Psi'}$. Thus, we find
\begin{equation}
\begin{split}
    \Tr(\rho R\rho R^\dag)&= \frac{1}{2^{N_p-1}}|\braket{\Psi|R|\Psi}|^2.
\end{split}
\end{equation}
The normalization factor in the denominator evaluates to $\Tr \rho^2= \frac{1}{2^{N_p-1}}$, so the ``Renyi-2'' distance 
\begin{equation}
    \frac{\Tr (\rho R\rho R^\dag)}{\Tr \rho^2}=|\braket{\Psi|R|\Psi}|^2.
\end{equation}
We note that the Renyi-2 observable only gives the modulus squares of the expectation values of $R$, which determine the representation of $R$ in the space up to an overall phase.

We have found that, within the coherent subspace, the two definitions of the expectation value of modular transformations are identical to those for the ground states up to overall normalization. Thus, in principle, we can read off the modular matrices for the decohered doubled Ising state from those of the ground states in the subspace Eq.~\eqref{eq: Ising degeneracy}. Below, we briefly outline how this is done for the $S$ matrix. 

The $S$ transformation for the doubled Ising theory is defined as~\cite{SET} 
\begin{equation}
    \ket{a_1\overline{a_2}}_y=\sum_{b_1,b_2} S_{a_1b_1}{S}_{a_2b_2}\ket{b_1\overline{b_2}}_x.
\end{equation}
Here, the labels $a_1,a_2,b_1,b_2$ take values in $1,\sigma,\psi$, and $S$ is the (chiral) Ising $S$ matrix. We have also used the fact that $S$ is real. The state $\ket{a_1\overline{a_2}}_{y}$ is defined such that a topological charge measurement performed around the cycle $x$ yields the measurement outcome $a_1\overline{a_2}$. Similarly, one can define $\ket{b_1\overline{b_2}}_x$. 

\begin{widetext}
Let us illustrate the calculation, using the transformation of $\frac{1}{\sqrt{2}}(\ket{11}_x+\ket{\psi\bar{\psi}}_x)$ to compute the first row of the $S$ matrix. The $S$ transformation maps $x$ to $y$, so the state becomes
\begin{equation}
\begin{split}
    \frac{1}{\sqrt{2}}(\ket{11}_y+\ket{\psi\bar{\psi}}_y)&=\sum_{b_1b_2}\frac{1}{\sqrt{2}}(S_{1b_1}S_{1b_2}+S_{\psi b_1}{S}_{\bar{\psi} b_2})\ket{b_1\overline{b_2}}_x\\
    &=\frac{1}{2\sqrt{2}}\sum_{b_1,b_2\in\{1,\psi\}}\ket{b_1\overline{b_2}}_x + \frac{1}{\sqrt{2}}\ket{\sigma\bar{\sigma}}_x\\
    &=\frac12 \frac{1}{\sqrt{2}}(\ket{11}_x+\ket{\psi\bar{\psi}}_x)  +
    \frac12\frac{1}{\sqrt{2}}(\ket{\psi 1}_x+\ket{1\bar{\psi}}_x) + \frac{1}{\sqrt{2}}\ket{\sigma\bar{\sigma}}_x,
\end{split}
\end{equation}
where we have used the known $S$ matrix for the Ising theory~\cite{Kitaev:2005hzj}.
\end{widetext}

Following similar steps for the other states in the subspace in Eq.~\eqref{eq: Ising degeneracy}, we find 
\begin{equation} \label{eq: decohered Ising S and T}
    S=\frac12
    \begin{pmatrix}
        1 & \sqrt{2} & 1\\
        \sqrt{2} & 0 & -\sqrt{2}\\
        1 & -\sqrt{2} & 1
    \end{pmatrix}, \quad
    T=\begin{pmatrix}
        1 & 0 & 0\\
        0 & 1 & 0\\
        0 & 0 & -1
    \end{pmatrix}.
\end{equation}
One can check that $S$ and $T$ satisfy $S^2=1$ and $(ST)^3=S^2$, so they indeed form a representation of the modular group, as expected.
We also point out that although $S$ is identical to the $S$ matrix of the (chiral) Ising TO, the $T$ matrices are different. The $T$ matrix of the Ising theory reads
\begin{equation}
    T_{\rm Ising}=\begin{pmatrix}
        1 & 0 & 0\\
        0 & e^{\frac{i\pi}{8}} & 0\\
        0 & 0 & -1
    \end{pmatrix}.
\end{equation}

Similar results can be obtained for higher-genus surfaces. This way, we find the modular data (i.e. unitary, finite-dimensional representation of the mapping class group on any closed oriented surfaces). Notably, the $S$ and $T$ matrices in Eq.~\eqref{eq: decohered Ising S and T} do not correspond to any (premodular) anyon theory. 

Lastly, we note that the $S$ and $T$ matrices in Eq.~\eqref{eq: decohered Ising S and T} can be derived in another way -- using the identification in Eq.~\eqref{eq: identify Ising and TC states}. Namely, the $\Z_2$ TC ground states are projected to the subspace invariant under the $e\leftrightarrow m$ symmetry. The modular matrices projected to this subspace are precisely those in Eq.~\eqref{eq: decohered Ising S and T}. More generally, one can define the modular data for any anyon theory projected to an invariant subspace under an anyon-permuting symmetry. In general, the resulting modular data does not correspond to any anyon theory. For example, by considering the $\Z_2$ anyon permutation symmetry in the $\mathrm{Spin}(2n)_1$ TO, we find the following modular matrices
\begin{equation} 
    S=\frac12
    \begin{pmatrix}
        1 & \sqrt{2} & 1\\
        \sqrt{2} & 0 & -\sqrt{2}\\
        1 & -\sqrt{2} & 1
    \end{pmatrix}, \:
    T=\begin{pmatrix}
        1 & 0 & 0\\
        0 & e^{\frac{i\pi n}{4}} & 0\\
        0 & 0 & -1
    \end{pmatrix}.
\end{equation}
Again, none of them corresponds to any anyon theory.

\section{Structure of premodular categories}
\label{premodular-cat}

In this appendix, we review the mathematical theory of premodular categories.
Let ${\cal C}$ is a premodular category and $\cal T$ be its transparent center. By Degline's theorem~\cite{Deligne}, $\cal T$ must be isomorphic to the category of finite-dimensional linear representations of some finite group $G$, denoted by $\Rep(G,z)$. Here, $z$ is an order-2 central element in $G$. Simple objects in $\Rep(G,z)$ are labeled by irreps $\pi$ of $G$. Fusion is given by the tensor product of representations.
The self statistics is determined by $z$:
\begin{equation}
 \theta({\pi})=\frac{\chi_\pi(z)}{\chi_\pi(1)}=\frac{\chi_\pi(z)}{\dim \pi}.   
\end{equation}
where $\chi_\pi(g)=\Tr \pi(g)$ is the character. Here, we use the fact that $\pi(1)=\id$ so that $\chi_\pi(1)=\dim \pi$, i.e., the dimension of the representation. Notice that {because $z$ is central, $\pi(z)$ commutes with every $\pi(g)$ for all $g\in G$. By Schur's lemma, $\pi(z)$ must be proportional to the identity. Then from $z^2=1$ it follows that $\pi(z)^2=\id$, so $\theta(\pi)=\pm 1$.} 

Physically, this means that the transparent center is isomorphic to the (bosonic or fermionic) gauge charges of a $G$ gauge theory.
In other words, as explained below, a premodular category can be uniquely associated with a (bosonic or fermionic) SET phase. 

First we consider the case when $z=1$, so all particles in $\cal T$ are bosonic. We can think of $\cal T$ as the bosonic gauge charge of some gauge group $G$, which can condense to yield a modular category $\tilde{\cal C}={\cal C}/{\cal T}$. Conversely, ${\cal C}$ can be thought of as ``$G$-symmetrizing'' (the mathematical parlance is $G$-equivariantizing) the theory $\tilde{\cal C}$. By $G$ symmetrizing, we mean projecting to the $G$-invariant states. It is closely related to gauging the $G$ symmetry, where one first introduces $G$ defects and then symmetrizes~\cite{SET}. Whereas, in this case, one directly symmetrizes the anyon theory without introducing $G$ flux anyons.

Let us illustrate with an example. The $\Z_2$ TC topological order has a $\Z_2$ anyonic symmetry that permutes the $e$ and $m$ anyons. Fully gauging the $\Z_2$ symmetry yields the ${\rm Ising}\boxtimes \overline{\rm Ising}$ topological order. Instead of gauging, if  the $\Z_2$ TC is ``symmetrized'', then we obtain a subset of ${\rm Ising}\boxtimes \overline{\rm Ising}$: $I, \psi , \bar{\psi}, \psi\bar{\psi}$ and $\sigma\bar{\sigma}$.  We can interpret $\psi\bar{\psi}$ as the $\Z_2$ gauge charge, $\psi $ and $\bar{\psi}$ as the $\psi$ anyon in the TC with even/odd gauge charges attached, and $\sigma\bar{\sigma}$ is $e+ m$ (where $+$ means direct sum), since the two particles are exchanged under the symmetry. They satisfy the fusion rule
\begin{equation}
    \sigma\bar{\sigma}\times\sigma\bar{\sigma} =  I+\psi+\bar{\psi}+\psi\bar{\psi}.
\end{equation}
This fusion rule can be understood from the identification $\sigma\bar{\sigma}\sim e+ m$:
\begin{equation}
    (e+ m)\times (e+ m)=e\times e + m\times m + e\times m+ m\times e.
\end{equation}
Naively, the result is $2(I+\psi)$. However, the $\Z_2$ symmetry has a nontrivial action on each of the 2-dimensional fusion space, and hence they should each decompose into a direct sum of $\Z_2$ even and odd representations. Since $\psi\bar{\psi}$ is the $\Z_2$ charge, $2 I$ should be $I+\psi\bar{\psi}$, and $\psi+\bar{\psi}$ for the other term.

As expected, the symmetrized TC anyons forms a premodular category ${\cal C}$, with ${\cal T}=\{I,\psi\bar{\psi}\}=\Rep(\Z_2,1)$. Condensing ${\cal T}$ in ${\cal C}$ returns the $\Z_2$ TC.  These are precisely the anyons in the decohered doubled Ising TO.

According to the general classification of symmetry-enriched topological phases~\cite{SET}, another possibility is that the symmetry fractionalizes on anyons even when no anyons are permuted. To give a simple example,  consider the semion topological order ${\cal C}=\{1,s\}$, where $s\times s=1$ and $\theta(s)=i$. Physically, it can be realized as a chiral spin liquid in certain spin-1/2 lattice models with full SO(3) spin rotation symmetry. Let us for now focus on a $\Z_2$ subgroup of SO(3), i.e. a $\pi$ rotation. The semion carries a half charge under the $\Z_2$ (a remanant of the spin-$1/2$ representation), which means fusing two semions yields a $\Z_2$ charge $b$. Symmetrizing the semion TO, we find the so-called $\Z_4^{(1)}$ anyon theory $\{1,s, b, sb\}$, where $b=s^2$, with the transparent center ${\cal T}=\{1, b\}$. This premodular category describes the decohered $\Z_4$ TC considered in Section~\ref{Z4TC}.

For a more complicated example of symmetry fractionalization, we consider the $\mathbb{D}_2=\Z_2\times\Z_2=\{1, X,Y,Z\}$ subgroup of SO(3). For this symmetry group in a semion TO, the semion transforms as a two-dimensional projective representation of the $\Z_2\times\Z_2$ symmetry. In this representation, $X$, $Y$, and $Z$ act as the Pauli matrices. After symmetrizing $\mathbb{D}_2$, the semion becomes a non-Abelian anyon with $d=2$, and satisfies the following fusion rule:
\begin{equation}
    s\times s=1+X+Y+Z,
\end{equation}
where here, $X$, $Y$, and $Z$ denote charged bosons of the corresponding symmetry. In other words, they are the three nontrivial one-dimensional representations of $\mathbb{D}_2$. The fusion rules are identical to those of the representation category of the order-8 quaternion group $Q_8$ (although with different $R$ symbols). Following Ref.~\cite{Wang:2016rzy}, we denote this category by $\Rep_s(Q_8)$.

Interestingly, there are three other premodular categories with the same fusion rules and topological spins, corresponding to the three other $\Z_2\times\Z_2$ projective representations on the semion. We will collectively denote them as $\Rep_s(D_8)$, where $D_8$ is the dihedral group of order 8. The difference between the $\Rep_s(D_8)$'s and $\Rep_s(Q_8)$ is that the latter admits a (minimal) modular extension (defined later), while the former do not. Equivalently, going back to the semion SET, the $\Z_2\times\Z_2$ symmetry has a nontrivial 't Hooft anomaly in each of the three $\Rep_s(D_8)$ categories, but for $\Rep_s(Q_8)$ the symmetry is non-anomalous (as the SET can be realized in 2D lattice models with on-site symmetry group). 

Now, we turn to the more general case with $z\neq 1$. We can still condense the maximal bosonic subcategory of $\Rep(G, z)$, i.e., the subcategory of irreps $\pi$ with $\pi(z)=\id$. Let us analyze the structure of the remaining category.

First, we show that the maximal bosonic subcategory is isomorphic to $\Rep(G_b,1)$ with $G_b=G/\{1,z\}$. To see this, first we show that an irrep $\pi$ of $G$ with $\pi(z)=\id$ is canonically isomorphic to a irrep of $G_b$. To this end, we choose an arbitrary lifting for $\tilde{g}\in G_b$ to $G$, denoted as $f(\tilde{g})$. Notice that $f(\tilde{g})f(\tilde{h})$ is equal to $f(\tilde{g}\tilde{h})$ up to $z$. Then we define $\tilde{\pi}$ as
\begin{equation}
    \tilde{\pi}(\tilde{g})=\pi(f(\tilde{g})).
\end{equation}
Because $\pi(z)=\id$, $\tilde{\pi}$ is well-defined and does not depend on the lifting.
We see that $\tilde{\pi}$ is a representation of $G_b$:
\begin{equation}
    \tilde{\pi}(\tilde{g})\tilde{\pi}(\tilde{h})=\pi (f(\tilde{g})f(\tilde{h}))=\pi (f(\tilde{g}\tilde{h}))=\tilde{\pi}(\tilde{g}\tilde{h}).
\end{equation}
Furthermore, since $\pi$ is irreducible, $\tilde{\pi}$ is too. Hence, we have shown that each $G$ irrep $\pi$ is also canonically a $G_b$ irrep. Moreover, if $\pi_1$ and $\pi_2$ are distinct irreps (but both satisfying $\pi(z)=\id$), so are $\tilde{\pi}_1$ and $\tilde{\pi}_2$.

Let us now prove that this captures all of the $G_b$ irreps. Through the orthogonality relation of characters: 
\begin{equation}
    \sum_{\pi}\chi_\pi(1)\chi^*_\pi(z)=\sum_{\pi}(\dim \pi)^2 \theta(\pi)=0,
\end{equation}
which implies that 
\begin{equation}
    \sum_{\theta(\pi)=1}(\dim \pi)^2=\frac{|G|}{2}=|G_b|.
\end{equation}
Therefore, we have found all the irreps of $G_b$, and this subcategory with $\pi(z)=1$ is identified as $\Rep(G_b,1)$. 

It is well-known that the $\Rep(G_b,1)$ category can be condensed. More precisely, we can form a condensable algebra object
\begin{equation}
    A=\bigoplus_{\pi \in \Rep(G_b,1)}(\dim \pi) \pi,
\end{equation}
whose quantum dimension is $d_A=|G_b|$. After condensation, the resulting category has total quantum dimension $\sqrt{|G|}/\sqrt{d_A}=\sqrt{2}$, which should be $\Z_2^{(1)}$.
  
  Note that $G_b$ and $G$ fit into the following short exact sequence:
\begin{equation}
    1\rightarrow \{1,z\}\rightarrow G\rightarrow G_b\rightarrow 1.
\end{equation}
In other words, $G$ is a central extension of $G_b$ by $\Z_2=\{1,z\}$. The central extension is uniquely determined by a 2-cocycle $\omega \in\H^2[G_b, \Z_2]$. One should think of this category as describing the symmetry of a fermionic system, where the fundamental fermion carries a projective representation of the symmetry group $G_b$, while bosonic excitations carry linear representations of $G_b$. Once $G_b$ and the fermion parity are both gauged, $\Rep(G,z)$ emerges as the subcategory of $G$ gauge charges.

Physically, the premodular category can be obtained from equivariantization of a fermionic modular tensor category, or in other words, a fermionic TO enriched by the $G_b$ symmetry group. In this case, $z$ should be identified as the fermion parity symmetry, and the fermion carries a projective representation of $G_b$ whose projective class is precisely $\omega$.

Essentially, Degline's theorem implies that all premodular categories arise as equivariantization of a finite unitary symmetry of a bosonic or fermionic SET phase. This also implies that a premodular category $\C$ can always be embedded (as a subcategory) into a modular category. For example, the modular category can always be taken as the Drinfeld center $\cZ(\C)$.

An important question is whether the premodular category admits a minimal modular extension. That is, a modular category $\cal{M}$ which contains $\cal{C}$ as a subcategory, and has the smallest quantum dimension among all such modular theories. In fact, the minimal modular extension should satisfy $\cal{D}_{\cal{M}}=|G|\cal{D}_{\cal{C}}$. When $\cal{T}$ is bosonic (i.e. $z=1$), the existence of a minimal modular extension is deeply related to the question of whether the associated SET phase obtained from condensing $\cal{T}$ has an 't Hooft anomaly. When the SET is non-anomalous, the minimal modular extension is obtained by gauging the $G$ symmetry. When there is a nontrivial 't Hooft anomaly, there exists no minimal modular extension. An example is the $\Rep_s(D_8)$ mentioned above, which admits no minimal modular extension due to the 't Hooft anomaly of the corresponding semion SET.

Similar results are expected to hold when $z\neq 1$, although the full details have not been worked out yet (see Ref.~\cite{johnson2024minimal} for discussions on this subject). If the associated fermionic SET is non-anomalous, a minimal modular extension can be found by gauging the entire $G$ symmetry, including both the bosonic global symmetry $G_b$ and the fermion parity. It was recently proven that when $\cal{T}=\Z_2^{(1)}$ a minimal modular extension always exists~\cite{johnson2024minimal}. 

Lastly, premodular categories have the following useful factorization property (Theorem 3.13 in 
 Ref. \cite{DGNO}):

{\theorem Let $\C$ be a premodular category, and $\mathcal{B}\subset \C$ is a modular subcategory. Then $\C=\cal{B}\boxtimes \cal{B}'$, where $\cal{B}'$ is the commutant of $\cal{B}$ in $\C$.}

\twocolumngrid

\bibliography{TO}

%merlin.mbs apsrev4-1.bst 2010-07-25 4.21a (PWD, AO, DPC) hacked
%Control: key (0)
%Control: author (0) dotless jnrlst
%Control: editor formatted (1) identically to author
%Control: production of article title (0) allowed
%Control: page (1) range
%Control: year (0) verbatim
%Control: production of eprint (0) enabled
\begin{thebibliography}{114}%
\makeatletter
\providecommand \@ifxundefined [1]{%
 \@ifx{#1\undefined}
}%
\providecommand \@ifnum [1]{%
 \ifnum #1\expandafter \@firstoftwo
 \else \expandafter \@secondoftwo
 \fi
}%
\providecommand \@ifx [1]{%
 \ifx #1\expandafter \@firstoftwo
 \else \expandafter \@secondoftwo
 \fi
}%
\providecommand \natexlab [1]{#1}%
\providecommand \enquote  [1]{``#1''}%
\providecommand \bibnamefont  [1]{#1}%
\providecommand \bibfnamefont [1]{#1}%
\providecommand \citenamefont [1]{#1}%
\providecommand \href@noop [0]{\@secondoftwo}%
\providecommand \href [0]{\begingroup \@sanitize@url \@href}%
\providecommand \@href[1]{\@@startlink{#1}\@@href}%
\providecommand \@@href[1]{\endgroup#1\@@endlink}%
\providecommand \@sanitize@url [0]{\catcode `\\12\catcode `\$12\catcode `\&12\catcode `\#12\catcode `\^12\catcode `\_12\catcode `\%12\relax}%
\providecommand \@@startlink[1]{}%
\providecommand \@@endlink[0]{}%
\providecommand \url  [0]{\begingroup\@sanitize@url \@url }%
\providecommand \@url [1]{\endgroup\@href {#1}{\urlprefix }}%
\providecommand \urlprefix  [0]{URL }%
\providecommand \Eprint [0]{\href }%
\providecommand \doibase [0]{http://dx.doi.org/}%
\providecommand \selectlanguage [0]{\@gobble}%
\providecommand \bibinfo  [0]{\@secondoftwo}%
\providecommand \bibfield  [0]{\@secondoftwo}%
\providecommand \translation [1]{[#1]}%
\providecommand \BibitemOpen [0]{}%
\providecommand \bibitemStop [0]{}%
\providecommand \bibitemNoStop [0]{.\EOS\space}%
\providecommand \EOS [0]{\spacefactor3000\relax}%
\providecommand \BibitemShut  [1]{\csname bibitem#1\endcsname}%
\let\auto@bib@innerbib\@empty
%</preamble>
\bibitem [{\citenamefont {Kitaev}(2006)}]{Kitaev:2005hzj}%
  \BibitemOpen
  \bibfield  {author} {\bibinfo {author} {\bibfnamefont {Alexei}\ \bibnamefont {Kitaev}},\ }\bibfield  {title} {\enquote {\bibinfo {title} {{Anyons in an exactly solved model and beyond}},}\ }\href {\doibase 10.1016/j.aop.2005.10.005} {\bibfield  {journal} {\bibinfo  {journal} {Annals Phys.}\ }\textbf {\bibinfo {volume} {321}},\ \bibinfo {pages} {2--111} (\bibinfo {year} {2006})},\ \Eprint {http://arxiv.org/abs/cond-mat/0506438} {arXiv:cond-mat/0506438} \BibitemShut {NoStop}%
\bibitem [{\citenamefont {Wen}(2015)}]{Wen_2015}%
  \BibitemOpen
  \bibfield  {author} {\bibinfo {author} {\bibfnamefont {Xiao-Gang}\ \bibnamefont {Wen}},\ }\bibfield  {title} {\enquote {\bibinfo {title} {A theory of 2+1d bosonic topological orders},}\ }\href {\doibase 10.1093/nsr/nwv077} {\bibfield  {journal} {\bibinfo  {journal} {National Science Review}\ }\textbf {\bibinfo {volume} {3}},\ \bibinfo {pages} {68–106} (\bibinfo {year} {2015})}\BibitemShut {NoStop}%
\bibitem [{\citenamefont {Johnson-Freyd}(2022)}]{Johnson_Freyd_2022}%
  \BibitemOpen
  \bibfield  {author} {\bibinfo {author} {\bibfnamefont {Theo}\ \bibnamefont {Johnson-Freyd}},\ }\bibfield  {title} {\enquote {\bibinfo {title} {On the classification of topological orders},}\ }\href {\doibase 10.1007/s00220-022-04380-3} {\bibfield  {journal} {\bibinfo  {journal} {Communications in Mathematical Physics}\ }\textbf {\bibinfo {volume} {393}},\ \bibinfo {pages} {989–1033} (\bibinfo {year} {2022})}\BibitemShut {NoStop}%
\bibitem [{\citenamefont {Wen}(2017)}]{WenRMP}%
  \BibitemOpen
  \bibfield  {author} {\bibinfo {author} {\bibfnamefont {Xiao-Gang}\ \bibnamefont {Wen}},\ }\bibfield  {title} {\enquote {\bibinfo {title} {Colloquium: Zoo of quantum-topological phases of matter},}\ }\href {\doibase 10.1103/RevModPhys.89.041004} {\bibfield  {journal} {\bibinfo  {journal} {Rev. Mod. Phys.}\ }\textbf {\bibinfo {volume} {89}},\ \bibinfo {pages} {041004} (\bibinfo {year} {2017})}\BibitemShut {NoStop}%
\bibitem [{\citenamefont {Wen}(2004)}]{wen2004quantum}%
  \BibitemOpen
  \bibfield  {author} {\bibinfo {author} {\bibfnamefont {Xiao-Gang}\ \bibnamefont {Wen}},\ }\href@noop {} {\emph {\bibinfo {title} {Quantum field theory of many-body systems: From the origin of sound to an origin of light and electrons}}}\ (\bibinfo  {publisher} {Oxford university press},\ \bibinfo {year} {2004})\BibitemShut {NoStop}%
\bibitem [{\citenamefont {Simon}(2023)}]{simon2023topological}%
  \BibitemOpen
  \bibfield  {author} {\bibinfo {author} {\bibfnamefont {Steven~H}\ \bibnamefont {Simon}},\ }\href@noop {} {\emph {\bibinfo {title} {Topological quantum}}}\ (\bibinfo  {publisher} {Oxford University Press},\ \bibinfo {year} {2023})\BibitemShut {NoStop}%
\bibitem [{\citenamefont {Zeng}\ \emph {et~al.}(2019)\citenamefont {Zeng}, \citenamefont {Chen}, \citenamefont {Zhou}, \citenamefont {Wen} \emph {et~al.}}]{zeng2019quantum}%
  \BibitemOpen
  \bibfield  {author} {\bibinfo {author} {\bibfnamefont {Bei}\ \bibnamefont {Zeng}}, \bibinfo {author} {\bibfnamefont {Xie}\ \bibnamefont {Chen}}, \bibinfo {author} {\bibfnamefont {Duan-Lu}\ \bibnamefont {Zhou}}, \bibinfo {author} {\bibfnamefont {Xiao-Gang}\ \bibnamefont {Wen}},  \emph {et~al.},\ }\href@noop {} {\emph {\bibinfo {title} {Quantum information meets quantum matter}}}\ (\bibinfo  {publisher} {Springer},\ \bibinfo {year} {2019})\BibitemShut {NoStop}%
\bibitem [{\citenamefont {Hastings}(2011)}]{HastingsFiniteT}%
  \BibitemOpen
  \bibfield  {author} {\bibinfo {author} {\bibfnamefont {Matthew~B.}\ \bibnamefont {Hastings}},\ }\bibfield  {title} {\enquote {\bibinfo {title} {Topological order at nonzero temperature},}\ }\href {\doibase 10.1103/PhysRevLett.107.210501} {\bibfield  {journal} {\bibinfo  {journal} {Phys. Rev. Lett.}\ }\textbf {\bibinfo {volume} {107}},\ \bibinfo {pages} {210501} (\bibinfo {year} {2011})}\BibitemShut {NoStop}%
\bibitem [{\citenamefont {Castelnovo}\ and\ \citenamefont {Chamon}(2007{\natexlab{a}})}]{CastelnovoPRB2007}%
  \BibitemOpen
  \bibfield  {author} {\bibinfo {author} {\bibfnamefont {Claudio}\ \bibnamefont {Castelnovo}}\ and\ \bibinfo {author} {\bibfnamefont {Claudio}\ \bibnamefont {Chamon}},\ }\bibfield  {title} {\enquote {\bibinfo {title} {Entanglement and topological entropy of the toric code at finite temperature},}\ }\href {\doibase 10.1103/PhysRevB.76.184442} {\bibfield  {journal} {\bibinfo  {journal} {Phys. Rev. B}\ }\textbf {\bibinfo {volume} {76}},\ \bibinfo {pages} {184442} (\bibinfo {year} {2007}{\natexlab{a}})}\BibitemShut {NoStop}%
\bibitem [{\citenamefont {Nussinov}\ and\ \citenamefont {Ortiz}(2008)}]{Nussinov2008thermalselfcorrecting}%
  \BibitemOpen
  \bibfield  {author} {\bibinfo {author} {\bibfnamefont {Zohar}\ \bibnamefont {Nussinov}}\ and\ \bibinfo {author} {\bibfnamefont {Gerardo}\ \bibnamefont {Ortiz}},\ }\bibfield  {title} {\enquote {\bibinfo {title} {Autocorrelations and thermal fragility of anyonic loops in topologically quantum ordered systems},}\ }\href {\doibase 10.1103/PhysRevB.77.064302} {\bibfield  {journal} {\bibinfo  {journal} {Phys. Rev. B}\ }\textbf {\bibinfo {volume} {77}},\ \bibinfo {pages} {064302} (\bibinfo {year} {2008})}\BibitemShut {NoStop}%
\bibitem [{\citenamefont {Lu}\ \emph {et~al.}(2020)\citenamefont {Lu}, \citenamefont {Hsieh},\ and\ \citenamefont {Grover}}]{Lu:2019owx}%
  \BibitemOpen
  \bibfield  {author} {\bibinfo {author} {\bibfnamefont {Tsung-Cheng}\ \bibnamefont {Lu}}, \bibinfo {author} {\bibfnamefont {Timothy~H.}\ \bibnamefont {Hsieh}}, \ and\ \bibinfo {author} {\bibfnamefont {Tarun}\ \bibnamefont {Grover}},\ }\bibfield  {title} {\enquote {\bibinfo {title} {{Detecting Topological Order at Finite Temperature Using Entanglement Negativity}},}\ }\href {\doibase 10.1103/PhysRevLett.125.116801} {\bibfield  {journal} {\bibinfo  {journal} {Phys. Rev. Lett.}\ }\textbf {\bibinfo {volume} {125}},\ \bibinfo {pages} {116801} (\bibinfo {year} {2020})},\ \Eprint {http://arxiv.org/abs/1912.04293} {arXiv:1912.04293 [cond-mat.str-el]} \BibitemShut {NoStop}%
\bibitem [{\citenamefont {Sang}\ \emph {et~al.}(2023)\citenamefont {Sang}, \citenamefont {Zou},\ and\ \citenamefont {Hsieh}}]{Sang:2023rsp}%
  \BibitemOpen
  \bibfield  {author} {\bibinfo {author} {\bibfnamefont {Shengqi}\ \bibnamefont {Sang}}, \bibinfo {author} {\bibfnamefont {Yijian}\ \bibnamefont {Zou}}, \ and\ \bibinfo {author} {\bibfnamefont {Timothy~H.}\ \bibnamefont {Hsieh}},\ }\bibfield  {title} {\enquote {\bibinfo {title} {{Mixed-state Quantum Phases: Renormalization and Quantum Error Correction}},}\ }\href@noop {} {\  (\bibinfo {year} {2023})},\ \Eprint {http://arxiv.org/abs/2310.08639} {arXiv:2310.08639 [quant-ph]} \BibitemShut {NoStop}%
\bibitem [{\citenamefont {Landon-Cardinal}\ and\ \citenamefont {Poulin}(2013)}]{Poulin2013thermal}%
  \BibitemOpen
  \bibfield  {author} {\bibinfo {author} {\bibfnamefont {Olivier}\ \bibnamefont {Landon-Cardinal}}\ and\ \bibinfo {author} {\bibfnamefont {David}\ \bibnamefont {Poulin}},\ }\bibfield  {title} {\enquote {\bibinfo {title} {Local topological order inhibits thermal stability in 2d},}\ }\href {\doibase 10.1103/PhysRevLett.110.090502} {\bibfield  {journal} {\bibinfo  {journal} {Phys. Rev. Lett.}\ }\textbf {\bibinfo {volume} {110}},\ \bibinfo {pages} {090502} (\bibinfo {year} {2013})}\BibitemShut {NoStop}%
\bibitem [{\citenamefont {Brown}\ \emph {et~al.}(2016)\citenamefont {Brown}, \citenamefont {Loss}, \citenamefont {Pachos}, \citenamefont {Self},\ and\ \citenamefont {Wootton}}]{Brown:2014idi}%
  \BibitemOpen
  \bibfield  {author} {\bibinfo {author} {\bibfnamefont {Benjamin~J.}\ \bibnamefont {Brown}}, \bibinfo {author} {\bibfnamefont {Daniel}\ \bibnamefont {Loss}}, \bibinfo {author} {\bibfnamefont {Jiannis~K.}\ \bibnamefont {Pachos}}, \bibinfo {author} {\bibfnamefont {Chris~N.}\ \bibnamefont {Self}}, \ and\ \bibinfo {author} {\bibfnamefont {James~R.}\ \bibnamefont {Wootton}},\ }\bibfield  {title} {\enquote {\bibinfo {title} {{Quantum memories at finite temperature}},}\ }\href {\doibase 10.1103/RevModPhys.88.045005} {\bibfield  {journal} {\bibinfo  {journal} {Rev. Mod. Phys.}\ }\textbf {\bibinfo {volume} {88}},\ \bibinfo {pages} {045005} (\bibinfo {year} {2016})},\ \Eprint {http://arxiv.org/abs/1411.6643} {arXiv:1411.6643 [quant-ph]} \BibitemShut {NoStop}%
\bibitem [{\citenamefont {Dennis}\ \emph {et~al.}(2002)\citenamefont {Dennis}, \citenamefont {Kitaev}, \citenamefont {Landahl},\ and\ \citenamefont {Preskill}}]{Dennis:2001nw}%
  \BibitemOpen
  \bibfield  {author} {\bibinfo {author} {\bibfnamefont {Eric}\ \bibnamefont {Dennis}}, \bibinfo {author} {\bibfnamefont {Alexei}\ \bibnamefont {Kitaev}}, \bibinfo {author} {\bibfnamefont {Andrew}\ \bibnamefont {Landahl}}, \ and\ \bibinfo {author} {\bibfnamefont {John}\ \bibnamefont {Preskill}},\ }\bibfield  {title} {\enquote {\bibinfo {title} {{Topological quantum memory}},}\ }\href {\doibase 10.1063/1.1499754} {\bibfield  {journal} {\bibinfo  {journal} {J. Math. Phys.}\ }\textbf {\bibinfo {volume} {43}},\ \bibinfo {pages} {4452--4505} (\bibinfo {year} {2002})},\ \Eprint {http://arxiv.org/abs/quant-ph/0110143} {arXiv:quant-ph/0110143} \BibitemShut {NoStop}%
\bibitem [{\citenamefont {Garratt}\ \emph {et~al.}(2023)\citenamefont {Garratt}, \citenamefont {Weinstein},\ and\ \citenamefont {Altman}}]{Garratt:2022ycp}%
  \BibitemOpen
  \bibfield  {author} {\bibinfo {author} {\bibfnamefont {Samuel~J.}\ \bibnamefont {Garratt}}, \bibinfo {author} {\bibfnamefont {Zack}\ \bibnamefont {Weinstein}}, \ and\ \bibinfo {author} {\bibfnamefont {Ehud}\ \bibnamefont {Altman}},\ }\bibfield  {title} {\enquote {\bibinfo {title} {Measurements conspire nonlocally to restructure critical quantum states},}\ }\href {\doibase 10.1103/PhysRevX.13.021026} {\bibfield  {journal} {\bibinfo  {journal} {Phys. Rev. X}\ }\textbf {\bibinfo {volume} {13}},\ \bibinfo {pages} {021026} (\bibinfo {year} {2023})}\BibitemShut {NoStop}%
\bibitem [{\citenamefont {Yang}\ \emph {et~al.}(2023)\citenamefont {Yang}, \citenamefont {Mao},\ and\ \citenamefont {Jian}}]{Yang:2023dol}%
  \BibitemOpen
  \bibfield  {author} {\bibinfo {author} {\bibfnamefont {Zhou}\ \bibnamefont {Yang}}, \bibinfo {author} {\bibfnamefont {Dan}\ \bibnamefont {Mao}}, \ and\ \bibinfo {author} {\bibfnamefont {Chao-Ming}\ \bibnamefont {Jian}},\ }\bibfield  {title} {\enquote {\bibinfo {title} {Entanglement in a one-dimensional critical state after measurements},}\ }\href {\doibase 10.1103/PhysRevB.108.165120} {\bibfield  {journal} {\bibinfo  {journal} {Phys. Rev. B}\ }\textbf {\bibinfo {volume} {108}},\ \bibinfo {pages} {165120} (\bibinfo {year} {2023})}\BibitemShut {NoStop}%
\bibitem [{\citenamefont {Sun}\ \emph {et~al.}(2023)\citenamefont {Sun}, \citenamefont {Yao},\ and\ \citenamefont {Jian}}]{Sun:2023alk}%
  \BibitemOpen
  \bibfield  {author} {\bibinfo {author} {\bibfnamefont {Xinyu}\ \bibnamefont {Sun}}, \bibinfo {author} {\bibfnamefont {Hong}\ \bibnamefont {Yao}}, \ and\ \bibinfo {author} {\bibfnamefont {Shao-Kai}\ \bibnamefont {Jian}},\ }\bibfield  {title} {\enquote {\bibinfo {title} {{New critical states induced by measurement}},}\ }\href@noop {} {\  (\bibinfo {year} {2023})},\ \Eprint {http://arxiv.org/abs/2301.11337} {arXiv:2301.11337 [quant-ph]} \BibitemShut {NoStop}%
\bibitem [{\citenamefont {Lee}\ \emph {et~al.}(2023)\citenamefont {Lee}, \citenamefont {Jian},\ and\ \citenamefont {Xu}}]{Lee:2023fsk}%
  \BibitemOpen
  \bibfield  {author} {\bibinfo {author} {\bibfnamefont {Jong~Yeon}\ \bibnamefont {Lee}}, \bibinfo {author} {\bibfnamefont {Chao-Ming}\ \bibnamefont {Jian}}, \ and\ \bibinfo {author} {\bibfnamefont {Cenke}\ \bibnamefont {Xu}},\ }\bibfield  {title} {\enquote {\bibinfo {title} {Quantum criticality under decoherence or weak measurement},}\ }\href {\doibase 10.1103/PRXQuantum.4.030317} {\bibfield  {journal} {\bibinfo  {journal} {PRX Quantum}\ }\textbf {\bibinfo {volume} {4}},\ \bibinfo {pages} {030317} (\bibinfo {year} {2023})}\BibitemShut {NoStop}%
\bibitem [{\citenamefont {Zou}\ \emph {et~al.}(2023)\citenamefont {Zou}, \citenamefont {Sang},\ and\ \citenamefont {Hsieh}}]{Zou:2023rmw}%
  \BibitemOpen
  \bibfield  {author} {\bibinfo {author} {\bibfnamefont {Yijian}\ \bibnamefont {Zou}}, \bibinfo {author} {\bibfnamefont {Shengqi}\ \bibnamefont {Sang}}, \ and\ \bibinfo {author} {\bibfnamefont {Timothy~H.}\ \bibnamefont {Hsieh}},\ }\bibfield  {title} {\enquote {\bibinfo {title} {{Channeling quantum criticality}},}\ }\href@noop {} {\  (\bibinfo {year} {2023})},\ \Eprint {http://arxiv.org/abs/2301.07141} {arXiv:2301.07141 [quant-ph]} \BibitemShut {NoStop}%
\bibitem [{\citenamefont {Ma}(2023)}]{Ma:2023tmy}%
  \BibitemOpen
  \bibfield  {author} {\bibinfo {author} {\bibfnamefont {Ruochen}\ \bibnamefont {Ma}},\ }\bibfield  {title} {\enquote {\bibinfo {title} {{Exploring critical systems under measurements and decoherence via Keldysh field theory}},}\ }\href@noop {} {\  (\bibinfo {year} {2023})},\ \Eprint {http://arxiv.org/abs/2304.08277} {arXiv:2304.08277 [quant-ph]} \BibitemShut {NoStop}%
\bibitem [{\citenamefont {de~Groot}\ \emph {et~al.}(2022)\citenamefont {de~Groot}, \citenamefont {Turzillo},\ and\ \citenamefont {Schuch}}]{deGroot2022}%
  \BibitemOpen
  \bibfield  {author} {\bibinfo {author} {\bibfnamefont {Caroline}\ \bibnamefont {de~Groot}}, \bibinfo {author} {\bibfnamefont {Alex}\ \bibnamefont {Turzillo}}, \ and\ \bibinfo {author} {\bibfnamefont {Norbert}\ \bibnamefont {Schuch}},\ }\bibfield  {title} {\enquote {\bibinfo {title} {{Symmetry Protected Topological Order in Open Quantum Systems}},}\ }\href {\doibase 10.22331/q-2022-11-10-856} {\bibfield  {journal} {\bibinfo  {journal} {Quantum}\ }\textbf {\bibinfo {volume} {6}},\ \bibinfo {pages} {856} (\bibinfo {year} {2022})}\BibitemShut {NoStop}%
\bibitem [{\citenamefont {Lee}\ \emph {et~al.}(2022)\citenamefont {Lee}, \citenamefont {You},\ and\ \citenamefont {Xu}}]{Lee:2022hog}%
  \BibitemOpen
  \bibfield  {author} {\bibinfo {author} {\bibfnamefont {Jong~Yeon}\ \bibnamefont {Lee}}, \bibinfo {author} {\bibfnamefont {Yi-Zhuang}\ \bibnamefont {You}}, \ and\ \bibinfo {author} {\bibfnamefont {Cenke}\ \bibnamefont {Xu}},\ }\bibfield  {title} {\enquote {\bibinfo {title} {{Symmetry protected topological phases under decoherence}},}\ }\href@noop {} {\  (\bibinfo {year} {2022})},\ \Eprint {http://arxiv.org/abs/2210.16323} {arXiv:2210.16323 [cond-mat.str-el]} \BibitemShut {NoStop}%
\bibitem [{\citenamefont {Ma}\ and\ \citenamefont {Wang}(2023)}]{Ma:2022pvq}%
  \BibitemOpen
  \bibfield  {author} {\bibinfo {author} {\bibfnamefont {Ruochen}\ \bibnamefont {Ma}}\ and\ \bibinfo {author} {\bibfnamefont {Chong}\ \bibnamefont {Wang}},\ }\bibfield  {title} {\enquote {\bibinfo {title} {Average symmetry-protected topological phases},}\ }\href {\doibase 10.1103/PhysRevX.13.031016} {\bibfield  {journal} {\bibinfo  {journal} {Phys. Rev. X}\ }\textbf {\bibinfo {volume} {13}},\ \bibinfo {pages} {031016} (\bibinfo {year} {2023})}\BibitemShut {NoStop}%
\bibitem [{\citenamefont {{Zhang}}\ \emph {et~al.}(2022)\citenamefont {{Zhang}}, \citenamefont {{Qi}},\ and\ \citenamefont {{Bi}}}]{ZhangQiBi2022}%
  \BibitemOpen
  \bibfield  {author} {\bibinfo {author} {\bibfnamefont {Jian-Hao}\ \bibnamefont {{Zhang}}}, \bibinfo {author} {\bibfnamefont {Yang}\ \bibnamefont {{Qi}}}, \ and\ \bibinfo {author} {\bibfnamefont {Zhen}\ \bibnamefont {{Bi}}},\ }\bibfield  {title} {\enquote {\bibinfo {title} {{Strange Correlation Function for Average Symmetry-Protected Topological Phases}},}\ }\href {\doibase 10.48550/arXiv.2210.17485} {\bibfield  {journal} {\bibinfo  {journal} {arXiv e-prints}\ ,\ \bibinfo {eid} {arXiv:2210.17485}} (\bibinfo {year} {2022})},\ \Eprint {http://arxiv.org/abs/2210.17485} {arXiv:2210.17485 [cond-mat.str-el]} \BibitemShut {NoStop}%
\bibitem [{\citenamefont {Molignini}\ and\ \citenamefont {Cooper}(2023)}]{Moligini2023topological}%
  \BibitemOpen
  \bibfield  {author} {\bibinfo {author} {\bibfnamefont {Paolo}\ \bibnamefont {Molignini}}\ and\ \bibinfo {author} {\bibfnamefont {Nigel~R.}\ \bibnamefont {Cooper}},\ }\bibfield  {title} {\enquote {\bibinfo {title} {Topological phase transitions at finite temperature},}\ }\href {\doibase 10.1103/PhysRevResearch.5.023004} {\bibfield  {journal} {\bibinfo  {journal} {Phys. Rev. Res.}\ }\textbf {\bibinfo {volume} {5}},\ \bibinfo {pages} {023004} (\bibinfo {year} {2023})}\BibitemShut {NoStop}%
\bibitem [{\citenamefont {Ma}\ \emph {et~al.}(2023)\citenamefont {Ma}, \citenamefont {Zhang}, \citenamefont {Bi}, \citenamefont {Cheng},\ and\ \citenamefont {Wang}}]{Ma:2023rji}%
  \BibitemOpen
  \bibfield  {author} {\bibinfo {author} {\bibfnamefont {Ruochen}\ \bibnamefont {Ma}}, \bibinfo {author} {\bibfnamefont {Jian-Hao}\ \bibnamefont {Zhang}}, \bibinfo {author} {\bibfnamefont {Zhen}\ \bibnamefont {Bi}}, \bibinfo {author} {\bibfnamefont {Meng}\ \bibnamefont {Cheng}}, \ and\ \bibinfo {author} {\bibfnamefont {Chong}\ \bibnamefont {Wang}},\ }\bibfield  {title} {\enquote {\bibinfo {title} {{Topological Phases with Average Symmetries: the Decohered, the Disordered, and the Intrinsic}},}\ }\href@noop {} {\  (\bibinfo {year} {2023})},\ \Eprint {http://arxiv.org/abs/2305.16399} {arXiv:2305.16399 [cond-mat.str-el]} \BibitemShut {NoStop}%
\bibitem [{\citenamefont {Ma}\ and\ \citenamefont {Turzillo}(2024)}]{Ma:2024kma}%
  \BibitemOpen
  \bibfield  {author} {\bibinfo {author} {\bibfnamefont {Ruochen}\ \bibnamefont {Ma}}\ and\ \bibinfo {author} {\bibfnamefont {Alex}\ \bibnamefont {Turzillo}},\ }\bibfield  {title} {\enquote {\bibinfo {title} {{Symmetry Protected Topological Phases of Mixed States in the Doubled Space}},}\ }\href@noop {} {\  (\bibinfo {year} {2024})},\ \Eprint {http://arxiv.org/abs/2403.13280} {arXiv:2403.13280 [quant-ph]} \BibitemShut {NoStop}%
\bibitem [{\citenamefont {Guo}\ \emph {et~al.}(2024)\citenamefont {Guo}, \citenamefont {Zhang}, \citenamefont {Yang},\ and\ \citenamefont {Bi}}]{guo2024locally}%
  \BibitemOpen
  \bibfield  {author} {\bibinfo {author} {\bibfnamefont {Yuchen}\ \bibnamefont {Guo}}, \bibinfo {author} {\bibfnamefont {Jian-Hao}\ \bibnamefont {Zhang}}, \bibinfo {author} {\bibfnamefont {Shuo}\ \bibnamefont {Yang}}, \ and\ \bibinfo {author} {\bibfnamefont {Zhen}\ \bibnamefont {Bi}},\ }\href@noop {} {\enquote {\bibinfo {title} {Locally purified density operators for symmetry-protected topological phases in mixed states},}\ } (\bibinfo {year} {2024}),\ \Eprint {http://arxiv.org/abs/2403.16978} {arXiv:2403.16978 [cond-mat.str-el]} \BibitemShut {NoStop}%
\bibitem [{\citenamefont {Xue}\ \emph {et~al.}(2024)\citenamefont {Xue}, \citenamefont {Lee},\ and\ \citenamefont {Bao}}]{Xue:2024bkt}%
  \BibitemOpen
  \bibfield  {author} {\bibinfo {author} {\bibfnamefont {Hanyu}\ \bibnamefont {Xue}}, \bibinfo {author} {\bibfnamefont {Jong~Yeon}\ \bibnamefont {Lee}}, \ and\ \bibinfo {author} {\bibfnamefont {Yimu}\ \bibnamefont {Bao}},\ }\bibfield  {title} {\enquote {\bibinfo {title} {{Tensor network formulation of symmetry protected topological phases in mixed states}},}\ }\href@noop {} {\  (\bibinfo {year} {2024})},\ \Eprint {http://arxiv.org/abs/2403.17069} {arXiv:2403.17069 [cond-mat.str-el]} \BibitemShut {NoStop}%
\bibitem [{\citenamefont {Fan}\ \emph {et~al.}(2023)\citenamefont {Fan}, \citenamefont {Bao}, \citenamefont {Altman},\ and\ \citenamefont {Vishwanath}}]{Fan:2023rvp}%
  \BibitemOpen
  \bibfield  {author} {\bibinfo {author} {\bibfnamefont {Ruihua}\ \bibnamefont {Fan}}, \bibinfo {author} {\bibfnamefont {Yimu}\ \bibnamefont {Bao}}, \bibinfo {author} {\bibfnamefont {Ehud}\ \bibnamefont {Altman}}, \ and\ \bibinfo {author} {\bibfnamefont {Ashvin}\ \bibnamefont {Vishwanath}},\ }\bibfield  {title} {\enquote {\bibinfo {title} {{Diagnostics of mixed-state topological order and breakdown of quantum memory}},}\ }\href@noop {} {\  (\bibinfo {year} {2023})},\ \Eprint {http://arxiv.org/abs/2301.05689} {arXiv:2301.05689 [quant-ph]} \BibitemShut {NoStop}%
\bibitem [{\citenamefont {Bao}\ \emph {et~al.}(2023)\citenamefont {Bao}, \citenamefont {Fan}, \citenamefont {Vishwanath},\ and\ \citenamefont {Altman}}]{Bao:2023zry}%
  \BibitemOpen
  \bibfield  {author} {\bibinfo {author} {\bibfnamefont {Yimu}\ \bibnamefont {Bao}}, \bibinfo {author} {\bibfnamefont {Ruihua}\ \bibnamefont {Fan}}, \bibinfo {author} {\bibfnamefont {Ashvin}\ \bibnamefont {Vishwanath}}, \ and\ \bibinfo {author} {\bibfnamefont {Ehud}\ \bibnamefont {Altman}},\ }\bibfield  {title} {\enquote {\bibinfo {title} {{Mixed-state topological order and the errorfield double formulation of decoherence-induced transitions}},}\ }\href@noop {} {\  (\bibinfo {year} {2023})},\ \Eprint {http://arxiv.org/abs/2301.05687} {arXiv:2301.05687 [quant-ph]} \BibitemShut {NoStop}%
\bibitem [{\citenamefont {Wang}\ \emph {et~al.}(2023)\citenamefont {Wang}, \citenamefont {Wu},\ and\ \citenamefont {Wang}}]{Wang:2023uoj}%
  \BibitemOpen
  \bibfield  {author} {\bibinfo {author} {\bibfnamefont {Zijian}\ \bibnamefont {Wang}}, \bibinfo {author} {\bibfnamefont {Zhengzhi}\ \bibnamefont {Wu}}, \ and\ \bibinfo {author} {\bibfnamefont {Zhong}\ \bibnamefont {Wang}},\ }\bibfield  {title} {\enquote {\bibinfo {title} {{Intrinsic Mixed-state Topological Order Without Quantum Memory}},}\ }\href@noop {} {\  (\bibinfo {year} {2023})},\ \Eprint {http://arxiv.org/abs/2307.13758} {arXiv:2307.13758 [quant-ph]} \BibitemShut {NoStop}%
\bibitem [{\citenamefont {Chen}\ and\ \citenamefont {Grover}(2023)}]{Chen:2023vxo}%
  \BibitemOpen
  \bibfield  {author} {\bibinfo {author} {\bibfnamefont {Yu-Hsueh}\ \bibnamefont {Chen}}\ and\ \bibinfo {author} {\bibfnamefont {Tarun}\ \bibnamefont {Grover}},\ }\bibfield  {title} {\enquote {\bibinfo {title} {{Separability transitions in topological states induced by local decoherence}},}\ }\href@noop {} {\  (\bibinfo {year} {2023})},\ \Eprint {http://arxiv.org/abs/2309.11879} {arXiv:2309.11879 [quant-ph]} \BibitemShut {NoStop}%
\bibitem [{\citenamefont {Li}\ and\ \citenamefont {Mong}(2024)}]{Li:2024rgz}%
  \BibitemOpen
  \bibfield  {author} {\bibinfo {author} {\bibfnamefont {Zhuan}\ \bibnamefont {Li}}\ and\ \bibinfo {author} {\bibfnamefont {Roger S.~K.}\ \bibnamefont {Mong}},\ }\bibfield  {title} {\enquote {\bibinfo {title} {{Replica topological order in quantum mixed states and quantum error correction}},}\ }\href@noop {} {\  (\bibinfo {year} {2024})},\ \Eprint {http://arxiv.org/abs/2402.09516} {arXiv:2402.09516 [quant-ph]} \BibitemShut {NoStop}%
\bibitem [{\citenamefont {Ellison}\ \emph {et~al.}(2023)\citenamefont {Ellison}, \citenamefont {Chen}, \citenamefont {Dua}, \citenamefont {Shirley}, \citenamefont {Tantivasadakarn},\ and\ \citenamefont {Williamson}}]{Ellison:2022web}%
  \BibitemOpen
  \bibfield  {author} {\bibinfo {author} {\bibfnamefont {Tyler~D.}\ \bibnamefont {Ellison}}, \bibinfo {author} {\bibfnamefont {Yu-An}\ \bibnamefont {Chen}}, \bibinfo {author} {\bibfnamefont {Arpit}\ \bibnamefont {Dua}}, \bibinfo {author} {\bibfnamefont {Wilbur}\ \bibnamefont {Shirley}}, \bibinfo {author} {\bibfnamefont {Nathanan}\ \bibnamefont {Tantivasadakarn}}, \ and\ \bibinfo {author} {\bibfnamefont {Dominic~J.}\ \bibnamefont {Williamson}},\ }\bibfield  {title} {\enquote {\bibinfo {title} {Pauli topological subsystem codes from {A}belian anyon theories},}\ }\href {\doibase 10.22331/q-2023-10-12-1137} {\bibfield  {journal} {\bibinfo  {journal} {{Quantum}}\ }\textbf {\bibinfo {volume} {7}},\ \bibinfo {pages} {1137} (\bibinfo {year} {2023})}\BibitemShut {NoStop}%
\bibitem [{\citenamefont {Lessa}\ \emph {et~al.}(2024{\natexlab{a}})\citenamefont {Lessa}, \citenamefont {Ma}, \citenamefont {Zhang}, \citenamefont {Bi}, \citenamefont {Cheng},\ and\ \citenamefont {Wang}}]{Lessa:2024gcw}%
  \BibitemOpen
  \bibfield  {author} {\bibinfo {author} {\bibfnamefont {Leonardo~A.}\ \bibnamefont {Lessa}}, \bibinfo {author} {\bibfnamefont {Ruochen}\ \bibnamefont {Ma}}, \bibinfo {author} {\bibfnamefont {Jian-Hao}\ \bibnamefont {Zhang}}, \bibinfo {author} {\bibfnamefont {Zhen}\ \bibnamefont {Bi}}, \bibinfo {author} {\bibfnamefont {Meng}\ \bibnamefont {Cheng}}, \ and\ \bibinfo {author} {\bibfnamefont {Chong}\ \bibnamefont {Wang}},\ }\bibfield  {title} {\enquote {\bibinfo {title} {{Strong-to-Weak Spontaneous Symmetry Breaking in Mixed Quantum States}},}\ }\href@noop {} {\  (\bibinfo {year} {2024}{\natexlab{a}})},\ \Eprint {http://arxiv.org/abs/2405.03639} {arXiv:2405.03639 [quant-ph]} \BibitemShut {NoStop}%
\bibitem [{\citenamefont {Cottrell}\ \emph {et~al.}(2019)\citenamefont {Cottrell}, \citenamefont {Freivogel}, \citenamefont {Hofman},\ and\ \citenamefont {Lokhande}}]{Cottrell:2018ash}%
  \BibitemOpen
  \bibfield  {author} {\bibinfo {author} {\bibfnamefont {William}\ \bibnamefont {Cottrell}}, \bibinfo {author} {\bibfnamefont {Ben}\ \bibnamefont {Freivogel}}, \bibinfo {author} {\bibfnamefont {Diego~M.}\ \bibnamefont {Hofman}}, \ and\ \bibinfo {author} {\bibfnamefont {Sagar~F.}\ \bibnamefont {Lokhande}},\ }\bibfield  {title} {\enquote {\bibinfo {title} {{How to Build the Thermofield Double State}},}\ }\href {\doibase 10.1007/JHEP02(2019)058} {\bibfield  {journal} {\bibinfo  {journal} {JHEP}\ }\textbf {\bibinfo {volume} {02}},\ \bibinfo {pages} {058} (\bibinfo {year} {2019})},\ \Eprint {http://arxiv.org/abs/1811.11528} {arXiv:1811.11528 [hep-th]} \BibitemShut {NoStop}%
\bibitem [{\citenamefont {Lucia}\ \emph {et~al.}(2023)\citenamefont {Lucia}, \citenamefont {P\'erez-Garc\'\i{}a},\ and\ \citenamefont {P\'erez-Hern\'andez}}]{Lucia:2021orn}%
  \BibitemOpen
  \bibfield  {author} {\bibinfo {author} {\bibfnamefont {Angelo}\ \bibnamefont {Lucia}}, \bibinfo {author} {\bibfnamefont {David}\ \bibnamefont {P\'erez-Garc\'\i{}a}}, \ and\ \bibinfo {author} {\bibfnamefont {Antonio}\ \bibnamefont {P\'erez-Hern\'andez}},\ }\bibfield  {title} {\enquote {\bibinfo {title} {{Thermalization in Kitaev\textquoteright{}s quantum double models via tensor network techniques}},}\ }\href {\doibase 10.1017/fms.2023.98} {\bibfield  {journal} {\bibinfo  {journal} {Forum Math. Sigma}\ }\textbf {\bibinfo {volume} {11}},\ \bibinfo {pages} {e107} (\bibinfo {year} {2023})},\ \Eprint {http://arxiv.org/abs/2107.01628} {arXiv:2107.01628 [quant-ph]} \BibitemShut {NoStop}%
\bibitem [{\citenamefont {Coser}\ and\ \citenamefont {P{\'{e}}rez-Garc{\'{i}}a}(2019)}]{Coser2019classificationof}%
  \BibitemOpen
  \bibfield  {author} {\bibinfo {author} {\bibfnamefont {Andrea}\ \bibnamefont {Coser}}\ and\ \bibinfo {author} {\bibfnamefont {David}\ \bibnamefont {P{\'{e}}rez-Garc{\'{i}}a}},\ }\bibfield  {title} {\enquote {\bibinfo {title} {Classification of phases for mixed states via fast dissipative evolution},}\ }\href {\doibase 10.22331/q-2019-08-12-174} {\bibfield  {journal} {\bibinfo  {journal} {{Quantum}}\ }\textbf {\bibinfo {volume} {3}},\ \bibinfo {pages} {174} (\bibinfo {year} {2019})}\BibitemShut {NoStop}%
\bibitem [{\citenamefont {Kato}\ and\ \citenamefont {Brand\~ao}(2019)}]{Kato:2016pgk}%
  \BibitemOpen
  \bibfield  {author} {\bibinfo {author} {\bibfnamefont {Kohtaro}\ \bibnamefont {Kato}}\ and\ \bibinfo {author} {\bibfnamefont {Fernando G. S.~L.}\ \bibnamefont {Brand\~ao}},\ }\bibfield  {title} {\enquote {\bibinfo {title} {{Quantum Approximate Markov Chains are Thermal}},}\ }\href {\doibase 10.1007/s00220-019-03485-6} {\bibfield  {journal} {\bibinfo  {journal} {Commun. Math. Phys.}\ }\textbf {\bibinfo {volume} {370}},\ \bibinfo {pages} {117--149} (\bibinfo {year} {2019})},\ \Eprint {http://arxiv.org/abs/1609.06636} {arXiv:1609.06636 [quant-ph]} \BibitemShut {NoStop}%
\bibitem [{\citenamefont {Sang}\ and\ \citenamefont {Hsieh}(2024)}]{Sang:2024vkl}%
  \BibitemOpen
  \bibfield  {author} {\bibinfo {author} {\bibfnamefont {Shengqi}\ \bibnamefont {Sang}}\ and\ \bibinfo {author} {\bibfnamefont {Timothy~H.}\ \bibnamefont {Hsieh}},\ }\bibfield  {title} {\enquote {\bibinfo {title} {{Stability of mixed-state quantum phases via finite Markov length}},}\ }\href@noop {} {\  (\bibinfo {year} {2024})},\ \Eprint {http://arxiv.org/abs/2404.07251} {arXiv:2404.07251 [quant-ph]} \BibitemShut {NoStop}%
\bibitem [{\citenamefont {Bauer}(2024)}]{Bauer:2024qpc}%
  \BibitemOpen
  \bibfield  {author} {\bibinfo {author} {\bibfnamefont {Andreas}\ \bibnamefont {Bauer}},\ }\bibfield  {title} {\enquote {\bibinfo {title} {{Low-overhead non-Clifford topological fault-tolerant circuits for all non-chiral abelian topological phases}},}\ }\href@noop {} {\  (\bibinfo {year} {2024})},\ \Eprint {http://arxiv.org/abs/2403.12119} {arXiv:2403.12119 [quant-ph]} \BibitemShut {NoStop}%
\bibitem [{\citenamefont {Nussinov}\ and\ \citenamefont {Ortiz}(2009{\natexlab{a}})}]{Nussinov2009thermalTQO}%
  \BibitemOpen
  \bibfield  {author} {\bibinfo {author} {\bibfnamefont {Zohar}\ \bibnamefont {Nussinov}}\ and\ \bibinfo {author} {\bibfnamefont {Gerardo}\ \bibnamefont {Ortiz}},\ }\bibfield  {title} {\enquote {\bibinfo {title} {A symmetry principle for topological quantum order},}\ }\href {\doibase https://doi.org/10.1016/j.aop.2008.11.002} {\bibfield  {journal} {\bibinfo  {journal} {Annals of Physics}\ }\textbf {\bibinfo {volume} {324}},\ \bibinfo {pages} {977--1057} (\bibinfo {year} {2009}{\natexlab{a}})}\BibitemShut {NoStop}%
\bibitem [{\citenamefont {Castelnovo}\ and\ \citenamefont {Chamon}(2007{\natexlab{b}})}]{Castelnovo2007}%
  \BibitemOpen
  \bibfield  {author} {\bibinfo {author} {\bibfnamefont {Claudio}\ \bibnamefont {Castelnovo}}\ and\ \bibinfo {author} {\bibfnamefont {Claudio}\ \bibnamefont {Chamon}},\ }\bibfield  {title} {\enquote {\bibinfo {title} {Topological order and topological entropy in classical systems},}\ }\href {\doibase 10.1103/PhysRevB.76.174416} {\bibfield  {journal} {\bibinfo  {journal} {Phys. Rev. B}\ }\textbf {\bibinfo {volume} {76}},\ \bibinfo {pages} {174416} (\bibinfo {year} {2007}{\natexlab{b}})}\BibitemShut {NoStop}%
\bibitem [{\citenamefont {Bravyi}\ \emph {et~al.}(2010)\citenamefont {Bravyi}, \citenamefont {Hastings},\ and\ \citenamefont {Michalakis}}]{Bravyi:2010ida}%
  \BibitemOpen
  \bibfield  {author} {\bibinfo {author} {\bibfnamefont {Sergey}\ \bibnamefont {Bravyi}}, \bibinfo {author} {\bibfnamefont {Matthew~B.}\ \bibnamefont {Hastings}}, \ and\ \bibinfo {author} {\bibfnamefont {Spyridon}\ \bibnamefont {Michalakis}},\ }\bibfield  {title} {\enquote {\bibinfo {title} {{Topological quantum order: Stability under local perturbations}},}\ }\href {\doibase 10.1063/1.3490195} {\bibfield  {journal} {\bibinfo  {journal} {J. Math. Phys.}\ }\textbf {\bibinfo {volume} {51}},\ \bibinfo {pages} {093512} (\bibinfo {year} {2010})},\ \Eprint {http://arxiv.org/abs/1001.0344} {arXiv:1001.0344 [quant-ph]} \BibitemShut {NoStop}%
\bibitem [{\citenamefont {Haah}(2021)}]{Haah2021classification}%
  \BibitemOpen
  \bibfield  {author} {\bibinfo {author} {\bibfnamefont {Jeongwan}\ \bibnamefont {Haah}},\ }\bibfield  {title} {\enquote {\bibinfo {title} {{Classification of translation invariant topological Pauli stabilizer codes for prime dimensional qudits on two-dimensional lattices}},}\ }\href {\doibase 10.1063/5.0021068} {\bibfield  {journal} {\bibinfo  {journal} {Journal of Mathematical Physics}\ }\textbf {\bibinfo {volume} {62}},\ \bibinfo {pages} {012201} (\bibinfo {year} {2021})},\ \Eprint {http://arxiv.org/abs/1812.11193} {arXiv:1812.11193 [quant-ph]} \BibitemShut {NoStop}%
\bibitem [{\citenamefont {Poulin}(2005)}]{PoulinPRL2005}%
  \BibitemOpen
  \bibfield  {author} {\bibinfo {author} {\bibfnamefont {David}\ \bibnamefont {Poulin}},\ }\bibfield  {title} {\enquote {\bibinfo {title} {Stabilizer formalism for operator quantum error correction},}\ }\href {\doibase 10.1103/PhysRevLett.95.230504} {\bibfield  {journal} {\bibinfo  {journal} {Phys. Rev. Lett.}\ }\textbf {\bibinfo {volume} {95}},\ \bibinfo {pages} {230504} (\bibinfo {year} {2005})}\BibitemShut {NoStop}%
\bibitem [{\citenamefont {Bombin}\ \emph {et~al.}(2009)\citenamefont {Bombin}, \citenamefont {Kargarian},\ and\ \citenamefont {Martin-Delgado}}]{BombinPRB2009}%
  \BibitemOpen
  \bibfield  {author} {\bibinfo {author} {\bibfnamefont {H.}~\bibnamefont {Bombin}}, \bibinfo {author} {\bibfnamefont {M.}~\bibnamefont {Kargarian}}, \ and\ \bibinfo {author} {\bibfnamefont {M.~A.}\ \bibnamefont {Martin-Delgado}},\ }\bibfield  {title} {\enquote {\bibinfo {title} {Interacting anyonic fermions in a two-body color code model},}\ }\href {\doibase 10.1103/PhysRevB.80.075111} {\bibfield  {journal} {\bibinfo  {journal} {Phys. Rev. B}\ }\textbf {\bibinfo {volume} {80}},\ \bibinfo {pages} {075111} (\bibinfo {year} {2009})}\BibitemShut {NoStop}%
\bibitem [{\citenamefont {Bombin}(2010)}]{BombinTSC}%
  \BibitemOpen
  \bibfield  {author} {\bibinfo {author} {\bibfnamefont {H.}~\bibnamefont {Bombin}},\ }\bibfield  {title} {\enquote {\bibinfo {title} {Topological subsystem codes},}\ }\href {\doibase 10.1103/PhysRevA.81.032301} {\bibfield  {journal} {\bibinfo  {journal} {Phys. Rev. A}\ }\textbf {\bibinfo {volume} {81}},\ \bibinfo {pages} {032301} (\bibinfo {year} {2010})}\BibitemShut {NoStop}%
\bibitem [{\citenamefont {Knill}\ \emph {et~al.}(2000)\citenamefont {Knill}, \citenamefont {Laflamme},\ and\ \citenamefont {Viola}}]{Knill2000noiseless}%
  \BibitemOpen
  \bibfield  {author} {\bibinfo {author} {\bibfnamefont {Emanuel}\ \bibnamefont {Knill}}, \bibinfo {author} {\bibfnamefont {Raymond}\ \bibnamefont {Laflamme}}, \ and\ \bibinfo {author} {\bibfnamefont {Lorenza}\ \bibnamefont {Viola}},\ }\bibfield  {title} {\enquote {\bibinfo {title} {Theory of quantum error correction for general noise},}\ }\href {\doibase 10.1103/PhysRevLett.84.2525} {\bibfield  {journal} {\bibinfo  {journal} {Phys. Rev. Lett.}\ }\textbf {\bibinfo {volume} {84}},\ \bibinfo {pages} {2525--2528} (\bibinfo {year} {2000})}\BibitemShut {NoStop}%
\bibitem [{\citenamefont {Kribs}\ \emph {et~al.}(2006)\citenamefont {Kribs}, \citenamefont {Laflamme}, \citenamefont {Poulin},\ and\ \citenamefont {Lesosky}}]{Kribs2006operator}%
  \BibitemOpen
  \bibfield  {author} {\bibinfo {author} {\bibfnamefont {David~W.}\ \bibnamefont {Kribs}}, \bibinfo {author} {\bibfnamefont {Raymond}\ \bibnamefont {Laflamme}}, \bibinfo {author} {\bibfnamefont {David}\ \bibnamefont {Poulin}}, \ and\ \bibinfo {author} {\bibfnamefont {Maia}\ \bibnamefont {Lesosky}},\ }\bibfield  {title} {\enquote {\bibinfo {title} {Operator quantum error correction},}\ }\href {\doibase 10.5555/2012086.2012092} {\bibfield  {journal} {\bibinfo  {journal} {Quantum Info. Comput.}\ }\textbf {\bibinfo {volume} {6}},\ \bibinfo {pages} {382–399} (\bibinfo {year} {2006})}\BibitemShut {NoStop}%
\bibitem [{\citenamefont {Zanardi}\ \emph {et~al.}(2004)\citenamefont {Zanardi}, \citenamefont {Lidar},\ and\ \citenamefont {Lloyd}}]{Zanardi2004Tensor}%
  \BibitemOpen
  \bibfield  {author} {\bibinfo {author} {\bibfnamefont {Paolo}\ \bibnamefont {Zanardi}}, \bibinfo {author} {\bibfnamefont {Daniel~A.}\ \bibnamefont {Lidar}}, \ and\ \bibinfo {author} {\bibfnamefont {Seth}\ \bibnamefont {Lloyd}},\ }\bibfield  {title} {\enquote {\bibinfo {title} {Quantum tensor product structures are observable induced},}\ }\href {\doibase 10.1103/PhysRevLett.92.060402} {\bibfield  {journal} {\bibinfo  {journal} {Phys. Rev. Lett.}\ }\textbf {\bibinfo {volume} {92}},\ \bibinfo {pages} {060402} (\bibinfo {year} {2004})}\BibitemShut {NoStop}%
\bibitem [{\citenamefont {Fattal}\ \emph {et~al.}(2004)\citenamefont {Fattal}, \citenamefont {Cubitt}, \citenamefont {Yamamoto}, \citenamefont {Bravyi},\ and\ \citenamefont {Chuang}}]{fattal2004entanglement}%
  \BibitemOpen
  \bibfield  {author} {\bibinfo {author} {\bibfnamefont {David}\ \bibnamefont {Fattal}}, \bibinfo {author} {\bibfnamefont {Toby~S.}\ \bibnamefont {Cubitt}}, \bibinfo {author} {\bibfnamefont {Yoshihisa}\ \bibnamefont {Yamamoto}}, \bibinfo {author} {\bibfnamefont {Sergey}\ \bibnamefont {Bravyi}}, \ and\ \bibinfo {author} {\bibfnamefont {Isaac~L.}\ \bibnamefont {Chuang}},\ }\href@noop {} {\enquote {\bibinfo {title} {Entanglement in the stabilizer formalism},}\ } (\bibinfo {year} {2004}),\ \Eprint {http://arxiv.org/abs/quant-ph/0406168} {arXiv:quant-ph/0406168 [quant-ph]} \BibitemShut {NoStop}%
\bibitem [{\citenamefont {Bombin}\ \emph {et~al.}(2012)\citenamefont {Bombin}, \citenamefont {Duclos-Cianci},\ and\ \citenamefont {Poulin}}]{BombinPoulin2012Universal}%
  \BibitemOpen
  \bibfield  {author} {\bibinfo {author} {\bibfnamefont {H}~\bibnamefont {Bombin}}, \bibinfo {author} {\bibfnamefont {Guillaume}\ \bibnamefont {Duclos-Cianci}}, \ and\ \bibinfo {author} {\bibfnamefont {David}\ \bibnamefont {Poulin}},\ }\bibfield  {title} {\enquote {\bibinfo {title} {Universal topological phase of two-dimensional stabilizer codes},}\ }\href {\doibase 10.1088/1367-2630/14/7/073048} {\bibfield  {journal} {\bibinfo  {journal} {New Journal of Physics}\ }\textbf {\bibinfo {volume} {14}},\ \bibinfo {pages} {073048} (\bibinfo {year} {2012})}\BibitemShut {NoStop}%
\bibitem [{\citenamefont {Levin}\ and\ \citenamefont {Wen}(2003)}]{Levin2003fermion}%
  \BibitemOpen
  \bibfield  {author} {\bibinfo {author} {\bibfnamefont {Michael}\ \bibnamefont {Levin}}\ and\ \bibinfo {author} {\bibfnamefont {Xiao-Gang}\ \bibnamefont {Wen}},\ }\bibfield  {title} {\enquote {\bibinfo {title} {Fermions, strings, and gauge fields in lattice spin models},}\ }\href {\doibase 10.1103/PhysRevB.67.245316} {\bibfield  {journal} {\bibinfo  {journal} {Phys. Rev. B}\ }\textbf {\bibinfo {volume} {67}},\ \bibinfo {pages} {245316} (\bibinfo {year} {2003})}\BibitemShut {NoStop}%
\bibitem [{\citenamefont {Kawagoe}\ and\ \citenamefont {Levin}(2020)}]{Kawagoe2020}%
  \BibitemOpen
  \bibfield  {author} {\bibinfo {author} {\bibfnamefont {Kyle}\ \bibnamefont {Kawagoe}}\ and\ \bibinfo {author} {\bibfnamefont {Michael}\ \bibnamefont {Levin}},\ }\bibfield  {title} {\enquote {\bibinfo {title} {Microscopic definitions of anyon data},}\ }\href {\doibase 10.1103/PhysRevB.101.115113} {\bibfield  {journal} {\bibinfo  {journal} {Phys. Rev. B}\ }\textbf {\bibinfo {volume} {101}},\ \bibinfo {pages} {115113} (\bibinfo {year} {2020})}\BibitemShut {NoStop}%
\bibitem [{\citenamefont {Bonderson}(2012)}]{Bonderson2012}%
  \BibitemOpen
  \bibfield  {author} {\bibinfo {author} {\bibfnamefont {Parsa~H.}\ \bibnamefont {Bonderson}},\ }\emph {\bibinfo {title} {Non-Abelian Anyons and Interferometry}},\ \href {\doibase 10.7907/5NDZ-W890} {Ph.D. thesis},\ \bibinfo  {school} {Caltech} (\bibinfo {year} {2012})\BibitemShut {NoStop}%
\bibitem [{\citenamefont {Chen}\ \emph {et~al.}(2018)\citenamefont {Chen}, \citenamefont {Kapustin},\ and\ \citenamefont {Radi\v{c}evi\'c}}]{Chen:2017fvr}%
  \BibitemOpen
  \bibfield  {author} {\bibinfo {author} {\bibfnamefont {Yu-An}\ \bibnamefont {Chen}}, \bibinfo {author} {\bibfnamefont {Anton}\ \bibnamefont {Kapustin}}, \ and\ \bibinfo {author} {\bibfnamefont {\DJ{}or\dj{}e}\ \bibnamefont {Radi\v{c}evi\'c}},\ }\bibfield  {title} {\enquote {\bibinfo {title} {{Exact bosonization in two spatial dimensions and a new class of lattice gauge theories}},}\ }\href {\doibase 10.1016/j.aop.2018.03.024} {\bibfield  {journal} {\bibinfo  {journal} {Annals Phys.}\ }\textbf {\bibinfo {volume} {393}},\ \bibinfo {pages} {234--253} (\bibinfo {year} {2018})},\ \Eprint {http://arxiv.org/abs/1711.00515} {arXiv:1711.00515 [cond-mat.str-el]} \BibitemShut {NoStop}%
\bibitem [{\citenamefont {Shi}\ \emph {et~al.}(2020)\citenamefont {Shi}, \citenamefont {Kato},\ and\ \citenamefont {Kim}}]{Shi2020fusion}%
  \BibitemOpen
  \bibfield  {author} {\bibinfo {author} {\bibfnamefont {Bowen}\ \bibnamefont {Shi}}, \bibinfo {author} {\bibfnamefont {Kohtaro}\ \bibnamefont {Kato}}, \ and\ \bibinfo {author} {\bibfnamefont {Isaac~H.}\ \bibnamefont {Kim}},\ }\bibfield  {title} {\enquote {\bibinfo {title} {Fusion rules from entanglement},}\ }\href {\doibase https://doi.org/10.1016/j.aop.2020.168164} {\bibfield  {journal} {\bibinfo  {journal} {Annals of Physics}\ }\textbf {\bibinfo {volume} {418}},\ \bibinfo {pages} {168164} (\bibinfo {year} {2020})}\BibitemShut {NoStop}%
\bibitem [{\citenamefont {Ellison}\ \emph {et~al.}(2022)\citenamefont {Ellison}, \citenamefont {Chen}, \citenamefont {Dua}, \citenamefont {Shirley}, \citenamefont {Tantivasadakarn},\ and\ \citenamefont {Williamson}}]{Ellison2021}%
  \BibitemOpen
  \bibfield  {author} {\bibinfo {author} {\bibfnamefont {Tyler~D.}\ \bibnamefont {Ellison}}, \bibinfo {author} {\bibfnamefont {Yu-An}\ \bibnamefont {Chen}}, \bibinfo {author} {\bibfnamefont {Arpit}\ \bibnamefont {Dua}}, \bibinfo {author} {\bibfnamefont {Wilbur}\ \bibnamefont {Shirley}}, \bibinfo {author} {\bibfnamefont {Nathanan}\ \bibnamefont {Tantivasadakarn}}, \ and\ \bibinfo {author} {\bibfnamefont {Dominic~J.}\ \bibnamefont {Williamson}},\ }\bibfield  {title} {\enquote {\bibinfo {title} {Pauli stabilizer models of twisted quantum doubles},}\ }\href {\doibase 10.1103/PRXQuantum.3.010353} {\bibfield  {journal} {\bibinfo  {journal} {PRX Quantum}\ }\textbf {\bibinfo {volume} {3}},\ \bibinfo {pages} {010353} (\bibinfo {year} {2022})}\BibitemShut {NoStop}%
\bibitem [{\citenamefont {Nussinov}\ and\ \citenamefont {Ortiz}(2009{\natexlab{b}})}]{Nussinov2009gaugelike}%
  \BibitemOpen
  \bibfield  {author} {\bibinfo {author} {\bibfnamefont {Zohar}\ \bibnamefont {Nussinov}}\ and\ \bibinfo {author} {\bibfnamefont {Gerardo}\ \bibnamefont {Ortiz}},\ }\bibfield  {title} {\enquote {\bibinfo {title} {A symmetry principle for topological quantum order},}\ }\href {\doibase https://doi.org/10.1016/j.aop.2008.11.002} {\bibfield  {journal} {\bibinfo  {journal} {Annals of Physics}\ }\textbf {\bibinfo {volume} {324}},\ \bibinfo {pages} {977--1057} (\bibinfo {year} {2009}{\natexlab{b}})}\BibitemShut {NoStop}%
\bibitem [{\citenamefont {Gaiotto}\ \emph {et~al.}(2015)\citenamefont {Gaiotto}, \citenamefont {Kapustin}, \citenamefont {Seiberg},\ and\ \citenamefont {Willett}}]{Gaiotto2015}%
  \BibitemOpen
  \bibfield  {author} {\bibinfo {author} {\bibfnamefont {Davide}\ \bibnamefont {Gaiotto}}, \bibinfo {author} {\bibfnamefont {Anton}\ \bibnamefont {Kapustin}}, \bibinfo {author} {\bibfnamefont {Nathan}\ \bibnamefont {Seiberg}}, \ and\ \bibinfo {author} {\bibfnamefont {Brian}\ \bibnamefont {Willett}},\ }\bibfield  {title} {\enquote {\bibinfo {title} {Generalized global symmetries},}\ }\href {\doibase 10.1007/jhep02(2015)172} {\bibfield  {journal} {\bibinfo  {journal} {J. High Energ. Phys.}\ }\textbf {\bibinfo {volume} {2015}} (\bibinfo {year} {2015}),\ 10.1007/jhep02(2015)172}\BibitemShut {NoStop}%
\bibitem [{\citenamefont {Qi}\ \emph {et~al.}(2021)\citenamefont {Qi}, \citenamefont {Radzihovsky},\ and\ \citenamefont {Hermele}}]{Qi2021higherform}%
  \BibitemOpen
  \bibfield  {author} {\bibinfo {author} {\bibfnamefont {Marvin}\ \bibnamefont {Qi}}, \bibinfo {author} {\bibfnamefont {Leo}\ \bibnamefont {Radzihovsky}}, \ and\ \bibinfo {author} {\bibfnamefont {Michael}\ \bibnamefont {Hermele}},\ }\bibfield  {title} {\enquote {\bibinfo {title} {Fracton phases via exotic higher-form symmetry-breaking},}\ }\href {\doibase https://doi.org/10.1016/j.aop.2020.168360} {\bibfield  {journal} {\bibinfo  {journal} {Annals of Physics}\ }\textbf {\bibinfo {volume} {424}},\ \bibinfo {pages} {168360} (\bibinfo {year} {2021})}\BibitemShut {NoStop}%
\bibitem [{\citenamefont {{Bu{\v{c}}a}}\ and\ \citenamefont {{Prosen}}(2012)}]{BucaProsen2012}%
  \BibitemOpen
  \bibfield  {author} {\bibinfo {author} {\bibfnamefont {Berislav}\ \bibnamefont {{Bu{\v{c}}a}}}\ and\ \bibinfo {author} {\bibfnamefont {Toma{\v{z}}}\ \bibnamefont {{Prosen}}},\ }\bibfield  {title} {\enquote {\bibinfo {title} {{A note on symmetry reductions of the Lindblad equation: transport in constrained open spin chains}},}\ }\href {\doibase 10.1088/1367-2630/14/7/073007} {\bibfield  {journal} {\bibinfo  {journal} {New Journal of Physics}\ }\textbf {\bibinfo {volume} {14}},\ \bibinfo {eid} {073007} (\bibinfo {year} {2012})},\ \Eprint {http://arxiv.org/abs/1203.0943} {arXiv:1203.0943 [quant-ph]} \BibitemShut {NoStop}%
\bibitem [{\citenamefont {Albert}\ and\ \citenamefont {Jiang}(2014)}]{AlbertJiang2014}%
  \BibitemOpen
  \bibfield  {author} {\bibinfo {author} {\bibfnamefont {Victor~V.}\ \bibnamefont {Albert}}\ and\ \bibinfo {author} {\bibfnamefont {Liang}\ \bibnamefont {Jiang}},\ }\bibfield  {title} {\enquote {\bibinfo {title} {Symmetries and conserved quantities in lindblad master equations},}\ }\href {\doibase 10.1103/PhysRevA.89.022118} {\bibfield  {journal} {\bibinfo  {journal} {Phys. Rev. A}\ }\textbf {\bibinfo {volume} {89}},\ \bibinfo {pages} {022118} (\bibinfo {year} {2014})}\BibitemShut {NoStop}%
\bibitem [{\citenamefont {{Albert}}(2018)}]{Albert2018}%
  \BibitemOpen
  \bibfield  {author} {\bibinfo {author} {\bibfnamefont {Victor~V.}\ \bibnamefont {{Albert}}},\ }\bibfield  {title} {\enquote {\bibinfo {title} {{Lindbladians with multiple steady states: theory and applications}},}\ }\href {\doibase 10.48550/arXiv.1802.00010} {\bibfield  {journal} {\bibinfo  {journal} {arXiv e-prints}\ ,\ \bibinfo {eid} {arXiv:1802.00010}} (\bibinfo {year} {2018})},\ \Eprint {http://arxiv.org/abs/1802.00010} {arXiv:1802.00010 [quant-ph]} \BibitemShut {NoStop}%
\bibitem [{\citenamefont {Lieu}\ \emph {et~al.}(2020)\citenamefont {Lieu}, \citenamefont {Belyansky}, \citenamefont {Young}, \citenamefont {Lundgren}, \citenamefont {Albert},\ and\ \citenamefont {Gorshkov}}]{Lieu2020}%
  \BibitemOpen
  \bibfield  {author} {\bibinfo {author} {\bibfnamefont {Simon}\ \bibnamefont {Lieu}}, \bibinfo {author} {\bibfnamefont {Ron}\ \bibnamefont {Belyansky}}, \bibinfo {author} {\bibfnamefont {Jeremy~T.}\ \bibnamefont {Young}}, \bibinfo {author} {\bibfnamefont {Rex}\ \bibnamefont {Lundgren}}, \bibinfo {author} {\bibfnamefont {Victor~V.}\ \bibnamefont {Albert}}, \ and\ \bibinfo {author} {\bibfnamefont {Alexey~V.}\ \bibnamefont {Gorshkov}},\ }\bibfield  {title} {\enquote {\bibinfo {title} {Symmetry breaking and error correction in open quantum systems},}\ }\href {\doibase 10.1103/PhysRevLett.125.240405} {\bibfield  {journal} {\bibinfo  {journal} {Phys. Rev. Lett.}\ }\textbf {\bibinfo {volume} {125}},\ \bibinfo {pages} {240405} (\bibinfo {year} {2020})}\BibitemShut {NoStop}%
\bibitem [{\citenamefont {Hsin}\ \emph {et~al.}(2023)\citenamefont {Hsin}, \citenamefont {Luo},\ and\ \citenamefont {Sun}}]{Hsin2023anomalies}%
  \BibitemOpen
  \bibfield  {author} {\bibinfo {author} {\bibfnamefont {Po-Shen}\ \bibnamefont {Hsin}}, \bibinfo {author} {\bibfnamefont {Zhu-Xi}\ \bibnamefont {Luo}}, \ and\ \bibinfo {author} {\bibfnamefont {Hao-Yu}\ \bibnamefont {Sun}},\ }\href@noop {} {\enquote {\bibinfo {title} {Anomalies of average symmetries: Entanglement and open quantum systems},}\ } (\bibinfo {year} {2023}),\ \Eprint {http://arxiv.org/abs/2312.09074} {arXiv:2312.09074 [cond-mat.str-el]} \BibitemShut {NoStop}%
\bibitem [{\citenamefont {Hastings}\ and\ \citenamefont {Wen}(2005)}]{Hastings:2005xm}%
  \BibitemOpen
  \bibfield  {author} {\bibinfo {author} {\bibfnamefont {M.~B.}\ \bibnamefont {Hastings}}\ and\ \bibinfo {author} {\bibfnamefont {Xiao-Gang}\ \bibnamefont {Wen}},\ }\bibfield  {title} {\enquote {\bibinfo {title} {{Quasi-adiabatic continuation of quantum states: The Stability of topological ground state degeneracy and emergent gauge invariance}},}\ }\href {\doibase 10.1103/PhysRevB.72.045141} {\bibfield  {journal} {\bibinfo  {journal} {Phys. Rev. B}\ }\textbf {\bibinfo {volume} {72}},\ \bibinfo {pages} {045141} (\bibinfo {year} {2005})},\ \Eprint {http://arxiv.org/abs/cond-mat/0503554} {arXiv:cond-mat/0503554} \BibitemShut {NoStop}%
\bibitem [{\citenamefont {Pace}\ and\ \citenamefont {Wen}(2023)}]{Pace2023}%
  \BibitemOpen
  \bibfield  {author} {\bibinfo {author} {\bibfnamefont {Salvatore~D.}\ \bibnamefont {Pace}}\ and\ \bibinfo {author} {\bibfnamefont {Xiao-Gang}\ \bibnamefont {Wen}},\ }\bibfield  {title} {\enquote {\bibinfo {title} {Exact emergent higher-form symmetries in bosonic lattice models},}\ }\href {\doibase 10.1103/PhysRevB.108.195147} {\bibfield  {journal} {\bibinfo  {journal} {Phys. Rev. B}\ }\textbf {\bibinfo {volume} {108}},\ \bibinfo {pages} {195147} (\bibinfo {year} {2023})}\BibitemShut {NoStop}%
\bibitem [{\citenamefont {{Kong}}\ \emph {et~al.}(2020)\citenamefont {{Kong}}, \citenamefont {{Lan}}, \citenamefont {{Wen}}, \citenamefont {{Zhang}},\ and\ \citenamefont {{Zheng}}}]{KongHolographicSym}%
  \BibitemOpen
  \bibfield  {author} {\bibinfo {author} {\bibfnamefont {Liang}\ \bibnamefont {{Kong}}}, \bibinfo {author} {\bibfnamefont {Tian}\ \bibnamefont {{Lan}}}, \bibinfo {author} {\bibfnamefont {Xiao-Gang}\ \bibnamefont {{Wen}}}, \bibinfo {author} {\bibfnamefont {Zhi-Hao}\ \bibnamefont {{Zhang}}}, \ and\ \bibinfo {author} {\bibfnamefont {Hao}\ \bibnamefont {{Zheng}}},\ }\bibfield  {title} {\enquote {\bibinfo {title} {{Algebraic higher symmetry and categorical symmetry: A holographic and entanglement view of symmetry}},}\ }\href {\doibase 10.1103/PhysRevResearch.2.043086} {\bibfield  {journal} {\bibinfo  {journal} {Physical Review Research}\ }\textbf {\bibinfo {volume} {2}},\ \bibinfo {eid} {043086} (\bibinfo {year} {2020})},\ \Eprint {http://arxiv.org/abs/2005.14178} {arXiv:2005.14178} \BibitemShut {NoStop}%
\bibitem [{\citenamefont {Cian}\ \emph {et~al.}(2022)\citenamefont {Cian}, \citenamefont {Hafezi},\ and\ \citenamefont {Barkeshli}}]{Cian:2022vjb}%
  \BibitemOpen
  \bibfield  {author} {\bibinfo {author} {\bibfnamefont {Ze-Pei}\ \bibnamefont {Cian}}, \bibinfo {author} {\bibfnamefont {Mohammad}\ \bibnamefont {Hafezi}}, \ and\ \bibinfo {author} {\bibfnamefont {Maissam}\ \bibnamefont {Barkeshli}},\ }\bibfield  {title} {\enquote {\bibinfo {title} {{Extracting Wilson loop operators and fractional statistics from a single bulk ground state}},}\ }\href@noop {} {\  (\bibinfo {year} {2022})},\ \Eprint {http://arxiv.org/abs/2209.14302} {arXiv:2209.14302 [cond-mat.str-el]} \BibitemShut {NoStop}%
\bibitem [{\citenamefont {Hsin}\ \emph {et~al.}(2019)\citenamefont {Hsin}, \citenamefont {Lam},\ and\ \citenamefont {Seiberg}}]{Hsin:2018vcg}%
  \BibitemOpen
  \bibfield  {author} {\bibinfo {author} {\bibfnamefont {Po-Shen}\ \bibnamefont {Hsin}}, \bibinfo {author} {\bibfnamefont {Ho~Tat}\ \bibnamefont {Lam}}, \ and\ \bibinfo {author} {\bibfnamefont {Nathan}\ \bibnamefont {Seiberg}},\ }\bibfield  {title} {\enquote {\bibinfo {title} {{Comments on One-Form Global Symmetries and Their Gauging in 3d and 4d}},}\ }\href {\doibase 10.21468/SciPostPhys.6.3.039} {\bibfield  {journal} {\bibinfo  {journal} {SciPost Phys.}\ }\textbf {\bibinfo {volume} {6}},\ \bibinfo {pages} {039} (\bibinfo {year} {2019})},\ \Eprint {http://arxiv.org/abs/1812.04716} {arXiv:1812.04716 [hep-th]} \BibitemShut {NoStop}%
\bibitem [{\citenamefont {Bomb{\'i}n}(2014)}]{Bombin2014structure}%
  \BibitemOpen
  \bibfield  {author} {\bibinfo {author} {\bibfnamefont {H{\'e}ctor}\ \bibnamefont {Bomb{\'i}n}},\ }\bibfield  {title} {\enquote {\bibinfo {title} {Structure of 2d topological stabilizer codes},}\ }\href {\doibase 10.1007/s00220-014-1893-4} {\bibfield  {journal} {\bibinfo  {journal} {Communications in Mathematical Physics}\ }\textbf {\bibinfo {volume} {327}},\ \bibinfo {pages} {387--432} (\bibinfo {year} {2014})}\BibitemShut {NoStop}%
\bibitem [{\citenamefont {Levin}\ and\ \citenamefont {Wen}(2005)}]{Levin:2004mi}%
  \BibitemOpen
  \bibfield  {author} {\bibinfo {author} {\bibfnamefont {Michael~A.}\ \bibnamefont {Levin}}\ and\ \bibinfo {author} {\bibfnamefont {Xiao-Gang}\ \bibnamefont {Wen}},\ }\bibfield  {title} {\enquote {\bibinfo {title} {{String net condensation: A Physical mechanism for topological phases}},}\ }\href {\doibase 10.1103/PhysRevB.71.045110} {\bibfield  {journal} {\bibinfo  {journal} {Phys. Rev. B}\ }\textbf {\bibinfo {volume} {71}},\ \bibinfo {pages} {045110} (\bibinfo {year} {2005})},\ \Eprint {http://arxiv.org/abs/cond-mat/0404617} {arXiv:cond-mat/0404617} \BibitemShut {NoStop}%
\bibitem [{\citenamefont {Lin}\ \emph {et~al.}(2021)\citenamefont {Lin}, \citenamefont {Levin},\ and\ \citenamefont {Burnell}}]{Lin:2020bak}%
  \BibitemOpen
  \bibfield  {author} {\bibinfo {author} {\bibfnamefont {Chien-Hung}\ \bibnamefont {Lin}}, \bibinfo {author} {\bibfnamefont {Michael}\ \bibnamefont {Levin}}, \ and\ \bibinfo {author} {\bibfnamefont {Fiona~J.}\ \bibnamefont {Burnell}},\ }\bibfield  {title} {\enquote {\bibinfo {title} {{Generalized string-net models: A thorough exposition}},}\ }\href {\doibase 10.1103/PhysRevB.103.195155} {\bibfield  {journal} {\bibinfo  {journal} {Phys. Rev. B}\ }\textbf {\bibinfo {volume} {103}},\ \bibinfo {pages} {195155} (\bibinfo {year} {2021})},\ \Eprint {http://arxiv.org/abs/2012.14424} {arXiv:2012.14424 [cond-mat.str-el]} \BibitemShut {NoStop}%
\bibitem [{\citenamefont {Heinrich}\ \emph {et~al.}(2016)\citenamefont {Heinrich}, \citenamefont {Burnell}, \citenamefont {Fidkowski},\ and\ \citenamefont {Levin}}]{HeinrichPRB2016}%
  \BibitemOpen
  \bibfield  {author} {\bibinfo {author} {\bibfnamefont {Chris}\ \bibnamefont {Heinrich}}, \bibinfo {author} {\bibfnamefont {Fiona}\ \bibnamefont {Burnell}}, \bibinfo {author} {\bibfnamefont {Lukasz}\ \bibnamefont {Fidkowski}}, \ and\ \bibinfo {author} {\bibfnamefont {Michael}\ \bibnamefont {Levin}},\ }\bibfield  {title} {\enquote {\bibinfo {title} {Symmetry-enriched string nets: Exactly solvable models for set phases},}\ }\href {\doibase 10.1103/PhysRevB.94.235136} {\bibfield  {journal} {\bibinfo  {journal} {Phys. Rev. B}\ }\textbf {\bibinfo {volume} {94}},\ \bibinfo {pages} {235136} (\bibinfo {year} {2016})}\BibitemShut {NoStop}%
\bibitem [{\citenamefont {Cheng}\ \emph {et~al.}(2017)\citenamefont {Cheng}, \citenamefont {Gu}, \citenamefont {Jiang},\ and\ \citenamefont {Qi}}]{ChengSET2017}%
  \BibitemOpen
  \bibfield  {author} {\bibinfo {author} {\bibfnamefont {Meng}\ \bibnamefont {Cheng}}, \bibinfo {author} {\bibfnamefont {Zheng-Cheng}\ \bibnamefont {Gu}}, \bibinfo {author} {\bibfnamefont {Shenghan}\ \bibnamefont {Jiang}}, \ and\ \bibinfo {author} {\bibfnamefont {Yang}\ \bibnamefont {Qi}},\ }\bibfield  {title} {\enquote {\bibinfo {title} {Exactly solvable models for symmetry-enriched topological phases},}\ }\href {\doibase 10.1103/PhysRevB.96.115107} {\bibfield  {journal} {\bibinfo  {journal} {Phys. Rev. B}\ }\textbf {\bibinfo {volume} {96}},\ \bibinfo {pages} {115107} (\bibinfo {year} {2017})}\BibitemShut {NoStop}%
\bibitem [{\citenamefont {Shao}(2023)}]{Shao:2023gho}%
  \BibitemOpen
  \bibfield  {author} {\bibinfo {author} {\bibfnamefont {Shu-Heng}\ \bibnamefont {Shao}},\ }\bibfield  {title} {\enquote {\bibinfo {title} {{What's Done Cannot Be Undone: TASI Lectures on Non-Invertible Symmetries}},}\ }\href@noop {} {\  (\bibinfo {year} {2023})},\ \Eprint {http://arxiv.org/abs/2308.00747} {arXiv:2308.00747 [hep-th]} \BibitemShut {NoStop}%
\bibitem [{\citenamefont {Zhang}\ \emph {et~al.}(2012)\citenamefont {Zhang}, \citenamefont {Grover}, \citenamefont {Turner}, \citenamefont {Oshikawa},\ and\ \citenamefont {Vishwanath}}]{Zhang:2011jd}%
  \BibitemOpen
  \bibfield  {author} {\bibinfo {author} {\bibfnamefont {Yi}~\bibnamefont {Zhang}}, \bibinfo {author} {\bibfnamefont {Tarun}\ \bibnamefont {Grover}}, \bibinfo {author} {\bibfnamefont {Ari}\ \bibnamefont {Turner}}, \bibinfo {author} {\bibfnamefont {Masaki}\ \bibnamefont {Oshikawa}}, \ and\ \bibinfo {author} {\bibfnamefont {Ashvin}\ \bibnamefont {Vishwanath}},\ }\bibfield  {title} {\enquote {\bibinfo {title} {Quasiparticle statistics and braiding from ground-state entanglement},}\ }\href {\doibase 10.1103/PhysRevB.85.235151} {\bibfield  {journal} {\bibinfo  {journal} {Phys. Rev. B}\ }\textbf {\bibinfo {volume} {85}},\ \bibinfo {pages} {235151} (\bibinfo {year} {2012})}\BibitemShut {NoStop}%
\bibitem [{\citenamefont {Kitaev}(2003)}]{Kitaev:1997wr}%
  \BibitemOpen
  \bibfield  {author} {\bibinfo {author} {\bibfnamefont {A.~Yu.}\ \bibnamefont {Kitaev}},\ }\bibfield  {title} {\enquote {\bibinfo {title} {{Fault tolerant quantum computation by anyons}},}\ }\href {\doibase 10.1016/S0003-4916(02)00018-0} {\bibfield  {journal} {\bibinfo  {journal} {Annals Phys.}\ }\textbf {\bibinfo {volume} {303}},\ \bibinfo {pages} {2--30} (\bibinfo {year} {2003})},\ \Eprint {http://arxiv.org/abs/quant-ph/9707021} {arXiv:quant-ph/9707021} \BibitemShut {NoStop}%
\bibitem [{\citenamefont {Liu}(2023)}]{Liu:2023opg}%
  \BibitemOpen
  \bibfield  {author} {\bibinfo {author} {\bibfnamefont {Shang}\ \bibnamefont {Liu}},\ }\bibfield  {title} {\enquote {\bibinfo {title} {{Efficient Preparation of Nonabelian Topological Orders in the Doubled Hilbert Space}},}\ }\href@noop {} {\  (\bibinfo {year} {2023})},\ \Eprint {http://arxiv.org/abs/2311.18497} {arXiv:2311.18497 [quant-ph]} \BibitemShut {NoStop}%
\bibitem [{\citenamefont {Barkeshli}\ \emph {et~al.}(2019)\citenamefont {Barkeshli}, \citenamefont {Bonderson}, \citenamefont {Cheng},\ and\ \citenamefont {Wang}}]{SET}%
  \BibitemOpen
  \bibfield  {author} {\bibinfo {author} {\bibfnamefont {Maissam}\ \bibnamefont {Barkeshli}}, \bibinfo {author} {\bibfnamefont {Parsa}\ \bibnamefont {Bonderson}}, \bibinfo {author} {\bibfnamefont {Meng}\ \bibnamefont {Cheng}}, \ and\ \bibinfo {author} {\bibfnamefont {Zhenghan}\ \bibnamefont {Wang}},\ }\bibfield  {title} {\enquote {\bibinfo {title} {Symmetry fractionalization, defects, and gauging of topological phases},}\ }\href {\doibase 10.1103/PhysRevB.100.115147} {\bibfield  {journal} {\bibinfo  {journal} {Phys. Rev. B}\ }\textbf {\bibinfo {volume} {100}},\ \bibinfo {pages} {115147} (\bibinfo {year} {2019})}\BibitemShut {NoStop}%
\bibitem [{\citenamefont {Teo}\ \emph {et~al.}(2015)\citenamefont {Teo}, \citenamefont {Hughes},\ and\ \citenamefont {Fradkin}}]{TeoSET2015}%
  \BibitemOpen
  \bibfield  {author} {\bibinfo {author} {\bibfnamefont {Jeffrey~C.Y.}\ \bibnamefont {Teo}}, \bibinfo {author} {\bibfnamefont {Taylor~L.}\ \bibnamefont {Hughes}}, \ and\ \bibinfo {author} {\bibfnamefont {Eduardo}\ \bibnamefont {Fradkin}},\ }\bibfield  {title} {\enquote {\bibinfo {title} {Theory of twist liquids: Gauging an anyonic symmetry},}\ }\href {\doibase 10.1016/j.aop.2015.05.012} {\bibfield  {journal} {\bibinfo  {journal} {Annals of Physics}\ }\textbf {\bibinfo {volume} {360}},\ \bibinfo {pages} {349–445} (\bibinfo {year} {2015})}\BibitemShut {NoStop}%
\bibitem [{\citenamefont {Cong}\ \emph {et~al.}(2017)\citenamefont {Cong}, \citenamefont {Cheng},\ and\ \citenamefont {Wang}}]{Cong:2017ffh}%
  \BibitemOpen
  \bibfield  {author} {\bibinfo {author} {\bibfnamefont {Iris}\ \bibnamefont {Cong}}, \bibinfo {author} {\bibfnamefont {Meng}\ \bibnamefont {Cheng}}, \ and\ \bibinfo {author} {\bibfnamefont {Zhenghan}\ \bibnamefont {Wang}},\ }\bibfield  {title} {\enquote {\bibinfo {title} {{Hamiltonian and Algebraic Theories of Gapped Boundaries in Topological Phases of Matter}},}\ }\href {\doibase 10.1007/s00220-017-2960-4} {\bibfield  {journal} {\bibinfo  {journal} {Commun. Math. Phys.}\ }\textbf {\bibinfo {volume} {355}},\ \bibinfo {pages} {645--689} (\bibinfo {year} {2017})},\ \Eprint {http://arxiv.org/abs/1707.04564} {arXiv:1707.04564 [cond-mat.str-el]} \BibitemShut {NoStop}%
\bibitem [{\citenamefont {Tarantino}\ \emph {et~al.}(2016)\citenamefont {Tarantino}, \citenamefont {Lindner},\ and\ \citenamefont {Fidkowski}}]{TarantinoSET2016}%
  \BibitemOpen
  \bibfield  {author} {\bibinfo {author} {\bibfnamefont {Nicolas}\ \bibnamefont {Tarantino}}, \bibinfo {author} {\bibfnamefont {Netanel~H}\ \bibnamefont {Lindner}}, \ and\ \bibinfo {author} {\bibfnamefont {Lukasz}\ \bibnamefont {Fidkowski}},\ }\bibfield  {title} {\enquote {\bibinfo {title} {Symmetry fractionalization and twist defects},}\ }\href {http://dx.doi.org/10.1088/1367-2630/18/3/035006} {\bibfield  {journal} {\bibinfo  {journal} {New J. Phys.}\ }\textbf {\bibinfo {volume} {18}},\ \bibinfo {pages} {035006} (\bibinfo {year} {2016})}\BibitemShut {NoStop}%
\bibitem [{\citenamefont {Chen}\ \emph {et~al.}(2015)\citenamefont {Chen}, \citenamefont {Burnell}, \citenamefont {Vishwanath},\ and\ \citenamefont {Fidkowski}}]{SETanomaly}%
  \BibitemOpen
  \bibfield  {author} {\bibinfo {author} {\bibfnamefont {Xie}\ \bibnamefont {Chen}}, \bibinfo {author} {\bibfnamefont {F.~J.}\ \bibnamefont {Burnell}}, \bibinfo {author} {\bibfnamefont {Ashvin}\ \bibnamefont {Vishwanath}}, \ and\ \bibinfo {author} {\bibfnamefont {Lukasz}\ \bibnamefont {Fidkowski}},\ }\bibfield  {title} {\enquote {\bibinfo {title} {Anomalous symmetry fractionalization and surface topological order},}\ }\href {\doibase 10.1103/PhysRevX.5.041013} {\bibfield  {journal} {\bibinfo  {journal} {Phys. Rev. X}\ }\textbf {\bibinfo {volume} {5}},\ \bibinfo {pages} {041013} (\bibinfo {year} {2015})}\BibitemShut {NoStop}%
\bibitem [{\citenamefont {Barkeshli}\ and\ \citenamefont {Cheng}(2020)}]{BarkeshliAnomaly1}%
  \BibitemOpen
  \bibfield  {author} {\bibinfo {author} {\bibfnamefont {Maissam}\ \bibnamefont {Barkeshli}}\ and\ \bibinfo {author} {\bibfnamefont {Meng}\ \bibnamefont {Cheng}},\ }\bibfield  {title} {\enquote {\bibinfo {title} {{Relative anomalies in (2+1)D symmetry enriched topological states}},}\ }\href {\doibase 10.21468/SciPostPhys.8.2.028} {\bibfield  {journal} {\bibinfo  {journal} {SciPost Phys.}\ }\textbf {\bibinfo {volume} {8}},\ \bibinfo {pages} {028} (\bibinfo {year} {2020})}\BibitemShut {NoStop}%
\bibitem [{\citenamefont {Bulmash}\ and\ \citenamefont {Barkeshli}(2020)}]{BarkeshliAnomaly2}%
  \BibitemOpen
  \bibfield  {author} {\bibinfo {author} {\bibfnamefont {Daniel}\ \bibnamefont {Bulmash}}\ and\ \bibinfo {author} {\bibfnamefont {Maissam}\ \bibnamefont {Barkeshli}},\ }\bibfield  {title} {\enquote {\bibinfo {title} {Absolute anomalies in (2+1)d symmetry-enriched topological states and exact (3+1)d constructions},}\ }\href {\doibase 10.1103/PhysRevResearch.2.043033} {\bibfield  {journal} {\bibinfo  {journal} {Phys. Rev. Res.}\ }\textbf {\bibinfo {volume} {2}},\ \bibinfo {pages} {043033} (\bibinfo {year} {2020})}\BibitemShut {NoStop}%
\bibitem [{\citenamefont {Aasen}\ \emph {et~al.}(2022)\citenamefont {Aasen}, \citenamefont {Bonderson},\ and\ \citenamefont {Knapp}}]{aasen2022torsorial}%
  \BibitemOpen
  \bibfield  {author} {\bibinfo {author} {\bibfnamefont {David}\ \bibnamefont {Aasen}}, \bibinfo {author} {\bibfnamefont {Parsa}\ \bibnamefont {Bonderson}}, \ and\ \bibinfo {author} {\bibfnamefont {Christina}\ \bibnamefont {Knapp}},\ }\href@noop {} {\enquote {\bibinfo {title} {{Torsorial actions on G-crossed braided tensor categories}},}\ } (\bibinfo {year} {2022}),\ \Eprint {http://arxiv.org/abs/2107.10270} {arXiv:2107.10270 [math.QA]} \BibitemShut {NoStop}%
\bibitem [{\citenamefont {Cheng}\ and\ \citenamefont {Williamson}(2020)}]{ChengPRR2020}%
  \BibitemOpen
  \bibfield  {author} {\bibinfo {author} {\bibfnamefont {Meng}\ \bibnamefont {Cheng}}\ and\ \bibinfo {author} {\bibfnamefont {Dominic~J.}\ \bibnamefont {Williamson}},\ }\bibfield  {title} {\enquote {\bibinfo {title} {Relative anomaly in ($1+1$)d rational conformal field theory},}\ }\href {\doibase 10.1103/PhysRevResearch.2.043044} {\bibfield  {journal} {\bibinfo  {journal} {Phys. Rev. Res.}\ }\textbf {\bibinfo {volume} {2}},\ \bibinfo {pages} {043044} (\bibinfo {year} {2020})}\BibitemShut {NoStop}%
\bibitem [{\citenamefont {Fidkowski}\ and\ \citenamefont {Vishwanath}(2017)}]{FidkowskiPRB2017}%
  \BibitemOpen
  \bibfield  {author} {\bibinfo {author} {\bibfnamefont {Lukasz}\ \bibnamefont {Fidkowski}}\ and\ \bibinfo {author} {\bibfnamefont {Ashvin}\ \bibnamefont {Vishwanath}},\ }\bibfield  {title} {\enquote {\bibinfo {title} {Realizing anomalous anyonic symmetries at the surfaces of three-dimensional gauge theories},}\ }\href {\doibase 10.1103/PhysRevB.96.045131} {\bibfield  {journal} {\bibinfo  {journal} {Physical Review B}\ }\textbf {\bibinfo {volume} {96}},\ \bibinfo {pages} {045131} (\bibinfo {year} {2017})}\BibitemShut {NoStop}%
\bibitem [{\citenamefont {Lan}\ \emph {et~al.}(2016)\citenamefont {Lan}, \citenamefont {Kong},\ and\ \citenamefont {Wen}}]{LanPRB2016}%
  \BibitemOpen
  \bibfield  {author} {\bibinfo {author} {\bibfnamefont {Tian}\ \bibnamefont {Lan}}, \bibinfo {author} {\bibfnamefont {Liang}\ \bibnamefont {Kong}}, \ and\ \bibinfo {author} {\bibfnamefont {Xiao-Gang}\ \bibnamefont {Wen}},\ }\bibfield  {title} {\enquote {\bibinfo {title} {Theory of (2+1)-dimensional fermionic topological orders and fermionic/bosonic topological orders with symmetries},}\ }\href {\doibase 10.1103/PhysRevB.94.155113} {\bibfield  {journal} {\bibinfo  {journal} {Phys. Rev. B}\ }\textbf {\bibinfo {volume} {94}},\ \bibinfo {pages} {155113} (\bibinfo {year} {2016})}\BibitemShut {NoStop}%
\bibitem [{\citenamefont {Walker}\ and\ \citenamefont {Wang}(2012)}]{WW}%
  \BibitemOpen
  \bibfield  {author} {\bibinfo {author} {\bibfnamefont {K.}~\bibnamefont {Walker}}\ and\ \bibinfo {author} {\bibfnamefont {Z.}~\bibnamefont {Wang}},\ }\href@noop {} {\bibfield  {journal} {\bibinfo  {journal} {Front. Phys. 7}\ ,\ \bibinfo {pages} {150}} (\bibinfo {year} {2012})}\BibitemShut {NoStop}%
\bibitem [{\citenamefont {von Keyserlingk}\ \emph {et~al.}(2013)\citenamefont {von Keyserlingk}, \citenamefont {Burnell},\ and\ \citenamefont {Simon}}]{Keyserlingk2013}%
  \BibitemOpen
  \bibfield  {author} {\bibinfo {author} {\bibfnamefont {C.~W.}\ \bibnamefont {von Keyserlingk}}, \bibinfo {author} {\bibfnamefont {F.~J.}\ \bibnamefont {Burnell}}, \ and\ \bibinfo {author} {\bibfnamefont {S.~H.}\ \bibnamefont {Simon}},\ }\bibfield  {title} {\enquote {\bibinfo {title} {Three-dimensional topological lattice models with surface anyons},}\ }\href {\doibase 10.1103/PhysRevB.87.045107} {\bibfield  {journal} {\bibinfo  {journal} {Phys. Rev. B}\ }\textbf {\bibinfo {volume} {87}},\ \bibinfo {pages} {045107} (\bibinfo {year} {2013})}\BibitemShut {NoStop}%
\bibitem [{\citenamefont {Bruillard}\ \emph {et~al.}(2017)\citenamefont {Bruillard}, \citenamefont {Galindo}, \citenamefont {Hagge}, \citenamefont {Ng}, \citenamefont {Plavnik}, \citenamefont {Rowell},\ and\ \citenamefont {Wang}}]{Bruillard_2017}%
  \BibitemOpen
  \bibfield  {author} {\bibinfo {author} {\bibfnamefont {Paul}\ \bibnamefont {Bruillard}}, \bibinfo {author} {\bibfnamefont {César}\ \bibnamefont {Galindo}}, \bibinfo {author} {\bibfnamefont {Tobias}\ \bibnamefont {Hagge}}, \bibinfo {author} {\bibfnamefont {Siu-Hung}\ \bibnamefont {Ng}}, \bibinfo {author} {\bibfnamefont {Julia~Yael}\ \bibnamefont {Plavnik}}, \bibinfo {author} {\bibfnamefont {Eric~C.}\ \bibnamefont {Rowell}}, \ and\ \bibinfo {author} {\bibfnamefont {Zhenghan}\ \bibnamefont {Wang}},\ }\bibfield  {title} {\enquote {\bibinfo {title} {Fermionic modular categories and the 16-fold way},}\ }\href {\doibase 10.1063/1.4982048} {\bibfield  {journal} {\bibinfo  {journal} {Journal of Mathematical Physics}\ }\textbf {\bibinfo {volume} {58}} (\bibinfo {year} {2017}),\ 10.1063/1.4982048}\BibitemShut {NoStop}%
\bibitem [{\citenamefont {Shi}(2020)}]{Shi2020Verlinde}%
  \BibitemOpen
  \bibfield  {author} {\bibinfo {author} {\bibfnamefont {Bowen}\ \bibnamefont {Shi}},\ }\bibfield  {title} {\enquote {\bibinfo {title} {Verlinde formula from entanglement},}\ }\href {\doibase 10.1103/PhysRevResearch.2.023132} {\bibfield  {journal} {\bibinfo  {journal} {Phys. Rev. Res.}\ }\textbf {\bibinfo {volume} {2}},\ \bibinfo {pages} {023132} (\bibinfo {year} {2020})}\BibitemShut {NoStop}%
\bibitem [{\citenamefont {Lessa}\ \emph {et~al.}(2024{\natexlab{b}})\citenamefont {Lessa}, \citenamefont {Cheng},\ and\ \citenamefont {Wang}}]{Lessa:2024wcw}%
  \BibitemOpen
  \bibfield  {author} {\bibinfo {author} {\bibfnamefont {Leonardo~A.}\ \bibnamefont {Lessa}}, \bibinfo {author} {\bibfnamefont {Meng}\ \bibnamefont {Cheng}}, \ and\ \bibinfo {author} {\bibfnamefont {Chong}\ \bibnamefont {Wang}},\ }\bibfield  {title} {\enquote {\bibinfo {title} {{Mixed-state quantum anomaly and multipartite entanglement}},}\ }\href@noop {} {\  (\bibinfo {year} {2024}{\natexlab{b}})},\ \Eprint {http://arxiv.org/abs/2401.17357} {arXiv:2401.17357 [cond-mat.str-el]} \BibitemShut {NoStop}%
\bibitem [{\citenamefont {Kim}\ \emph {et~al.}(2023)\citenamefont {Kim}, \citenamefont {Levin}, \citenamefont {Lin}, \citenamefont {Ranard},\ and\ \citenamefont {Shi}}]{Kim:2023ydi}%
  \BibitemOpen
  \bibfield  {author} {\bibinfo {author} {\bibfnamefont {Isaac~H.}\ \bibnamefont {Kim}}, \bibinfo {author} {\bibfnamefont {Michael}\ \bibnamefont {Levin}}, \bibinfo {author} {\bibfnamefont {Ting-Chun}\ \bibnamefont {Lin}}, \bibinfo {author} {\bibfnamefont {Daniel}\ \bibnamefont {Ranard}}, \ and\ \bibinfo {author} {\bibfnamefont {Bowen}\ \bibnamefont {Shi}},\ }\bibfield  {title} {\enquote {\bibinfo {title} {{Universal Lower Bound on Topological Entanglement Entropy}},}\ }\href {\doibase 10.1103/PhysRevLett.131.166601} {\bibfield  {journal} {\bibinfo  {journal} {Phys. Rev. Lett.}\ }\textbf {\bibinfo {volume} {131}},\ \bibinfo {pages} {166601} (\bibinfo {year} {2023})},\ \Eprint {http://arxiv.org/abs/2302.00689} {arXiv:2302.00689 [quant-ph]} \BibitemShut {NoStop}%
\bibitem [{\citenamefont {Zini}\ and\ \citenamefont {Wang}(2021)}]{Zini:2021lte}%
  \BibitemOpen
  \bibfield  {author} {\bibinfo {author} {\bibfnamefont {Modjtaba~Shokrian}\ \bibnamefont {Zini}}\ and\ \bibinfo {author} {\bibfnamefont {Zhenghan}\ \bibnamefont {Wang}},\ }\bibfield  {title} {\enquote {\bibinfo {title} {{Mixed-state TQFTs}},}\ }\href@noop {} {\  (\bibinfo {year} {2021})},\ \Eprint {http://arxiv.org/abs/2110.13946} {arXiv:2110.13946 [math.QA]} \BibitemShut {NoStop}%
\bibitem [{\citenamefont {Lan}\ \emph {et~al.}(2018)\citenamefont {Lan}, \citenamefont {Kong},\ and\ \citenamefont {Wen}}]{Lan3DTO1}%
  \BibitemOpen
  \bibfield  {author} {\bibinfo {author} {\bibfnamefont {Tian}\ \bibnamefont {Lan}}, \bibinfo {author} {\bibfnamefont {Liang}\ \bibnamefont {Kong}}, \ and\ \bibinfo {author} {\bibfnamefont {Xiao-Gang}\ \bibnamefont {Wen}},\ }\bibfield  {title} {\enquote {\bibinfo {title} {Classification of $\mathbf{(}3+1\mathbf{)}\mathrm{D}$ bosonic topological orders: The case when pointlike excitations are all bosons},}\ }\href {\doibase 10.1103/PhysRevX.8.021074} {\bibfield  {journal} {\bibinfo  {journal} {Phys. Rev. X}\ }\textbf {\bibinfo {volume} {8}},\ \bibinfo {pages} {021074} (\bibinfo {year} {2018})}\BibitemShut {NoStop}%
\bibitem [{\citenamefont {Lan}\ and\ \citenamefont {Wen}(2019)}]{Lan3DTO2}%
  \BibitemOpen
  \bibfield  {author} {\bibinfo {author} {\bibfnamefont {Tian}\ \bibnamefont {Lan}}\ and\ \bibinfo {author} {\bibfnamefont {Xiao-Gang}\ \bibnamefont {Wen}},\ }\bibfield  {title} {\enquote {\bibinfo {title} {Classification of $3+1\mathrm{D}$ bosonic topological orders (ii): The case when some pointlike excitations are fermions},}\ }\href {\doibase 10.1103/PhysRevX.9.021005} {\bibfield  {journal} {\bibinfo  {journal} {Phys. Rev. X}\ }\textbf {\bibinfo {volume} {9}},\ \bibinfo {pages} {021005} (\bibinfo {year} {2019})}\BibitemShut {NoStop}%
\bibitem [{\citenamefont {Johnson-Freyd}(2020)}]{JF2020}%
  \BibitemOpen
  \bibfield  {author} {\bibinfo {author} {\bibfnamefont {Theo}\ \bibnamefont {Johnson-Freyd}},\ }\href {https://arxiv.org/abs/2011.11165} {\enquote {\bibinfo {title} {(3+1)d topological orders with only a $\mathbb{Z}_2$-charged particle},}\ } (\bibinfo {year} {2020}),\ \Eprint {http://arxiv.org/abs/arXiv:2011.11165} {arXiv:2011.11165} \BibitemShut {NoStop}%
\bibitem [{\citenamefont {Chen}\ and\ \citenamefont {Hsin}(2021)}]{Chen:2021xks}%
  \BibitemOpen
  \bibfield  {author} {\bibinfo {author} {\bibfnamefont {Yu-An}\ \bibnamefont {Chen}}\ and\ \bibinfo {author} {\bibfnamefont {Po-Shen}\ \bibnamefont {Hsin}},\ }\bibfield  {title} {\enquote {\bibinfo {title} {{Exactly Solvable Lattice Hamiltonians and Gravitational Anomalies}},}\ }\href@noop {} {\  (\bibinfo {year} {2021})},\ \Eprint {http://arxiv.org/abs/2110.14644} {arXiv:2110.14644 [cond-mat.str-el]} \BibitemShut {NoStop}%
\bibitem [{\citenamefont {Fidkowski}\ \emph {et~al.}(2022)\citenamefont {Fidkowski}, \citenamefont {Haah},\ and\ \citenamefont {Hastings}}]{Fidkowski:2021unr}%
  \BibitemOpen
  \bibfield  {author} {\bibinfo {author} {\bibfnamefont {Lukasz}\ \bibnamefont {Fidkowski}}, \bibinfo {author} {\bibfnamefont {Jeongwan}\ \bibnamefont {Haah}}, \ and\ \bibinfo {author} {\bibfnamefont {Matthew~B.}\ \bibnamefont {Hastings}},\ }\bibfield  {title} {\enquote {\bibinfo {title} {{Gravitational anomaly of (3+1)-dimensional Z2 toric code with fermionic charges and fermionic loop self-statistics}},}\ }\href {\doibase 10.1103/PhysRevB.106.165135} {\bibfield  {journal} {\bibinfo  {journal} {Phys. Rev. B}\ }\textbf {\bibinfo {volume} {106}},\ \bibinfo {pages} {165135} (\bibinfo {year} {2022})},\ \Eprint {http://arxiv.org/abs/2110.14654} {arXiv:2110.14654 [cond-mat.str-el]} \BibitemShut {NoStop}%
\bibitem [{\citenamefont {Sohal}\ and\ \citenamefont {Prem}(2024)}]{Sohal:2024qvq}%
  \BibitemOpen
  \bibfield  {author} {\bibinfo {author} {\bibfnamefont {Ramanjit}\ \bibnamefont {Sohal}}\ and\ \bibinfo {author} {\bibfnamefont {Abhinav}\ \bibnamefont {Prem}},\ }\bibfield  {title} {\enquote {\bibinfo {title} {{A Noisy Approach to Intrinsically Mixed-State Topological Order}},}\ \ }(\bibinfo {year} {2024})\ \Eprint {http://arxiv.org/abs/2403.13879} {arXiv:2403.13879 [cond-mat.str-el]} \BibitemShut {NoStop}%
\bibitem [{\citenamefont {Lessa}\ \emph {et~al.}()\citenamefont {Lessa}, \citenamefont {Sang}, \citenamefont {Lu}, \citenamefont {Hsieh},\ and\ \citenamefont {Wang}}]{Lessa_unpub}%
  \BibitemOpen
  \bibfield  {author} {\bibinfo {author} {\bibfnamefont {Leonardo}\ \bibnamefont {Lessa}}, \bibinfo {author} {\bibfnamefont {Shengqi}\ \bibnamefont {Sang}}, \bibinfo {author} {\bibfnamefont {Tsung-Cheng}\ \bibnamefont {Lu}}, \bibinfo {author} {\bibfnamefont {Tim}\ \bibnamefont {Hsieh}}, \ and\ \bibinfo {author} {\bibfnamefont {Chong}\ \bibnamefont {Wang}},\ }\href@noop {} {}\bibinfo {note} {To appear}\BibitemShut {NoStop}%
\bibitem [{\citenamefont {Zhang}\ \emph {et~al.}()\citenamefont {Zhang}, \citenamefont {Xu}, \citenamefont {Zhang}, \citenamefont {Xu}, \citenamefont {Bi},\ and\ \citenamefont {Luo}}]{Luo_unpub}%
  \BibitemOpen
  \bibfield  {author} {\bibinfo {author} {\bibfnamefont {Carolyn}\ \bibnamefont {Zhang}}, \bibinfo {author} {\bibfnamefont {Yichen}\ \bibnamefont {Xu}}, \bibinfo {author} {\bibfnamefont {Jian-Hao}\ \bibnamefont {Zhang}}, \bibinfo {author} {\bibfnamefont {Cenke}\ \bibnamefont {Xu}}, \bibinfo {author} {\bibfnamefont {Zhen}\ \bibnamefont {Bi}}, \ and\ \bibinfo {author} {\bibfnamefont {Zhu-Xi}\ \bibnamefont {Luo}},\ }\href@noop {} {\enquote {\bibinfo {title} {Strong-to-weak spontaneous breaking of 1-form symmetry and intrinsically mixed topological order},}\ }\bibinfo {note} {To appear}\BibitemShut {NoStop}%
\bibitem [{\citenamefont {Cheng}(2019)}]{FermionLSM}%
  \BibitemOpen
  \bibfield  {author} {\bibinfo {author} {\bibfnamefont {Meng}\ \bibnamefont {Cheng}},\ }\bibfield  {title} {\enquote {\bibinfo {title} {{Fermionic Lieb-Schultz-Mattis theorems and weak symmetry-protected phases}},}\ }\href {\doibase 10.1103/PhysRevB.99.075143} {\bibfield  {journal} {\bibinfo  {journal} {Phys. Rev. B}\ }\textbf {\bibinfo {volume} {99}},\ \bibinfo {pages} {075143} (\bibinfo {year} {2019})}\BibitemShut {NoStop}%
\bibitem [{\citenamefont {Deligne}(2002)}]{Deligne}%
  \BibitemOpen
  \bibfield  {author} {\bibinfo {author} {\bibfnamefont {P.}~\bibnamefont {Deligne}},\ }\bibfield  {title} {\enquote {\bibinfo {title} {Categories tensorielles},}\ }\href@noop {} {\bibfield  {journal} {\bibinfo  {journal} {Moscow Math. J.}\ }\textbf {\bibinfo {volume} {2}},\ \bibinfo {pages} {227–248} (\bibinfo {year} {2002})}\BibitemShut {NoStop}%
\bibitem [{\citenamefont {Wang}\ and\ \citenamefont {Chen}(2017)}]{Wang:2016rzy}%
  \BibitemOpen
  \bibfield  {author} {\bibinfo {author} {\bibfnamefont {Zitao}\ \bibnamefont {Wang}}\ and\ \bibinfo {author} {\bibfnamefont {Xie}\ \bibnamefont {Chen}},\ }\bibfield  {title} {\enquote {\bibinfo {title} {{Twisted gauge theories in three-dimensional Walker-Wang models}},}\ }\href {\doibase 10.1103/PhysRevB.95.115142} {\bibfield  {journal} {\bibinfo  {journal} {Phys. Rev. B}\ }\textbf {\bibinfo {volume} {95}},\ \bibinfo {pages} {115142} (\bibinfo {year} {2017})},\ \Eprint {http://arxiv.org/abs/1611.09334} {arXiv:1611.09334 [cond-mat.str-el]} \BibitemShut {NoStop}%
\bibitem [{\citenamefont {Johnson-Freyd}\ and\ \citenamefont {Reutter}(2024)}]{johnson2024minimal}%
  \BibitemOpen
  \bibfield  {author} {\bibinfo {author} {\bibfnamefont {Theo}\ \bibnamefont {Johnson-Freyd}}\ and\ \bibinfo {author} {\bibfnamefont {David}\ \bibnamefont {Reutter}},\ }\bibfield  {title} {\enquote {\bibinfo {title} {Minimal nondegenerate extensions},}\ }\href@noop {} {\bibfield  {journal} {\bibinfo  {journal} {Journal of the American Mathematical Society}\ }\textbf {\bibinfo {volume} {37}},\ \bibinfo {pages} {81--150} (\bibinfo {year} {2024})}\BibitemShut {NoStop}%
\bibitem [{\citenamefont {Drinfeld}\ \emph {et~al.}(2010)\citenamefont {Drinfeld}, \citenamefont {Gelaki}, \citenamefont {Nikshych},\ and\ \citenamefont {Ostrik}}]{DGNO}%
  \BibitemOpen
  \bibfield  {author} {\bibinfo {author} {\bibfnamefont {Vladimir}\ \bibnamefont {Drinfeld}}, \bibinfo {author} {\bibfnamefont {Shlomo}\ \bibnamefont {Gelaki}}, \bibinfo {author} {\bibfnamefont {Dmitri}\ \bibnamefont {Nikshych}}, \ and\ \bibinfo {author} {\bibfnamefont {Victor}\ \bibnamefont {Ostrik}},\ }\href@noop {} {\enquote {\bibinfo {title} {{On braided fusion categories I}},}\ } (\bibinfo {year} {2010}),\ \Eprint {http://arxiv.org/abs/arXiv:0906.0620} {arXiv:0906.0620} \BibitemShut {NoStop}%
\end{thebibliography}%
\end{document}